\documentstyle[10pt, makeidx]{report}

\def \Oeuvres{O$\!$euvres}
\def \Plv{Pain\-le\-v\'e}
\def \KdV {Kor\-te\-weg-de Vries}

\def\CRAS{C.~R.~Acad.~Sc.~Paris}

\def\PTP{Prog.~Theor.~Phys.~}
\def\SAM{Stud.~Appl.~Math.~}
\def \ie {i.~e.~} 

\def\pasq{q} 
\def\pas{h}  
\def\Pas{H}  

\def \eds{\'equations diff\'erentielles}
\def \D {\hbox{d}}
\def \Log {\mathop{\rm Log}\nolimits}

\def \sinh{\mathop{\rm sinh}\nolimits}
\def \cotanh{\mathop{\rm cotanh}\nolimits}
\def \sech{\mathop{\rm sech}\nolimits}

\def \sn  {\mathop{\rm sn}\nolimits}

\def \grad{\mathop{\rm grad}\nolimits}
\def \mod#1{\vert #1 \vert}

\def \bfA {{\bf A}}
\def \bfE {{\bf E}}
\def \bfK {{\bf K}}

\def \bfP {{\bf P}}
\def \bfQ {{\bf Q}}
\def \bfR {{\bf R}}
\def \bfU {{\bf U}}

\def \bfu {{\bf u}}
\def \bfx {{\bf x}}
\def \bfp {{\bf p}}
\def \bfq {{\bf q}}
\def \bfv {{\bf v}}

\def \bsr {(b,\sigma,r)}

\newtheorem{Exercice}  {Exercise}  [chapter]

\def\Solution  {\smallskip \noindent{\it Solution}}

\def\Lecons {{\it Le\c cons}}

\makeindex

\begin{document}
\pagenumbering{arabic} 

\begin{center}
{\bf THE PAINLEV\'E APPROACH TO NONLINEAR ORDINARY DIFFERENTIAL EQUATIONS}
\end{center}

\vskip 0.5 truecm

{\bf R.~Conte}

Service de physique de l'\'etat condens\'e

Commissariat \`a l'\'energie atomique, Saclay

F--91191 Gif-sur-Yvette Cedex

\vskip 0.5 truecm

\begin{center}
{\it Proceedings of the Carg\`ese school (3--22 June 1996)}
\\
{\it La propri\'et\'e de Painlev\'e, un si\`ecle apr\`es}
\\
{\it The Painlev\'e property, one century later}
\\
\end{center}




\vskip 0.5 truecm
{\it Abstract}.
The ``Painlev\'e analysis'' is quite often perceived as a collection of tricks
reserved to experts.
The aim of this course is to demonstrate the contrary
and to unveil the simplicity and the beauty of a subject which is in fact
{\it the} theory of the (explicit) integration of nonlinear differential
equations.

To achieve our goal,
we will {\it not} start the exposition with a more or less precise
``Painlev\'e test''.
On the contrary, we will finish with it, after a gradual introduction
to the rich world of singularities of nonlinear differential equations,
so as to remove any cooking recipe.

The emphasis is put on embedding each method of the test into the well
known theorem of perturbations of Poincar\'e.
A summary can be found at the beginning of each chapter.

\vfill
{\it The Painlev\'e property, one century later},
ed.~R.~Conte, 
CRM series in mathematical physics (Springer--Verlag, Berlin, 1998)
\hfill 24~October~1997

\hfill S97/103 \hskip 1 truecm solv-int/9710020

\tableofcontents

\chapter{Introduction}
\indent

This course is a major revision of a previous one delivered in Chamonix
\cite{Chamonix1994}.
Let us start with a few realistic applications.

\section{A few elementary examples}
\indent

\subsection{Linearization of the Riccati equation}
\indent

\index{Riccati equation}
The Riccati equation
\begin{equation}
\label{eqRiccatiG} 
 u'=a_2(x) u^2 + a_1(x) u + a_0(x),\ a_2\not=0,\
\hbox{ with } '={\D \over \D x}
\end{equation}
is known to be linearizable,
but how to retrieve the explicit formula which maps it onto a linear equation?
One just looks for the {\it first} coefficient $u_0$ of an expansion for $u$
which describes the dependence on the integration constant (a simple pole),
\index{Laurent series}
\ie\ a Laurent series in some expansion function $\varphi(x)$~:
$u=\varphi^{-1} (u_0 + u_1 \varphi + \dots)$.
It is given by balancing the two lowest degree terms
$ - u_0 \varphi' \varphi^{-2} = a_2 (u_0 \varphi^{-1})^2$,
and the linearizing transformation is then simply
the change of function $u \to \varphi$ defined by
the {\it singular part transformation} \index{singular part transformation}
$u=u_0 \varphi^{-1}$
\begin{equation}
\label{eqRiccatiL} 
 u=-{\varphi' \over a_2 \varphi} ,\ 
\varphi'' - \left[{a_2' \over a_2} + a_1 \right] \varphi' + a_0 a_2 \varphi=0.
\end{equation}
This will be justified at the end of this course,
in section \ref{sectionSingularPartTransfo}.

\subsection{A first integral of the Lorenz model}
\indent

\index{Lorenz model}
The Lorenz model of atmospheric circulation
\begin{eqnarray}
& & 
{\D x \over \D t} = \sigma (y-x),\
{\D y \over \D t} = r x - y - x z,\
{\D z \over \D t} = x y - b z (x-y)
\label{eqLorenz}
\end{eqnarray}
admits for $\bsr=(0,1/3,\hbox{arbitrary})$ the first integral
\begin{equation}
\left[- {3 \over 4} x^4 + {4 \over 3} x (y-x) + (z-r+1) x^2\right]
 e^{4 t / 3}=K, 
\end{equation}
but can one go further,
\ie can one obtain more first integrals or even explicitly integrate?
The answer is yes \cite{Segur}.
By elimination of $(y,z)$, one first builds the second order equation for 
$x(t)$
\begin{equation}
{\D^2 x \over \D t^2} = {1 \over x} \left({\D x \over \D t}\right)^2
 - {x^3 \over 4}
 - {K \over 3 x} e^{-4 t / 3}.
\end{equation}
For $K=0$ this equation admits the first integral
\begin{equation}
{1 \over x^2} \left({\D x \over \D t}\right)^2 + {x^2 \over 4} = A^2,\
\end{equation}
and the general solution $x=(1/(2A)) \cosh(t-t_0)$.
For $K\not=0$ this equation for $x(t)$ is equivalent, as shown below,
to the following equation for $X(T)$
\begin{equation}
X''
=
{X'^2 \over X} - {X' \over T} + {\alpha X^2 + \beta \over T} + \gamma X^3
+ {\delta \over X},\ (\alpha,\beta,\gamma,\delta) \hbox{ constant},
\end{equation}
which is the third of six irreducible equations
discovered between 1900 and 1906 by Paul Prudent Painlev\'e and his
student Bertrand Gambier
and, according to the theory of Painlev\'e developed in these lectures,
the integration is then achieved (``parfaite'', says Painlev\'e).
Two words may not be familiar to the reader~: ``equivalent'' and 
``irreducible''.
``Equivalent'' means that the transformation law from the physical variables
$(x,t)$ to the variables $(X,T)$ which satisfy (P3) should not result from a 
good guess,
but should be looked for within a precise set of transformations 
(mathematically the homographic transformations 
(\ref{eqGroupHomographicContinuous}) defined section \ref{sectionGroups})
designed so as not to alter the structure of singularities
(poles, branch points, \dots).
In this case, one finds that the transformation
\begin{equation}
x=a(t) X,\ T=\tau(t),\
\hbox{ with }
a={2 i c \over 3} e^{-{t/3}},\
\tau=c e^{-{t/3}},\
c^4={27 \over 4} K,\
\end{equation}
maps the equation for $x(t)$ to the equation (P3) for $X(T)$
with the parameter values for (P3) $\alpha= \beta=0,\ \gamma=\delta=1$.

\index{irreducibility}
As to ``irreducible'',
it means that there exist no transformation,
again within a precise class 
(Drach, Umemura, see section \ref{sectionIrreducibility}),
reducing any of the six (Pn) equations either to a linear equation
or to a first order equation.
Consequently,
the general solution of (Pn) has no ``explicit expression'',
it is just {\it defined by the equation itself}.
There is absolutely no difference between
defining the ``exponential function'' from the general solution of $u'=u$ and
defining the ``P3 function'' from the general solution of the equation (P3).

\subsection{A reduction of the Boussinesq equation}
\indent

The Boussinesq equation of fluid mechanics
\begin{equation}
 u_{tt} + \left[u^2 + {1 \over 3} u_{xx} \right]_{xx}=0,
\end{equation}
in the stationary case where $u$ does not depend on time,
reduces to an ordinary differential equation (EDO) 
which admits two first integrals
\begin{equation}
\label{eqBoussinesqStationary}
u'' + 3 u^2 + K_1 x + K_2 =0
\end{equation}
and, depending on $K_1$, this ODE is equivalent either to the (P1) equation
\begin{equation}
u'' = 6 u^2 + x
\end{equation}
or, after one more integration, to an equation introduced by Weierstrass,
the {\it elliptic equation} \index{elliptic equation}
\begin{equation}
\label{eqWeierstrass}
u'^2 = 4 u^3 - g_2 u - g_3,\ (g_2,g_3) \hbox{ complex constants}.
\end{equation}
Both equations have a general solution single valued in the whole complex plane.

\section{``Solvable'' models, ``integrable'' equations and so on}
\indent

Two main fields contributed to the recent interest for the Painlev\'e theory.
The first one is statistical physics.
When Ising solved his one-dimensional model and found the partition function
$F=-(1 / \beta) \Log (2 \cosh(\beta J)),
\beta=k_{B} T$,
the result was {\it a posteriori} not surprising.
But, when Barouch, McCoy and Wu \cite{ItzyksonDrouffe}
expressed the correlation function of the
two-dimensional Ising model with a (P3) function,
this strongly contributed to revive the interest for these six functions,
which now appear in any ``solvable model'' of statistical physics
(see Di Francesco, this volume).
Retrospectively, the $\cosh$ function of Ising is a quite elementary output
of the Painlev\'e theory.
\medskip

The second field is that of partial differential equations (PDE),
as shown by the above Boussinesq example.
After the extension of the Fourier transform to nonlinear PDEs \cite{GGKM},
called {\it inverse spectral transform} (IST),
\index{inverse spectral transform}
Ablowitz and Segur \cite{AS1977}
noticed a link between those ``IST--integrable PDEs''
and the theory of Painlev\'e,
link expressed by Ablowitz, Ramani and Segur \cite{ARS1980} 
as the conjecture~: 
``Every ODE obtained by an exact reduction of a nonlinear PDE solvable
by the IST method has the Painlev\'e property''.
\index{Painlev\'e property}
For more details, see the book by Ablowitz and Clarkson
\cite{AblowitzClarkson}
and \cite{Musette}.

\section{Insufficiency of quadratures, the need for a theory}
\label{sectionPendulum}
\indent

Let us return to our main subject, the explicit, analytic integration of ODEs.
An exceptionally clear introduction {\it ad usum Delphini} is the
{\it Le\c con d'ouverture} (\Oeuvres\ vol.~I p.~199)
given by Painlev\'e in 1895 before starting the
{\it Le\c cons de Stockholm}.
For centuries, the question of integration has been formulated as~:
find enough first integrals in order to reduce the problem to a sequence of
{\it quadratures}. \index{quadratures}
But even the simple example of the pendulum shows the insufficiency of this
point of view.
Its motion is reducible to a quadrature defined by the integral
\begin{equation}
t - t_0 = \int_{u_0}^u {\D u \over \sqrt{(1-u^2)(1-k^2 u^2)}},\
k \hbox{ constant},
\end{equation}
giving the time $t$ as a ``function'' of the position $u$.
However, this {\it elliptic integral} does not provide the desired result,
\ie the position as a function of time,
and, worse, nothing ensures the existence of such an expression.
This classical problem (the inversion of the elliptic integral) could be
solved by Abel and Jacobi only by going to the complex domain,
leading to a unique value $u(t,t_0,u_0)=u_0 + \sn(t-t_0,k)$.
The symbol $\sn$ does deserve the name of {\it function} \index{function}
(this is one of the twelve Jacobi elliptic functions, equivalent to the unique
Weierstrass function)
because, for any complex $k$, the application $t \to \sn(t,k)$ is
single valued.

Following an idea of Briot and Bouquet, this led Painlev\'e to remark
({\it Le\c con} no.~1)~:
``Mais l'importance de cette classe d'\'equations 
[\`a solution g\'en\'erale uniforme]
appara\^\i t mieux encore si on observe que 
la plupart des transcendantes
auxiliaires,
dont le r\^ole est si consid\'erable
(fonctions exponentielle, elliptiques, fuchsiennes, etc),
int\`egrent des \'equations diff\'erentielles alg\'ebriques tr\`es simples.
Les \eds\ apparaissent donc comme la source des transcendantes uniformes 
les plus remarquables,
susceptibles notamment de contribuer \`a l'int\'egration d'autres \eds\
dont l'int\'egrale n'est plus uniforme.''

This is the famous ``double interest'' of differential equations~:
one may consider them either as the source for defining new functions,
or as a class of equations to be integrated with the existing functions
available.

\section{What can ``to integrate'' mean? The Painle\-v\'e property}
\label{sectionWhatIntegrate}
\indent

Any converging Taylor series defined on some part of the real line,
representing for instance a solution of an ODE on some interval
$-R <x <R$,
defines in fact an analytic function inside the disk $\mod{x}<R$.
Therefore, even when their variables are real,
differential equations and their solutions
are naturally defined in the complex plane.

To integrate an ODE is to acquire a global knowledge of its general solution,
not only the local knowledge ensured by the existence theorem of Cauchy.
So, the most demanding possible definition for the ``integrability'' of an ODE
is the single valuedness of its general solution,
so as to adapt this solution to any kind of initial conditions.
Since even linear equations may fail to have this property,
e.g.~$2 u' + x u=0,\ u=c x^{-1/2}$,
a more reasonable definition is the following one.

\index{uniformization}
\index{Painlev\'e property}
{\it Definition}.
The {\it Painlev\'e property} (PP) of an ODE is the uniformizability of
its general solution.

\index{stable equation}\index{stability}
Following Bureau \cite{Bureau1939},
we will call {\it stable} an equation with the PP.
In the above example, uniformization is achieved for instance by removing
from the complex plane any line joining the two branch points $0$ and 
$\infty$.

\section{Singularities of ordinary differential equations}
\indent

``Les fonctions, comme les \^etres vivants, sont caract\'eris\'ees par leurs
singularit\'es'' (Paul Montel).
Singularities are responsible for the limitation of the domain of validity of
Taylor or Laurent expansions, 
so their study is mandatory.

There exits a deep difference between the singularities of solutions of 
differential equations according as whether these equations are linear or
nonlinear.
In the linear case, the general solution (GS) has no other singularities than
those of the coefficients of the equation once solved for the highest
derivative.
These singularities have a location independent of the arbitrary coefficients
of integration and they are called {\it fixed}. \index{fixed singularity}

On the contrary,
solutions of nonlinear ODEs may have other singularities,
then called {\it movable}, \index{movable singularity}
at locations depending on the arbitrary coefficients.
Thus, the equations \cite{PaiLecons} 
\begin{eqnarray}
& &
{\D u \over \D x} + {u \over x^2} =0,\
u=c e^{1 / x},\
\\
& &
{\D u \over \D x} + {u^2 \over x} =0,\
u={1 \over c + \Log x},\
\\
& &
{\D u \over \D x} - {\sqrt{1 - u^2} \over x} =0,\
u=\sin (c + \Log x),\
\end{eqnarray}
where $c$ is the arbitrary constant of integration,
all have a fixed singularity in their general solution at $x=0$
(isolated essential singular point for the first one,
logarithmic branch point for the two others).
In addition,
among the last two ones, which are nonlinear,
the second one has movable simple poles,
and the third one has no movable singularity.
All three have the PP.

The possible singularities of differential equations have been classified by
Mittag-Leffler~: 
in addition to the familiar ones
(poles, branch points, essential singular points),
there can exist essential singular lines, analytic or not,
or perfect sets of singular points,
as illustrated by the Fuchsian and Kleinean functions of Poincar\'e.
One example is Chazy's equation of class III
\begin{eqnarray}
& &
u''' - 2 u u'' + 3 u'^2=0,
\label{eqChazyIII}
\end{eqnarray}
whose general solution is only defined inside or outside a circle
characterized by the three initial conditions (two for the center, one for the
radius);
this solution is holomorphic in its domain of definition and cannot be
analytically continued beyond it.
This equation therefore has the PP,
and the only singularity is a movable analytic essential singular line
which is a natural boundary.

\section{Outline and basic references}
\indent

For the outline, we refer the reader to the detailed table of contents;
the choice made is to develop the construction of necessary conditions,
at the expense of the explicit integration methods,
only briefly introduced in chapter \ref{chapterSufficiency}.

The basic texts are due to \Plv\ and his students and we abbreviate their
references as
{\it Le\c cons} (\cite{PaiLecons} {\it Le\c cons de Stockholm}, 1895,
 a high level course on nonlinear ODEs),
BSMF (\cite{PaiBSMF} first memoir, 1900, on first and second order ODEs),
Acta (\cite{PaiActa} second memoir, 1902,
 on second and higher order ODEs),
CRAS (\cite{PaiCRAS1906}, 1906, an addendum after Gambier discovered the 
functions (P4), (P5), (P6)),
Gambier (\cite{GambierThese} th\`ese, 1909, on second order ODEs),
Chazy   (\cite{ChazyThese}   th\`ese, 1910, on third and higher order ODEs),
Garnier (\cite{GarnierThese} th\`ese, 1911, on higher order ODEs).
Most \Plv\ works are reprinted in O$\!$euvres (\cite{PaiOeuvres} three volumes
1973, 74, 76, again available from CNRS-\'Editions,
e-mail editions@edition.cnrs.fr).
For a global overview of these results, see the book of Hille \cite{Hille}.
preferably to the one of Ince \cite{Ince}.
For a detailed exposition (indeed, in the classical period, it was kind of
fashionable to avoid details) and additional results,
see the three memoirs of Bureau 
M.~I \cite{BureauMI}, M.~II \cite{BureauMII}, M.~III \cite{BureauMIII}.
Peter Clarkson maintains an extensive bibliography \cite{ClarksonList}
covering both the classical and the recent period,
reproduced in \cite{AblowitzClarkson}.

\chapter{The meromorphy assumption}
\indent

\section{Specificity of the elliptic function}
\indent

A very deep result of L.~Fuchs, 
Poincar\'e (\cite{Poincare1885}, cf.~\Oeuvres\ de Painlev\'e III p.~189)
and Painlev\'e ({\it Le\c con}\ no.~7 p.~107)
is that
the class of first order ODEs $F(u',u,x)=0$,
with $F$ polynomial in $u'$ and $u$, analytic in $x$,
defines one and only one function,
from the general solution of (\ref{eqWeierstrass}).
This function is not historically new since this is precisely
the {\it elliptic function} \index{elliptic function}
$\wp$ introduced earlier by Weierstrass,
\ie the particular solution of
\begin{equation}
\wp'^2
= 4 \wp^3 - g_2 \wp - g_3
=4(\wp-e_1)(\wp-e_2)(\wp-e_3),\
(g_2,g_3,e_{\alpha}) \in {\cal C},
\end{equation}
which admits a pole at the origin
\begin{equation}
\wp(x,g_2,g_3)=x^{-2} + {g_2 \over 20} x^2 + {g_3 \over 28} x^4 + O(x^6).
\label{eqwpLaurent}
\end{equation}

The novelty of $\wp$ is elsewhere~:
this is the transcendental dependence of the general solution
$\wp(x-x_0,g_2,g_3)$ on the arbitrary constant $x_0$,
which makes impossible the reducibility of the elliptic equation to a linear
equation.
Among the many nice properties of elliptic functions
(see e.g.~\cite{AbramowitzStegun}),
the most interesting to us is their structure of singularities.
These are doubly periodic meromorphic functions
(which is their usual definition),
and there exists an {\it entire} \index{entire} function $\sigma$,
\ie without any singularity at a finite distance,
whose $-\wp$ is the second logarithmic derivative
\begin{equation}
\wp=-{\D \over \D x} \zeta,\
\zeta={\D \over \D x} \Log \sigma,\
\zeta''^2 + 4 \zeta'^3 - g_2 \zeta' + g_3 =0.
\end{equation}
Therefore the only singularities of the general solution $\wp(x-x_0,g_2,g_3)$ 
of (\ref{eqWeierstrass})
come from the zeroes of $\sigma$
and are a lattice of movable double poles located at
$x_0 + 2 m \omega + 2 n \omega',$
with $m$ and $n$ integers, $\omega,\omega'$ the two half-periods.

\section{The meromorphy assumption}
\label{sectionHoyer}
\indent

\index{meromorphy assumption}
\index{Laurent series}

The Laurent expansion (\ref{eqwpLaurent}) certainly motivated two students
of Weierstrass, Paul Hoyer and Sophie Kowalevski,
to investigate further the possibility for the GS of an ODE
to be represented by a Laurent series with a finite {\it principal part},
\index{principal part of a Laurent series}
so as to exclude essential singularities.
This meromorphy assumption, briefly said, consists in checking the existence
of the Laurent series and its ability to represent the GS,
\ie to contain enough arbitrary parameters.
But, since a Laurent series is only defined inside its annulus of convergence,
this study is only local and it cannot dispense of a further study in order to
explicitly integrate, using completely different means.

The first attempt is due to Hoyer in 1879 \cite{Hoyer} with the system
\begin{equation}
{\D \over \D t}
\pmatrix{x_1 \cr x_2 \cr x_3 \cr}
=
\pmatrix
{
a_1 & a_2 & a_3 \cr 
b_1 & b_2 & b_3 \cr 
c_1 & c_2 & c_3 \cr
} 
\pmatrix{x_2 x_3 \cr x_3 x_1 \cr x_1 x_2 \cr},
\end{equation}
under the restriction that neither the determinant nor any of its first or
second order diagonal minors vanishes;
he even generalized the assumption to the Puiseux series
$$
x_i=\sum_{j=0}^{+ \infty} A_{ij} \{(t-t_0)^{1 \over r}\}^{-n+j},\ i=1,2,3,
$$
with $n$ and $r$ positive integers and $(A_{10},A_{20},A_{30})\not=(0,0,0)$,
but in fact the numerous cases of integrability by elliptic functions which
he discovered were found by a direct Ansatz,
and not as necessary conditions for the Laurent series to exist.
Continued by Kowalevski with a quite similar system
except that it is six-dimensional, 
which will be seen section \ref{sectionToupie},
the method will only get its final shape with Gambier in 1910
(see pages 9 and 49 of his thesis).

\section{A flavor of the meromorphy test}
\label{sectionCasScalaire}
\indent

\index{meromorphy test}

We must warn the reader that this section is {\it not} the algorithm to apply,
but just a flavor of it;
the final algorithm will only be given section \ref{sectionTest}.

Let us start with a single equation;
the case of a system is not different, apart from technical complications.

Consider the equation
\begin{equation}
E(x,u) \equiv - {\D^2 u \over \D x^2} + 6 u^2 + g(x) = 0,
\label{eqCompleteP1}
\end{equation}
with $g$ analytic.

Assume that $u$ has a polar behavior at some location $x_0$ distinct from any
of the possible singularities of the coefficients of the equation, 
here $g(x)$;
such a pole is therefore movable.
\index{Laurent series}
One has to check the existence of {\it all} possible Laurent series 
with a finite principal part
\begin{equation}
u=\sum_{j=0}^{+ \infty} u_j \chi^{j+p},\ 
\chi=x-x_0,\
u_0 \not = 0,
\label{eqPoleExpansion1}
\end{equation}
in which
$-p $ is the order of the pole, which must be an integer,
and the coefficients $u_j$ are independent of $x$.

After insertion of this series in (\ref{eqCompleteP1}),
which is polynomial in $u$ and its derivatives,
and replacement of $g(x)$ by its Taylor series in the neighborhood of $x_0$,
the left-hand side,
as a sum of Laurent series,
is itself a Laurent series with a finite principal part
\begin{eqnarray}
E &=&
   [p(p-1)u_0 \chi^{p-2} + (p+1)p u_1 \chi^{p-1}+\dots]
+6 [u_0^2 \chi^{2p} + 2 u_0 u_1 \chi^{2p+1}+\dots]
\nonumber \\ & &
+  [g(x_0) + g'(x_0) \chi +\dots],
\end{eqnarray}
which we denote more generally 
\begin{equation}
E=\sum_{j=0}^{+ \infty} E_j \chi^{j+q},
\label{eqLaurentE}
\end{equation}
$q$ being the smallest integer of the list $(p-2,2p,0)$.
The method consists of expressing the conditions for this series to
identically vanish~: $\forall j \in {\cal N}\ :\ E_j=0$.
\medskip

{\it First step}.
Determine all possible {\it families of movable singularities} $(u_0,p)$.

This is expressed with three conditions~:
\begin{itemize}
\item
1) (condition $u_0 \not=0$)
equality of at least two elements of the list $(p-2,2p,0)$
($q$ denotes their common value),
the involved terms of $E$ being called {\it dominant} \index{dominant terms}
and denoted $\hat E$,

\item
2) (dominance condition)
inferiority of $q$ to the other elements of the list,

\item
3) (vanishing Laurent series condition)
vanishing of the coefficient $E_0$ of the lowest power $\chi^q,$
which involves only the dominant terms
\begin{equation}
E_0 \equiv \lim_{\chi \to 0} \chi^{-q} \hat E(x, u_0 \chi^p)=0,\ u_0\not=0,
\label{equ0}
\end{equation}

\end{itemize}
\ie respectively~:
one linear equation for $p$ by pair of terms considered,
several linear inequations for $p$,
one algebraic equation for $(u_0,p)$.

A necessary condition to prevent multivaluedness is then
\begin{itemize}
\item {\bf C0}. All possible values for $p$ are integer.
\end{itemize}

If there exists no family which is truly singular ($p$ negative),
the method stops without concluding.

Here,
the various possibilities for these linear equations and inequations are
\begin{eqnarray}
&
q=p-2=2p 
&
\hbox{ and } q\le 0,
\\
&
q=2p=0
&
\hbox{ and } q\le p-2,
\\
&
q=p-2=0
&
\hbox{ and } q\le 2p.
\end{eqnarray}
Their geometric representation is known as the
{\it Puiseux diagram}  \index{Puiseux diagram} 
or
{\it Newton's polygon} \index{Newton's polygon} 
(ref.~\cite{Hille} sec.~3.3, \cite{Ince} sec.~12.61).
The two solutions $(p,q)=(-2,-4)$, $(2,0)$ satisfy the condition {\bf C0} 
and the second one must be rejected, as being nonsingular.
So, the dominant part is here
$\hat E=-u'' + 6 u^2$.
The algebraic equation $E_0=0$
\begin{equation}
E_0 \equiv - 6 u_0 + 6 u_0^2 = 0,\ u_0 \not=0,
\end{equation}
has only one root $u_0=1$.

For $j=1,2,\dots$, each successive equation $E_j=0$ has then the form
\begin{equation}
\forall j\ge 1:\
E_j \equiv P(u_0,j) u_j + Q_j(\{u_l\ \vert \ l<j \}) = 0,
\label{eqRecurr0}
\end{equation}
here
\begin{eqnarray}
& &
P(u_0,j)=-(j-2)(j-3)+12 u_0=-(j+1)(j-6),
\\
& &
Q_1=0,\
Q_2=6 u_1^2,\
Q_3=12 u_1 u_2,
\\
\forall j\ge 4:\
& &
Q_j={g^{(j-4)}(x_0) \over (j-4)!} + 6 \sum_{k=1}^{j-1} u_k u_{j-k}. 
\end{eqnarray}
So the sequence $E_j=0, j \ge 1$, is just {\it one} linear equation with 
different right-hand sides,
and it can be solved recursively for $u_j$.
Whenever the positive integer $j$ is a zero of $P$, two subcases occur~:
either 
$Q_j$ does not vanish and the Laurent series does not exist,
or
$Q_j$ vanishes and $u_j$ is arbitrary.
Since $x_0$ is already arbitrary,
in order to represent the GS,
one wants $N-1$ additional arbitraries to enter the expansion,
where $N$ is the order of the ODE.
\index{index $-1$}
Let us admit for a moment that the value $j=-1$ is always a zero of $P$,
a result whose general proof (given section \ref{sectionMethodAlpha})
needs prerequisite notions of perturbation theory;
this value $j=-1$ will be seen to represent the arbitrary location of $x_0$.
Hence the following steps.
\medskip

{\it Second step}.
For each family,
determine the polynomial $P$ (do not compute $Q_j$ yet)
and require the necessary conditions~:

\label{pageC1}
\begin{itemize}
\item {\bf C1}. The polynomial $P$ has degree $N$.
\item {\bf C2}. $N-1$ zeroes of $P$ are positive integers.
\item {\bf C3}. The $N$ zeroes of $P$ are simple (\ie of multiplicity one).
\end{itemize}

If either {\bf C1}, {\bf C2} or {\bf C3} is violated, 
the method stops and one concludes to a failure
because the general solution cannot be meromorphic.

The zeroes of $P$ are called {\it indices} 
and $P=0$ itself is the {\it indicial equation}. 
\index{indicial equation}
Indeed, anticipating on the exposition of the general theory
sections \ref{sectionMethodBureau} and \ref{sectionMethodPerturbativeFuchsian},
they are the Fuchs indices $i$ near $\chi=0$
\index{Fuchs indices}
of a linear equation introduced by Darboux \cite{Darboux1883}
under the name ``\'equation auxiliaire'',
so the indicial equation is computed as follows \cite{FP1991}.
Take the derivative of $\hat E(x,u)$ with respect to $u$
\begin{equation}
\forall v:\
\hat E'(x,u) v \equiv \lim_{\lambda \to 0}
{\hat E(x, u+ \lambda v) - \hat E(x,u) \over \lambda},
\end{equation}
here
\begin{equation}
\hat E'(x,u) \equiv - \partial_x^2 + 12 u;
\end{equation}
evaluate this linear operator at point $u=u_0 \chi^p$,
which defines the ``auxiliary equation''
(\ie the linearized equation at the leading term) 
\index{linearized equation}
\begin{equation}
\forall v\ : \hat E'(x,u_0 \chi^p) v =0,
\label{eqAuxiliaryEquation}
\end{equation}
here
\begin{equation}
\forall v\ : \hat E'(x,u_0 \chi^{-2}) v 
  \equiv (- \partial_x^2 + 12 u_0 \chi^{-2}) v =0;
\end{equation}
establish the indicial equation of this linear ODE near its Fuchsian
singularity $\chi=0$
\begin{equation}
P(i) = \lim_{\chi \to 0} \chi^{-i-q} \hat E'(x,u_0 \chi^p) \chi^{i+p}=0,
\label{eqPolynomeCaracteristique}
\end{equation}
here
\begin{eqnarray}
P(i)& = & \lim_{\chi \to 0}
           \chi^{-i+4} (- \partial_x^2 + 12 u_0 \chi^{-2}) \chi^{i-2}
\nonumber \\
    & = & -(i-2)(i-3) + 12 u_0
\nonumber \\
    & = & -(i+1) (i-6).
\end{eqnarray}
The shift $i \to i+p$ in the above equation is just a convention
aimed at not producing an unfortunate difference between the Fuchs index $i$
and the index $j$ of the recursion relation $E_j=0$.

Now, one just has to check the existence of the Laurent series.

{\it Third step}.
For every positive integer zero $i$ of $P$ (a Fuchs index),
require the condition

\begin{itemize}
\item {\bf C4}. 

\begin{equation}
\forall i \in {\cal N},\ P(i)=0\ : Q_i=0.
\end{equation}

\end{itemize}

This is done by successively solving the recursion relation
up to the greatest positive integer Fuchs index.
As soon as a {\bf C4} condition is violated, one stops and concludes to
a failure~: the ODE has not the PP.
After the greatest positive integer Fuchs index has been checked,
the method is finished.

Here, one finds
\begin{equation}
 u_0=1,\ u_1=u_2=u_3=0,\ u_4=-{g_0 \over 10},\ u_5=-{g_0' \over 6},
\label{eqP1Laurent}
\end{equation}
and the condition {\bf C4} at index $i=6$ is
\begin{equation}
\label{eqMethodPole7}
Q_6 \equiv g_0''/2 =0.
\end{equation}
i.e., since $x_0$ is arbitrary, $g''=0$.
The ODE (\ref{eqCompleteP1}) is restricted to be (\ref{eqBoussinesqStationary})
which has been seen to have a meromorphic GS,
so in this case the generated necessary conditions are sufficient.

{\it Remarks}.
\begin{enumerate}

\item
We have retained the classical vocabulary
(``famille'' is used by Gambier, p.~38 of his thesis 
\cite{GambierThese},
``indices'' is used by Gambier, Chazy \cite{ChazyThese}
 and Bureau \cite{Bureau1939}),
rather than the one more recently introduced \cite{ARS1978,ARS1980}
(``branch'', ``resonances'').
Indeed, ``branch'' has another meaning in classical analysis,
where it denotes a determination of a multivalued application,
which may create some confusion.
As to ``resonance'', its identification with a basic notion of a linear theory,
the Fuchs indices, makes useless the introduction of such a term.

\item
Conditions $Q_i=0$ at Fuchs indices $i$ are often referred to as 
``no-logarithm conditions'' because,
if some of them are not satisfied,
there exists a generalization of the Laurent series,
called $\psi-$series (\cite{Hille} chap.~7),
\index{$\psi-$series}
which is a double expansion in $\chi$ and $\Log \chi$.
This series contains no logarithms (\ie reduces to the Laurent series)
iff all $Q_i$ vanish.

\item
One must prove that the radius $R$ of the punctured disk $\mod{x-x_0}<R$
in which the series converges is nonzero.

\item
As indicated by Gambier \cite{GambierThese} p.~50,
there is no need to expand the coefficients $g(x)$ of the equation 
around $x_0$.
This is achieved \cite{Conte1992a} 
by taking for the expansion variable not $x-x_0$,
but a mute variable $\chi$ with the only property $\chi_x=1$.
Coefficients $u_j$ in eq.~(\ref{eqP1Laurent}) become dependent on $x$
instead of $x_0$~:
\begin{eqnarray}
& &
 u_0=1,\ u_1=u_2=u_3=0,\ u_4=-{g(x) \over 10},\ u_5=-{g'(x) \over 15},\ 
\nonumber
\\
& &
Q_6 \equiv {g''(x) \over 2}=0.
\end{eqnarray}
\label{pageRemarkGambier}
\end{enumerate}

Let us insist again on the danger of using the present test as it is.
An example of Chazy makes evident the necessity for a more reliable test~:
the equation with a single valued general solution
(\cite{ChazyThese} p.~360)
\begin{eqnarray}
& &
(u''' -2 u' u'')^2 + 4 u''^2 (u'' - u'^2 - 1)=0,\
u=e^{c_1 x + c_2}/c_1 + {c_1^2-4 \over 4 c_1} x + c_3.
\nonumber
\end{eqnarray}
possesses a logarithmic family $u \sim - \Log(x-x_0)$.

\begin{Exercice}
Handle the equation
\begin{equation}
2 u u'' - 3 u'^2=0,\
u=c_1 / (x-c_2)^2.
\end{equation}
\end{Exercice}

\Solution. 
\begin{equation}
2p-2=2p-2,\
E_0 \equiv 2 u_0^2 p (p-1) -3 u_0^2 p^2 = 0.
\hbox{ Hence } p=-2,\ u_0 \hbox{ arbitrary.} \Box
\nonumber
\end{equation}

\section{Extension to a system}
\label{sectionCasSysteme}
\indent

If the differential equation is defined by a system
\begin{equation}
\bfE(x,\bfu)=0,
\label{eqDEgeneral}
\end{equation}
(boldface characters represent multicomponent quantities),
the scalar equations of section \ref{sectionCasScalaire}
become systems~:
a linear system for the components of $\bfp$,
an algebraic system for the components of $\bfu_0$,
a linear system with a rhs for $\bfu_j$,
a determinant for the indicial equation, etc.
Take the example of the Euler system (diagonal Hoyer system)
\begin{equation}
E_1 \equiv {\D x_1 \over \D t} - \alpha x_2 x_3 = 0,\
E_2 \equiv {\D x_2 \over \D t} - \beta  x_3 x_1 = 0,\
E_3 \equiv {\D x_3 \over \D t} - \gamma x_1 x_2 = 0.
\label{eqDiagonalHoyer}
\end{equation}


{\it First step}.
The necessary condition on $(\bfp,\bfu_0)$ is
\begin{itemize}
\item {\bf C0}. All components of $\bfp$ are integer,
                all components of $\bfu_0$ are nonzero.
\end{itemize}

Of course, if a component of $\bfu_0$ is zero,
one must increase by one the associated component of $\bfp$
until the new component of $\bfu_0$ becomes nonzero.
It there exists no truly singular family
(at least one component of $\bfp$ negative), 
the method stops without concluding.

Here,
the unique solution $\bfp$ of the linear system
$q_1=p_1-1=p_2+p_3$ and cyclically
is thus
$p_1=p_2=p_3=-1,q_1=q_2=q_3=-2$.
The algebraic system for $\bfu_0$
\begin{equation}
\bfE_0 \equiv 
\lim_{\chi \to 0} \chi^{-\bfq} \hat\bfE(x,\bfu_0 \chi^{\bfp})=0,\ 
\bfu_0\not={\bf 0},
\end{equation}
is written
\begin{equation}
E_{1,0} \equiv - x_{1,0} - \alpha x_{2,0} x_{3,0}=0
\hbox{ and cyclically},
\end{equation}
and it defines four families
\begin{equation}
x_{1,0}^2={1 \over \beta  \gamma},\
x_{2,0}^2={1 \over \gamma \alpha},\
x_{3,0}^2={1 \over \alpha \beta},\
x_{1,0} x_{2,0} x_{3,0} = - {1 \over \alpha \beta \gamma},
\end{equation}
which we gather under the unique algebraic writing
$x_{1,0}=a,x_{2,0}=b,x_{3,0}=c$.

{\it Second step}.
The linear system
\begin{equation}
\forall j\ge 1:\
\bfE_j \equiv \bfP(j) \bfu_j + \bfQ_j(\{\bfu_l\ \vert \ l<j \}) = 0
\label{eqMethodPole5}
\end{equation}
generates the indicial equation
\begin{equation}
\det \bfP(i)=0,\
\bfP(i)=
\lim_{\chi \to 0} 
\chi^{-i-\bfq} \hat \bfE'(x,\bfu_0 \chi^{\bfp}) \chi^{i+{\bfp}},
\label{eqMatriceSysteme}
\end{equation}
here
\begin{equation}
\bfP(i)
=
\pmatrix
{
       i-1 & - \alpha c & - \alpha b \cr 
- \beta  c &        i-1 & - \beta  a \cr 
- \gamma b & - \gamma a &        i-1 \cr
},\ 
\det \bfP(i) = (i+1)(i-2)^2=0.
\end{equation}

Classical results from linear algebra on the resolution of the matrix
equation $A X = B$ give the necessary conditions

\begin{itemize}
\item {\bf C1}. The polynomial  $\det \bfP$ has degree $N$.
\item {\bf C2}. $N-1$ zeroes of $\det \bfP$ are positive integers.
\item {\bf C3}. Every positive zero $i$ of $\det \bfP$ has a multiplicity
 equal to the dimension of the kernel of  $\det \bfP(i)$.
\end{itemize}

Here, 
each of the four families has the same indices $(-1,2,2)$
and, for the double index $i=2$,
the three rows of matrix $\bfP(2)$ are proportional,
so its kernel has dimension two.

{\it Third step}.
For every positive integer zero $i$ of $\bfP$ (a Fuchs index),
require the condition

\begin{itemize}
\item {\bf C4}. 
\begin{eqnarray}
\forall i \in {\cal N},\ \det \bfP(i)=0\ :
& &
\hbox{the vector } \bfQ_i 
\hbox{ is orthogonal to the kernel}
\nonumber \\ & &
\hbox{of the adjoint of operator } \bfP(i).
\end{eqnarray}

\end{itemize}

Here, the condition {\bf C4} is satisfied at index two,
and the Laurent series are finally
\begin{eqnarray}
& &
x_1=a \chi^{-1} + a_2 \chi + O(\chi^2),\ 
\chi=t-t_0,
\\
& &
x_2=b \chi^{-1} + b_2 \chi + O(\chi^2),\ 
\\
& &
x_3=c \chi^{-1} + c_2 \chi + O(\chi^2),\ a_2 + b_2 + c_2 = 0,
\end{eqnarray}
with $(t_0,b_2,c_2)$ arbitrary.

\section{Motion of a rigid body around a fixed point}
\label{sectionToupie}
\indent

It is ruled by the system
\begin{eqnarray}
& &
A {\D \omega_1 \over \D t}
+(C-B) \omega_2 \omega_3 + (x_3 k_2 - x_2 k_3)=0,\
{\D k_1 \over \D t} - \omega_3 k_2 + \omega_2 k_3=0,
\nonumber
\\
& &
B {\D \omega_2 \over \D t}
+(A-C) \omega_3 \omega_1 + (x_1 k_3 - x_3 k_1)=0,\
{\D k_2 \over \D t} - \omega_1 k_3 + \omega_3 k_1=0,
\\
& &
C {\D \omega_3 \over \D t}
+(B-A) \omega_1 \omega_2 + (x_2 k_1 - x_1 k_2)=0,\
{\D k_3 \over \D t} - \omega_2 k_1 + \omega_1 k_2=0,
\nonumber
\end{eqnarray}
depending on six parameters~:
the components $(A,B,C)$, positive, of the diagonal inertia momentum $I$
and the components $(x_1,x_2,x_3)$, real, of the vector $\overrightarrow{OG}$
linking the fixed point $O$ to the center of mass $G$.
Because it admits the three first integrals
\begin{eqnarray}
& &
K_1=
(I \overrightarrow \Omega) . \overrightarrow \Omega
- 2 
\overrightarrow{OG} . \overrightarrow k
=
A \omega_1^2 + B \omega_2^2 + C \omega_3^2 - 2 (x_1 k_1 + x_2 k_2 + x_3 k_3),
\nonumber \\
& &
K_2 =
(I \overrightarrow \Omega) . \overrightarrow k
=
A \omega_1 k_1 + B \omega_2 k_2 + C \omega_3 k_3,
\nonumber \\
& &
K_3 = 
\overrightarrow k . \overrightarrow k
=
k_1^2 + k_2^2 + k_3^2,
\nonumber
\end{eqnarray}
and a last Jacobi multiplier equal to $1$,
\begin{equation}
 \sum_{j=1}^3  \partial_{\omega_j} \left({\D \omega_j \over \D t}\right)
+\sum_{j=1}^3  \partial_{k_j} \left({\D k_j \over \D t}\right)
=0,
\end{equation}
a sufficient condition of reducibility to quadratures
(\ie to separation of variables,
which implies neither meromorphy nor single valuedness)
is the existence of a single additional first integral independent of time.

Before Kowalevski, the only such known cases were
\begin{itemize}
\item
the isotropy case, with
\begin{equation}
A=B=C \ : \
K_4= \overrightarrow{OG} . \overrightarrow \Omega
   =x_1 \omega_1 + x_2 \omega_2 + x_3 \omega_3,
\end{equation}

\item
the case of Euler (1750) and Poinsot(1851)
$G$ at the fixed point $O$ ($x_1=x_2=x_3=0$)
with
\begin{equation}
G=O \ : \
K_4=  \mod{I \overrightarrow \Omega}^2
   = A^2 \omega_1^2 + B^2 \omega_2^2 + C^2 \omega_3^2,
\label{eqEulerPoinsot}
\end{equation}

\item
the case of Lagrange (1788) and Poisson (1813)
\begin{equation}
A=B,x_1=x_2=0 \ : \
K_4= \omega_3,
\end{equation}
\end{itemize}
and for these three cases the general solution is elliptic,
hence meromorphic \cite{Golubev}.

Let us denote a family as
($t-t_0$ is abbreviated as $t$)
\begin{eqnarray}
& &
\omega_l
=
\sum_{j=0}^{+ \infty} \omega_{l,j} t^{n_l + j},\ 
k_l
=
\sum_{j=0}^{+ \infty} k_{l,j} t^{m_l + j},\ 
\ \omega_{1,0} \omega_{2,0} \omega_{3,0} k_{1,0} k_{2,0} k_{3,0} \not=0,
\nonumber
\end{eqnarray}
with $l=1,2,3$ and 
$(\omega_{l,j},k_{l,j})$ complex.
There exist numerous families, some of them with $n_l-n_k,m_l-m_k$ not integer.
To shorten, let us restrict to the case where all the differences 
$n_l-n_k,m_l-m_k$ are integer,
and redefine a family as
\begin{eqnarray}
& &
\omega_l
=
t^{n} \sum_{j=0}^{+ \infty} \omega_{l,j} t^{j},\ 
(\omega_{1,0},\omega_{2,0},\omega_{3,0}) \not=(0,0,0),\
\nonumber 
\\
& &
k_l
=
t^{m} \sum_{j=0}^{+ \infty} k_{l,j} t^{j},\ 
(k_{1,0}, k_{2,0}, k_{3,0}) \not=(0,0,0).
\end{eqnarray}
One such family is defined \cite{Kowa1889} by the exponents
$n_l=-1,m_l=-2$,
and the sextuplets $(\omega_{l,0}, k_{l,0})$ solutions of the algebraic system
\begin{eqnarray}
& &
{\hskip -0.7 truecm}
A \omega_{1,0} +(B-C) \omega_{2,0} \omega_{3,0} +x_2 k_{3,0} - x_3 k_{2,0}=0,\
2 k_{1,0} + \omega_{3,0} k_{2,0} - \omega_{2,0} k_{3,0}=0,
\nonumber \\
& &
{\hskip -0.7 truecm}
B \omega_{2,0} +(C-A) \omega_{3,0} \omega_{1,0} +x_3 k_{1,0} - x_1 k_{3,0}=0,\
2 k_{2,0} + \omega_{1,0} k_{3,0} - \omega_{3,0} k_{1,0}=0,
\nonumber \\
& &
{\hskip -0.7 truecm}
C \omega_{3,0} +(A-B) \omega_{1,0} \omega_{2,0} +x_1 k_{2,0} - x_2 k_{1,0}=0,\
2 k_{3,0} + \omega_{2,0} k_{1,0} - \omega_{1,0} k_{2,0}=0,
\nonumber
\end{eqnarray}
and the linear system for $j\ge 1$ is
\begin{eqnarray}
& &
{\hskip -0.8 truecm}
\pmatrix
{
(j-1) A & (C-B) \omega_{3,0} & (C-B) \omega_{2,0} & 0 & x_3 & - x_2 \cr
(A-C) \omega_{3,0} & (j-1) B & (A-C) \omega_{1,0} & - x_3 & 0 & x_1 \cr
(B-A) \omega_{2,0} & (B-A) \omega_{1,0} & (j-1) C & x_2 & - x_1 & 0 \cr
0 & k_{3,0} & - k_{2,0} & j-2 & - \omega_{3,0} & \omega_{2,0} \cr 
- k_{3,0} & 0 & k_{1,0} & \omega_{3,0} & j-2 & - \omega_{1,0} \cr 
k_{2,0} & - k_{1,0} & 0 & - \omega_{2,0} & \omega_{1,0} & j-2 \cr 
} 
\pmatrix{\omega_{1,j} \cr \omega_{2,j} \cr \omega_{3,j} \cr 
         k_{1,j} \cr k_{2,j} \cr k_{3,j} \cr}
\nonumber
\\
& &
{\hskip -0.8 truecm}
+ \bfQ_j=0.
\end{eqnarray}
The determinant $\det \bfP$ must have five positive zeroes.

In the generic case $(A,B,C)$ all different and $G\not=O$,
there exists a unique solution to the algebraic system,
depending on one arbitrary parameter and the root of an eighth degree
equation \cite{Kowa1890},
but the determinant
\begin{equation}
\det \bfP=A B C (j+1) j (j-2) (j-4) (j^2-j - \mu),
\end{equation}
where $\mu$ is an algebraic expression of $(A,B,C,x_1,x_2,x_3)$,
has five positive integer zeroes iff $\mu=0$,
which corresponds to inadmissible values for the six parameters
($A,B,C$ must be real positive, $x_1,x_2,x_3$ real).

A thorough discussion of the nongeneric cases of this family $n_l=-1,m_l=-2$
led Kowalevski to retrieve the three known cases, as expected,
and finally to find the subcase 
\begin{eqnarray}
& &
A=B,\
(x_1,x_2)\not=(0,0),\
\omega_{1,0}^2+\omega_{2,0}^2=0,\
\end{eqnarray}
for which the unique solution is 
\begin{eqnarray}
& &
\omega_{1,0} = - {i C \over 2 (x_1 + i x_2) \lambda},\
\omega_{2,0} = i \omega_{1,0},\
\omega_{3,0} = 2 i,\
i^2=-1,
\nonumber
\\
& &
k_{1,0} = - {2 C \over x_1 + i x_2},\
k_{2,0} = i k_{1,0},\
k_{3,0} = 0,
\\
& &
\det \bfP = A B C (j+1) (j-2) (j-3) (j-4) (j + 1 - 2 C/A) (j - 2 + 2 C/A),
\nonumber
\end{eqnarray}
in which $\lambda$ is defined by the relation
\begin{equation}
2 C - A - 4 \lambda x_3 = 0,\ \lambda\not=0.
\end{equation}
There exist five positive integer indices iff $A=2C,x_3=0$,
and the first integral 
\begin{equation}
A=B=2 C,\ x_3=0 \ : \
K_4= 
\mod{C(\omega_1 + i \omega_2)^2 + (x_1 + i x_2) (k_1 + i k_2)}^2
\end{equation}
terminates the proof of reducibility to quadratures.
Sophie Kowalevski then managed to explicitly integrate with hyperelliptic
integrals and to prove the meromorphy of the general solution,
a feat which won her an instantaneous fame.

{\it Remark}.
Neither Hoyer nor Kowalevski enforced conditions {\bf C3} and {\bf C4}.
This was done for the first time by Appelrot \cite{Appelrot},
who found another family $n_l=-1,m_l=0$ with the indices $(-1,-1,0,1,2,2)$
and, despite the absence of five positive integer indices,
computed the next terms and found at the double index $2$ the no-log condition
\begin{equation}
x_1 \sqrt{A(C-B)} + x_2 \sqrt{B(A-C)} + x_3 \sqrt{C(B-A)} = 0,
\label{eqAppelrotConditionComplexe}
\end{equation}
whose real and imaginary parts yield, with the convention $A>B>C$,
\begin{equation}
x_2=0,\
x_1 \sqrt{A(B-C)} + x_3 \sqrt{C(A-B)} = 0.
\label{eqAppelrotCondition}
\end{equation}
Nekrasov \cite{Nekrasov} and Lyapunov then proved the multivaluedness of this 
case by exhibiting yet another family with complex exponents.

\section{Insufficiency of the meromorphy}
\label{sectionInsufficiency}
\indent

Here is the opinion of Painlev\'e 
(\cite{PaiActa} pp.~10, 83, \Oeuvres\ III pp.~196, 269)~:
``${\rm M}^{\rm e}$ Kowalevski
se propose de trouver tous les cas o\`u le mouvement du 
solide est d\'efini par des
{\it fonctions m\'eromorphes de $t$ qui poss\`edent effectivement des p\^oles}.
Son proc\'ed\'e laisse \'echapper les cas o\`u ces fonctions seraient 
uniformes sans avoir de p\^oles,
soit qu'elles fussent {\it holomorphes},
soit que toutes leurs singularit\'es fussent transcendantes.

De plus, apr\`es avoir form\'e les conditions pour qu'il existe des p\^oles
mobiles,
${\rm M}^{\rm e}$ Kowalevski
remarque que ces conditions entra\^\i nent l'int\'egrabilit\'e des 
\'equations du mouvement,
ce qui lui permet de mener la question jusqu'au bout.
Mais cette remarque laisse \'echapper un cas o\`u il existe des p\^oles
et qui n'est pas un cas d'int\'egration.
Toutefois les g\'eom\`etres Russes ont montr\'e, par la suite, que,
dans ce cas, les \'equations du mouvement n'ont pas leur int\'egrale
uniforme.

Les r\'esultats de ${\rm M}^{\rm e}$ Kowalevski subsistent donc en fait.
Mais, si int\'eressante que soit la voie suivie par 
${\rm M}^{\rm e}$ Kowalevski, il \'etait d\'esirable de reprendre la question
d'une fa\c con plus rationnelle.
C'est ce que permettent les proc\'ed\'es que j'ai employ\'es pour les
\'equations du second ordre~:
ils fournissent de la mani\`ere la plus naturelle et la plus simple les
conditions n\'ecessaires pour que
{\it ce mouvement soit repr\'esent\'e par des fonctions uniformes de $t,$}
sans qu'il soit besoin de faire aucune hypoth\`ese sur ces fonctions.
Les conditions auxquelles on parvient ainsi ne diff\`erent pas d'ailleurs de
celles de ${\rm M}^{\rm e}$ Kowalevski.
Pour ce probl\`eme particulier, on n'arrive donc pas \`a des cas nouveaux.''

Painlev\'e thus insists that poles are not privileged~:
they are just one kind of singularity among many possible others.

``Une discussion qui \'ecarterait d'avance certaines singularit\'es comme
invraisemblables serait {\it inexistante}.''
(Painlev\'e, \cite{PaiActa} p.~6, \Oeuvres\ III p.~192).

``In the statement of the problem, {\it poles are not mentioned};
if in the final result the particular integrals prove to be meromorphic,
it is a {\it result} of the research.
Likewise, no mention is made of one or another type of critical or singular
point.''
(Bureau \cite{Bureau1992} p.~105).
\index{critical}

Thus, definitely, the meromorphy assumption has to be waived as a {\it global}
property,
although it may be, and indeed is, quite useful at the {\it local} level.
The only relevant property based on singularities is the Painlev\'e property
as defined in section \ref{sectionWhatIntegrate},
\index{Painlev\'e property}
and the goal is to build a rigorous theory without any {\it a priori} on the
movable singularities.
Despite the pessimistic opinion of Picard who thought the task impossible,
Painlev\'e built that theory by a clever application of the theorem of
perturbations of Poincar\'e and Lyapunov.

\section{A few examples to be settled}
\indent

\index{Painlev\'e test}
The theory to come and the resulting ``Painlev\'e test'' should be able
to handle the following differential systems for which the meromorphy test
of section \ref{sectionCasScalaire} is inconclusive or even erroneous.
We give the location where the solution can be found.

\begin{enumerate}

\item
(``Soit qu'elles fussent {\it holomorphes}'')
Extend the test to handle the equation $2 u u'' - u'^2=0$,
with general solution 
$u=(c_1 x + c_2)^2$.
Solution section \ref{sectionMethodPole}.

\item
(``Soit que toutes leurs singularit\'es fussent transcendantes'').
Extend the test to handle the equations
$u u'' - u'^2=0$
and
$2 u^2 u' u''' - 3 u^2 u''^2 + u'^4=0$,
whose general solution is, respectively,
$u=e^{c_1 x + c_2}$
and
$u=c_1 e^{1 / (c_2 x + c_3)}$.
One may notice that the second equation is the Schwarzian derivative of
\index{Schwarzian derivative}
$\Log u$.

\item
The ``uncoupled'' system with a meromorphic general solution
\begin{equation}
{\D u \over \D x} + u^2 = 0,\
{\D v \over \D x} + v^2 = 0.
\end{equation}
admits two families (modulo the exchange of $u$ and $v$)
\begin{eqnarray}
(F1)~:& & u \sim \chi^{-1},\ v \sim \chi^{-1},\ \hbox{indices } (-1,-1)
\\
(F2)~:& & u \sim \chi^{-1},\ v \sim v_0 \chi^{0},\ v_0 \hbox{ arbitrary},\
\hbox{indices } (-1,0),
\end{eqnarray}
of which the first one fails the condition {\bf C2}.
Solution section \ref{sectionMethodPole}.

\item
(``Un cas o\`u il existe des p\^oles et qui n'est pas un cas
d'int\'egration'').
The Bianchi IX cosmological model
\begin{equation}
(\Log A)'' = A^2 - (B-C)^2 \hbox{ and cyclically},\
'=\D / \D \tau,
\label{eqBianchiIX}
\end{equation}
admits for $B=C$ a particular four-parameter meromorphic solution \cite{Taub}
\begin{eqnarray}
& &
A={k_1 \over \sinh k_1 (\tau-\tau_1)},\
B=C={k_2^2 \sinh k_1 (\tau-\tau_1) \over k_1 \sinh^2 k_2 (\tau-\tau_2)}.
\label{eqTaubNoMetric}
\end{eqnarray}
Prove the absence of the PP by studying the family $\bfp=(0,-2,-2)$
which has only four Fuchs indices $(-1,0,1,2)$.
Solution section \ref{sectionBianchi}.

\item
The Bianchi IX model (\ref{eqBianchiIX})
admits a family $\bfp=(-1,-1,-1)$ with the indices $(-1,-1,-1,2,2,2)$.
Prove the absence of the PP by studying this family.
Solution section \ref{sectionBianchiClosedForms}.

\item
The Chazy's equation of class III (\ref{eqChazyIII})
admits a Laurent series which terminates
$u=u_0/(x-x_0)^2-6/(x-x_0)$, with $(x_0,u_0)$ arbitrary.
From the study of this family,
decide about the meromorphy of the general solution.
Solution section \ref{sectionChazy}.

\item
In a problem in geometry of surfaces, Darboux \cite{Darboux1878}
encountered the system
\index{Darboux-Halphen system}
\begin{equation}
\D x_1 / \D t=x_2 x_3 - x_1 (x_2 + x_3)
\hbox{ and cyclically},
\end{equation}
explicitly excluded by Hoyer, cf.~section \ref{sectionHoyer},
and found the two-parameter meromorphic solution
\begin{equation}
x_1=c/(t-t_0)^2 + 1/(t-t_0), x_2=x_3=1/(t-t_0),\
(t_0,c) \hbox{ arbitrary}.
\end{equation}
For $c=0$, the Fuchs indices are $(-1,-1,-1)$.
Extend the test to build no-log conditions at this triple $-1$ index.
Solution section \ref{sectionMethodPerturbativeFuchsian}.

\item
The equations
\begin{eqnarray}
\label{eqChazy1909Order2} & & -2 u u'' + 3 u'^2 + d_3 u^3=0,\ d_3 \not=0, \\
\label{eqChazy1918Order3} & & u''' + u u'' - 2 u'^2=0, \\
\label{eqChazy1918Order4} & & u'''' + 2 u u'' - 3 u'^2=0,
\end{eqnarray}
have no dominant behaviour.
Prove the absence of the PP for each of them \cite{Chazy1918}.
Solution section \ref{Miscellaneous perturbations}.

\end{enumerate}

\chapter{The true problems}
\indent

In this chapter,
we state the true problems and manage logically to the only
correct definition for the Painlev\'e property (PP)~:
\index{critical}
``absence of movable critical points in the general solution'',
equivalent to that already given in section \ref{sectionWhatIntegrate}.
This includes

\begin{description} 

\item -- 
the two classifications of singularities of differential equations
(fixed or movable, critical or noncritical),

\item -- 
the two differences between linear and nonlinear
(movable singularity, singular solution),
\index{singular solution}

\item -- 
the statement of the ambitious program proposed by Painlev\'e,
a first, quick look at the method of resolution
(the ``double method'' and the ``double interest'')
and the results for first order (equation of Riccati, function of Weierstrass)
and second order (classification of Gambier, the six Painlev\'e functions).
\index{Riccati equation}

\end{description}

All the ODEs considered are defined on ${\cal C}$ or on the
Riemann sphere (i.e.~the complex plane compactified by addition of the
unique point at infinity).

Firstly, a more precise definition of the term ``to integrate'' is required.

\medskip
\index{integrate}
\label{pageToIntegrate}
\label{DefToIntegrate}
{\it Definition}. 
{\it To integrate} an ODE, in the ``modern sense'' advocated by Painlev\'e,
is to find for the general solution a finite expression, 
possibly multivalued,
in a finite number of functions,
valid in the whole domain of definition.

The important terms in this definition are ``finite'' and ``function''.

{\it Example} 1 (nonintegrated ODE)~: the Taylor series
$u=u_0 \sum_{j=0}^{+ \infty} [- (x - x_0) u_0]^j$ 
for the Cauchy solution does represent
the general solution of the ODE $u'+u^2=0$ but this representation is local
and the integration cannot be considered as achieved until one has~:
found the radius of convergence,
performed the summation,
analytically continued the sum everywhere this is possible,
and identified the analytic continuation with the meromorphic function
$(x-x_1)^{-1}$.

{\it Example} 2 (integrated ODE)~: 
the ODE $2uu'-1=0$ has for general solution $u=(x-x_0)^{1/2},$
a multivalued finite expression built from the ``multivalued function''
(see below) $z \to z^{1/2}$.

Representations by an integral, a series or an infinite product are
acceptable iff they amount to a {\it global}, as opposed to {\it local},
knowledge of the solution.

A prerequisite to the integration in the sense of the above definition is 
therefore to extend the set of available functions,
to serve as a r\'eservoir from which to build finite expressions.
At this stage, one must go back to the term ``function''.

\medskip
{\it Definition} \cite{Bourbaki}.
\index{function}
A {\it function} is an application of a set of objects into a set of 
images which applies a given object onto one {\it and only one} image.

In other words, a function is characterized by its single valuedness,
and terms such as ``multivalued function'' should be carefully avoided.
In our context, 
a function is a single valued application of the Riemann sphere onto itself.

\medskip
\index{critical}
{\it Definition} (Painlev\'e \cite{PaiBSMF} p.~206). 
A {\it critical point} of an application of the Riemann sphere onto 
itself is any singular point, isolated or not, 
around which at least two determinations are permuted.
Common synonyms are~: for critical point, branch point, point of ramification;
for determination, branch.
Such a point is an obstacle for an application to be a function.

{\it Examples}~: 
the applications $x \to \sqrt{x-a}$ and $x \to \Log(x-a)$ both have
exactly two critical points, $a$ and $\infty$.
Around each of them are permuted respectively 
two determinations and a countable infinity of determinations.

{\it Remark}.
\index{essential singularity}
An {\it essential singular point} is not necessarily a critical point,
since essential singularities, isolated or not, can be critical or not.
Examples of critical essential singularities are~:
$x=0$ for $\tan(\Log x)$ (nonisolated)
or $\sin(C+\Log x)$ 
(isolated and transcendental,
{\it Le\c cons de Stockholm} pp.~5--6, \cite{PaiActa}, 
\cite{Ince} \S 14.1 p.~317).
Examples of noncritical essential singularities are~:
$x=\infty$ for $e^x$ 
or equivalently
$x=x_0$ for $e^{1/(x-x_0)}$ (isolated),
$x=\infty$ for $\tan x$ (nonisolated).
Although, according to a classical theorem of Picard, 
an analytic function can take any value but at most two 
($\infty$ and another one) in the neighborhood of
an isolated essential singularity, a noncritical essential singularity is
{\it not} an obstacle to single valuedness.

\section{First classification of singularities, unifor\-mi\-zation}
\label{sectionM1}
\indent

{\it Definition}.
The {\it first classification of singularities}
is the distinction critical or noncritical between singular points of
applications.
Note that it does not involve differential equations.

Consider a multivalued application of the Riemann sphere onto itself.
There exist two classical methods,
called {\it uniformizations},
to define from it a single valued application, i.e.~a function.

\index{uniformization}
The first one is to restrict the object space by subtracting some lines,
called {\it cuts},
so as to forbid local turns around critical points;
for the above two examples, one removes any line joining the two points
$a$ and $\infty$.

The second method is to extend the object space to a {\it Riemann surface},
made of several copies, called sheets, of the Riemann sphere, cut and pasted.
A point of the image space may then have several antecedents on the
Riemann surface defining the object space.
Example~: two sheets for $x \to \sqrt{x},$
a countable infinity of sheets for $x \to \Log x.$

As a consequence, to fill the r\'eservoir of functions,
one accepts all uniformizable applications,
at the price of either restricting the object space by cuts,
or defining a Riemann surface in the object space.


{\it Theorem}.
The general solution of a linear ODE is uniformizable.

{\it Proof}.
Let 
\begin{equation}
\label{eqODELinear}
 E \equiv \sum_{k=0}^{N} a_k(x) {\D^{(k)}u \over \D x^k}=0,\ a_N(x)=1
\end{equation}
be such an $N^{\rm th}$ order ODE.
Its general solution
\begin{equation}
 u= \sum_{j=1}^{N} c_j u_j,\ c_j \hbox{ arbitrary constant},
\end{equation}
has for only singularities those of the $N$ independent
particular solutions $u_j,$
a subset of the singularities of the coefficients $a_k$ (\cite{Ince} \S 15.1).
One knows where to make cuts or to paste the sheets of a Riemann surface
in order to uniformize the general solution.
QED.

This has important consequences.
Firstly, any linear ODE defines a function 
(Airy, Bessel, Gauss, Legendre, Whittaker, \dots).
Secondly, in the needed r\'eservoir of functions, 
one can put all the solutions of all the linear ODEs.
Thirdly, a nonlinear ODE is considered as integrated if it is linearizable 
(of course {\it via} a finite linearizing expression).
Fourthly, in order to extend the list of known functions by means of ODEs,
it is necessary to consider nonlinear ODEs.

Hence the problem stated by L.~Fuchs and Poincar\'e.

{\it Problem}. 
Define new functions by means of ODEs, necessarily nonlinear.

\section{Second classification of singularities, different kinds of solutions}
\label{sectionM2}
\indent

There exist two features of nonlinear ODEs without counterpart in the linear
case,
they concern the location of singularities of solutions
and the possible existence of solutions additional to the general solution.

The singularities of the solutions of nonlinear
ODEs may be located at {\it a priori} unknown locations,
which depend on the constants of integration.

{\it Definition} (already given in the chapter Introduction).
A singular point of a solution of an ODE is called {\it movable}
(resp.~{\it fixed}) if its location in the complex plane depends
(resp.~does not depend) on the integration constants.

The point at $\infty$ is to be considered as fixed.
A linear ODE has no movable singularities,
the zeroes of its general solution depend on the integration constants and are
sometimes called for this reason movable zeroes.

{\it Definition}.
The {\it second classification of singularities}
is the distinction movable or fixed between singularities of 
solutions of ODEs.

Among the four structures (critical or noncritical) and (fixed or movable) of
singularities of solutions of ODEs,
only one is an obstacle for this solution to be
uniformizable and hence to define a function.
\index{uniformization}
This is the presence of singularities at the same time critical and movable.
Indeed, in such a case, one knows neither where to make cuts nor where to
paste the Riemann sheets,
and uniformization is impossible.

But let us come to the second distinction between linear and nonlinear ODEs~:
contrary to linear ODEs, nonlinear ODEs may have several kinds of solutions.

{\it Definition}.
The {\it general solution} (GS) of an ODE of order $N$
is the set of all solutions mentioned in the existence theorem of 
Cauchy (section \ref{sectionTwoTheorems}),
i.e.~determined by the initial value.
It depends on $N$ arbitrary independent constants.

{\it Definition}.
A {\it particular solution} is any solution obtained from the
general solution by giving values to the arbitrary constants.
A synonym in English is special solution.

{\it Definition}.
A {\it singular solution} is any solution which is not particular.
Linear ODEs have no singular solution.
\index{singular solution}

{\it Example}. The Clairaut type equation
$ 2 u'^2 - x u' + u=0$
has the general solution $cx-2c^2$,
a particular solution $x-2$,
and the singular solution $x^2/8$.

A singular solution can only exist when the ODE 
\begin{equation}
E(u^{(N)},u^{(N-1)},\dots,u',u,x)=0,
\end{equation}
considered as an equation for the highest derivative $u^{(N)}$,
possesses at least two determinations (branches),
whose coincidence may define a singular solution.
This is a generalization of the notion of envelope of a one-parameter family
of curves.
A practical criterium to detect the singular solutions will be given in
section \ref{sectionRemovalSS}.

Painlev\'e stated the following programme
(\cite{PaiBSMF} p.~201, \Oeuvres\ vol.~III p.~123;
 \cite{PaiActa} p.~2,   \Oeuvres\ vol.~III p.~188)~:
\par \noindent
``D\'e\-ter\-mi\-ner toutes les \'equations diff\'erentielles alg\'ebriques du
premier ordre, puis du second ordre, puis du troisi\`eme ordre, etc., dont
l'int\'egrale g\'en\'erale est uniforme''.

\index{singular solution}
One notices that singular solutions are excluded from this statement.
Indeed, they present no interest at all for the theory of integration,
for, according to the above theorem,
they satisfy an ODE of a strictly lower order than the ODE under consideration
and have therefore been encountered at a lower order in the systematic
programme stated by Painlev\'e.

This problem (single valuedness of the general solution) splits into two
successive problems 
whose methods of solution are completely different~: 
absence of movable critical points,
then absence of fixed critical points.
Hence the final statement.

{\it Problem}.
Determine all the algebraic differential equations of first order, 
then second order, then third order, etc.,
whose general solution has no movable critical points.

This class of equations is often denoted ``with fixed critical points''.
Let us prove that it coincides with the definition of the PP given section 
\ref{sectionWhatIntegrate}.
Out of the four configurations of singularities 
(critical or noncritical) and (fixed or movable),
only the configuration (critical and movable) prevents uniformizability~:
one does not know where to put the cut since the point is movable.

We have now reached the usual definition,
equivalent to the one of section \ref{sectionWhatIntegrate}.

\index{Painlev\'e property}
{\it Definition}.
One calls {\it Painlev\'e property} of an ODE the absence of movable critical
singularities in its general solution.

\section{Groups of invariance of the PP}
\label{sectionGroups}
\indent

In the fulfillment of the programme of Painlev\'e,
it is sufficient to take one representative equation by class of equivalence 
of the PP.
There exist two relations of equivalence for the PP,
defined in sections
\ref{sectionGroupHomographic} and
\ref{sectionGroupBirational}.
Other relations of equivalence are defined in section \ref{sectionCartan},
but they violate the PP.

\subsection{The homographic group}
\label{sectionGroupHomographic}
\indent

\index{homographic group} 

{\it Theorem}. 
The only bijections (one to one mappings)
of the Riemann sphere are the homographic transformations
\begin{equation}
 z \to {\alpha z + \beta \over \gamma z + \delta},\
\alpha \delta - \beta \gamma \not=0,\
(\alpha, \beta, \gamma, \delta) \hbox{ arbitrary complex constants}.
\end{equation}

{\it Proof}. See any textbook.
These transformations define a six-parameter group ${\cal H}$ called
M\"obius group, also denoted ${\rm PSL}(2,{\cal C}).$
This group plays a fundamental r\^ole in the present theory.
Given two triplets of points, there exists a unique homographic 
transformation
applying one triplet onto the other one.

{\it Theorem}. 
The PP of an ODE $E(u,x)=0$ is invariant under
an arbitrary homographic transformation of the dependent variable $u$ 
and an arbitrary holomorphic 
change of the independent variable $x$
\begin{equation}
\label{eqGroupHomographicContinuous}
 (u,x) \to (U,X): u = {\alpha(x)U+\beta(x) \over \gamma(x)U+\delta(x)},
  X=\xi(x), \qquad \alpha \delta - \beta \gamma \not= 0, 
\end{equation}
where $\alpha, \beta, \gamma, \delta, \xi$ denote arbitrary analytic
(synonym~: holomorphic)
functions.

{\it Proof}. 
Let $x_0$ be a regular point of $(\alpha, \beta, \gamma, \delta, \xi),$
and $X_0=\xi(x_0)$ its transform.
In some neighborhood of $x_0,$
the transformation between $u$ and $U$ is 
close to a homographic transformation with constant coefficients
and, according to the previous theorem,
the first classification (critical, noncritical) is invariant~:
if $x_0$ is critical (resp.~noncritical) for $u,$
then $X_0$ is critical (resp.~noncritical) for $U,$
and {\it vice versa}.
Since $(\alpha, \beta, \gamma, \delta, \xi)$ do not depend on $x_0,$
the second classification (fixed or movable) is also invariant.
Thus the PP, which only depends on these two classifications, is invariant.
QED.

An element of this {\it homographic group}
(\ref{eqGroupHomographicContinuous}) will be denoted 
$T(\alpha, \beta, \gamma, \delta; \xi)$
or simply $T(\alpha, \beta; \xi)$ in the case $(\gamma=0, \delta=1).$
The representative equation is chosen so as to ``simplify'' some expression,
e.g.~a three-pole rational fraction the poles of which can be set at 
predefined locations like $(\infty,0,1).$

\begin{Exercice}
Choose a representative for the Riccati equation (\ref{eqRiccatiG}) in its
class of equivalence under the homographic group.
\end{Exercice}
\index{Riccati equation}

\Solution.
Two coefficients can be made numeric and the equation reduced to
\begin{equation}
\D U / \D X + U^2 + S(X)/2=0,
\label{eqRiccatiNormal}
\end{equation}
under the linear transformation $T(\alpha, \beta; \xi)$
\begin{equation}
u=\alpha U + \beta,\ X=x,\
\alpha=-1/a_2,\ \beta=-(a_2' + a_1 a_2)/(2 a_2^2).
\Box
\end{equation}
This canonical form, in which $S$ is called the {\it Schwarzian},
\index{Schwarzian}
will be encountered again in section \ref{sectionThreeODEs}.

\subsection{The birational group}
\label{sectionGroupBirational}
\indent

\index{birational group}

The PP is also invariant under a larger group \cite{PaiLecons,GambierThese},
namely the  group of birational transformations,
in short the {\it birational group},
\begin{eqnarray}
& &
{\hskip -2,0 truemm}
(u,x) \to (U,X) :\
u=r(x,U,\D U / \D X,\dots,\D^{N-1} U / \D X^{N-1})=0,\
x=\Xi(X),
\nonumber
\\
& &
{\hskip -2,0 truemm}
\label{eqGroupBirationalContinuous}
\\
& &
{\hskip -2,0 truemm}
(U,X) \to (u,x) :\
U=R(X,u,\D u / \D x,\dots,\D^{N-1} u / \D x^{N-1})=0,\
X=\xi(x),\
\nonumber
\end{eqnarray}
($N$ order of the equation,
$r$ and $R$ rational in $U,u$ and their derivatives,
analytic in $x,X$).

For instance, given the ODE $u'' - 2 u^3=0$
and the new dependent variable $U=u' + u^2$,
the algebraic elimination of $(u',u'')$ among these two equations and the
derivative of the second one yields the inverse transformation $u=U' / (2 U)$,
which, once inserted in the direct transformation, yields the 
transformed equation $U U'' - U'^2/2 - 2 U^3=0$.

\subsection{Groups of point transformations (Cartan equivalen\-ce classes)}
\label{sectionCartan}
\indent

\index{Cartan equivalence classes}

The definition of {\it to integrate} as given page \pageref{DefToIntegrate}
allows transformations outside the above two groups,
which therefore may alter the PP.
For instance, the unstable ODE $2uu'-1=0$ is made stable by the change
$u^2 \to U$.
One such group of point transformations,
studied by Roger Liouville \cite{Liouville}, 
Tresse \cite{Tresse} and Cartan \cite{Cartan},
is defined as (it includes hodograph transformations)
\index{hodograph transformations}
\begin{eqnarray}
& &
(u,x) \leftrightarrow (U,X)\ :\
u=f(X,U),\ x=g(X,U),\
U=F(x,u),\ X=G(x,u),
\label{eqGroupPoint}
\end{eqnarray}
and the variables $u$ and $x$ are two equivalent geometrical coordinates.
This geometric approach, in which
provides a complementary insight to that of Painlev\'e,
which forbids to exchange the dependent and the independent variables, 
see section \ref{sectionPendulum}.

The subgroup of fiber-preserving transformations
\begin{eqnarray}
& &
(u,x) \leftrightarrow (U,X)\ :\
u=f(X,U),\ x=g(X),\
U=F(x,u),\ X=G(x),
\label{eqGroupFiberPreserving}
\end{eqnarray}
whose equivalence classes are called {\it Cartan equivalence classes},
has been extensively studied by Kamran {\it et al.},
see e.g.~\cite{HsuKamran}.

\section{The double interest of differential equations}
\label{sectionDoubleInterest}
\indent

Let us return to the above problem.
At each differential order of the programme,
the results are twofold (this is the ``double interest'' of differential
equations)~:

\begin{enumerate}
\item
some {\it new functions }
(defined from the general solution of a stable ODE which is not
reducible to a lower order nor to a linear equation),

\item
an exhaustive list (\ie a {\it classification}) of stable ODEs,
which includes the ones defining new functions.

\end{enumerate}

Of course, each equation is characterized by one representative in its 
equivalence class.
Thus, as seen in the introduction,
the ODE for $x(t)$ in the case $b=0, \sigma=1/3$ of the Lorenz model
is not distinct, under the homographic group, 
from the (P3) equation in the case $\alpha= \beta=0,\ \gamma=\delta=1$.

For instance,
the result for order one and degree one
\index{degree}
(the {\it degree} of an algebraic ODE is the polynomial degree in the
highest derivative)
is~:
no new function,
one and only one stable equation which is the Riccati equation
(\ref{eqRiccatiG}).
\index{Riccati equation}

\section{The question of irreducibility}
\label{sectionIrreducibility}
\indent

\index{irreducibility}
The classical definition of irreducibility as given by the 
``groupe de rationalit\'e'' of Jules Drach
(Drach, in \Oeuvres\ vol.~III p.~14, \cite{PaiBSMF} p.~246, \cite{Pommaret})
had some weaknesses pointed out by Roger (not Joseph) Liouville in a 
passionating discussion with Painlev\'e in the {\it Comptes Rendus}
(see \Oeuvres\ vol.~III).
This is only after further mathematical developments,
namely the differential Galois theory,
that a precise definition of irreducibility could be given by Umemura 
\cite{Umemura}, see Okamoto, this volume.
This definition shares many features with the algorithm of Risch and Norman
in computer algebra (which decides if the primitive of a class of expressions,
e.g.~rational fractions, is inside or outside the class).

\section{The double method of Painlev\'e}
\label{sectionDoubleMethod}
\indent

To solve his problem, Painlev\'e split it into two parts
(this is the ``double m\'ethode'' \cite{PaiActa} p.~11)~:

\begin{itemize}
\item{[1]} (a local study) construction of {\it necessary} conditions for
stability,
\item{[2]} (a global study) proof of their {\it sufficiency},
either by expressing the general solution as a finite expression of a finite 
number of elementary functions (solutions of linear equations, \dots),
or by proving the irreducibility of the general solution and its freedom from
movable critical points.
\end{itemize}

The methods pertaining to each part are different.
If some necessary condition is violated in the first part, one stops and
proceeds to the next equations.
If one has exhausted the construction of necessary conditions,
or if one believes so 
(indeed, this process, although probably finite, is sometimes not bounded),
one turns to the explicit proof of sufficiency,
i.e.~practically one tries to integrate (no irreducible equation has been
discovered since 1906).

For a good presentation of ideas, see the book of E.~Hille \cite{Hille}.

{\it Remark}.
\index{essential singularity}
The reason why movable essential singularities create difficulties
lies in the inexistence of methods to express conditions that they be
noncritical ({\it Le\c cons} pp.~519 sq.).

\section{The physicist's point of view}
\label{sectionPhysicist}
\indent

The physicist is not interested in establishing a classification nor in
finding new functions.
Usually, some differential system, whether ordinary or partial,
is imposed by physics,
and the problem is to ``integrate'' it in some loose sense.
By the way, this loose objective is certainly the main responsible for
the numerous, of course divergent, interpretations of ``integrable''
to be found in the physicists' world.

The best applicability of the present theory arises when one knows nothing
or very little on the possible analytical results~:
first integral, conservation laws of PDEs,
particular solutions, \dots
Then, the first part of the double method
of section \ref{sectionDoubleMethod} happens to be a precious
{\it integrability detector}.
We have already seen a rough version of it~: 
the meromorphy test of section \ref{sectionCasScalaire},
the final version of which will be the Painlev\'e test section
\ref{sectionTest}.

The loose objective of the physicist implies performing the test to its end,
even if at some point it fails and should be stopped.
One thus gathers a lot of information in the form of necessary conditions for 
a piece of local single valuedness to exist.
This {\it partial integrability detector}
can be called the ``partial Painlev\'e test'' and will be 
examplified in section \ref{sectionTestPartial}.

Examining each condition separately, \ie independently of the others,
or simultaneously,
one {\it may} then find pieces of global information like
a first integral or a particular closed form solution.

\chapter{The classical results (L.~Fuchs, Poincar\'e, Painlev\'e)
\label{sectionClassical}}
\indent

The problem of determining all stable equations has been completely studied 
for several classes of ODEs,
while others are still unfinished.
We review here the main results achieved to date.

\section{ODEs of order one}
\indent

The completely studied class is (L.~Fuchs, Poincar\'e, Painlev\'e)
\begin{equation}
 E \equiv P(u',u,x)=0,\
P \hbox{ polynomial in } (u',u), \hbox{ analytic in } x.
\end{equation}

When the degree is one, i.e.~for the class $u'=R(u,x)$ with $R$ rational 
in $u$ and analytic in $x$,
one finds one and only one stable equation, the Riccati equation
(\ref{eqRiccatiG}).
Since it is linearizable, this case defines no new function.

When the degree is greater than one,
one finds one and only one new function, 
the elliptic function $\wp$ of Weierstrass, defined from 
eq.~(\ref{eqWeierstrass}).
The stable equations are~:
all the ODEs whose general solution has an algebraic dependence on the 
arbitrary constant,
\index{binomial equations}
plus five binomial equations with constant coefficients
(\ie $u'^n=P_{m}(u),\ (m,n) \in {\cal N},\ P_{m}$ polynomial of degree $m$).
Historically found by Briot and Bouquet,
these binomial equations have the following solution
(see e.g.~\cite{Murphy} Table 1 p.~73)
\label{pageBriotBouquet}
\begin{eqnarray}
u'^n
&=&(u-a)^{n+1}(u-b)^{n-1},\ n\ge2,\
{u-b \over u-a} = \left[{b-a \over n} (x-x_0) \right]^n
\\
u'^2
&=&(u-a)^2(u-b)(u-c),\
{1 \over u-a}=A \cosh\left[B (x-x_0) \right] + C
\label{eqBB2211}
\\
u'^2
&=&(u-a)(u-b)(u-c)(u-d),\
{1 \over u-a}= A \wp(x-x_0,g_2,g_3) + B
\label{eqBB21111}
\\
u'^3
&=&\left[(u-a)(u-b)(u-c)\right]^2,\
{1 \over u-a}= A \wp'(x-x_0,0,g_3) + B
\\
u'^4
&=&(u-a)^3(u-b)^3(u-c)^2,\
{1 \over u-c} - {1 \over a-c}= A \wp^2(x-x_0,g_2,0)
\\
u'^6
&=&(u-a)^5(u-b)^4(u-c)^3,\
{1 \over u-a}= A \wp^3(x-x_0,0,g_3) + B,
\end{eqnarray}
in which 
$a,b,c,d$ are complex constants
and $(A,B,C,g_2,g_3)$ algebraic expressions of $(a,b,c,d)$.

{\it Remarks}.

\begin{enumerate}
\item
Equation (\ref{eqBB2211}) is a degeneracy of (\ref{eqBB21111}).
The $T$ transformation $u \to a_i + u^{-1}$
($a_i$ zero of $P_{2n}$)
generates eleven other equations with $m < 2n,$
among them the Weierstrass equation (\ref{eqWeierstrass}).

\item
If the Weierstrass equation had not been known, it would have been discovered
at this order one of the systematic process of \Plv.

\end{enumerate}

\section{ODEs of order two, degree one}
\indent

The study of the class
\begin{equation}
\label{eqClassODE21}
 u''=R(u',u,x),\
R \hbox{ rational in } u', \hbox{ algebraic in } u, \hbox{ analytic in } x
\end{equation}
was started by Painlev\'e \cite{PaiBSMF,PaiActa,PaiCRAS1906} 
and finished by his student Gambier \cite{GambierThese}.

This class provides six new functions, the functions of \Plv,
defined by the ODEs
\begin{eqnarray*}
({\rm P1})\ 
u''
&=&
6 u^2 + x
\\
({\rm P2})\ 
u''
&=&
2 u^3 + x u + \alpha
\\
({\rm P3})\ 
u''
&=&
{u'^2 \over u} - {u' \over x} + {\alpha u^2 + \beta \over x} + \gamma u^3
+ {\delta \over u},\ 
\\
({\rm P4})\ 
u''
&=&
{u'^2 \over 2 u} + {3 \over 2} u^3 + 4 x u^2 + 2 (x^2 - \alpha) u 
+ {\beta \over u}
\\
({\rm P5})\ 
u''
&=&
\left[{1 \over 2 u} + {1 \over u-1} \right] u'^2
- {u' \over x}
+ {(u-1)^2 \over x^2} \left[ \alpha u + {\beta \over u} \right]
+ \gamma {u \over x}
+ \delta {u(u+1) \over u-1},
\\
({\rm P6})\
u''
&=&
{1 \over 2} \left[{1 \over u} + {1 \over u-1} + {1 \over u-x} \right] u'^2
- \left[{1 \over x} + {1 \over x-1} + {1 \over u-x} \right] u'
\\
& &
+ {u (u-1) (u-x) \over x^2 (x-1)^2} 
  \left[\alpha + \beta {x \over u^2} + \gamma {x-1 \over (u-1)^2} 
        + \delta {x (x-1) \over (u-x)^2} \right],
\end{eqnarray*}
depending on respectively 0,1,2,2,3,4 complex parameters,
since the homogra\-phic group allows to restrict to 
$\gamma (\gamma-1)=0,\ \delta (\delta-1)=0$ for (P3),
and to $\delta (\delta-1)=0$ for (P5).

The stable equations (\ref{eqClassODE21}) define $53$ equivalence classes
under the homographic group, including of course the six above ones.
They split into $50$ with $R$ rational in $u$ and $3$ with $R$ algebraic in 
$u$,
and their list can be found in~:
the original articles of Gambier \cite{Gambier1906,Gambier19061907},
\cite{PaiCRAS1906},
Gambier Th\`ese 1910,
Ince 1926 \cite{Ince}
(caution~: the numbers 5, 6, 48, 49, 50 of Gambier are changed to
6, 5, 49, 50, 48 in Ince),
Murphy 1960 \cite{Murphy},
Davis 1961 \cite{Davis},
Bureau M.~I 1964,
Cosgrove 1993 \cite{CosPDEhyper}.
In fact, the historical list of Gambier mixes two notions on purpose,
\index{irreducibility}
namely the irreducibility and the homographic group,
which makes this number $53$ rather arbitrary;
for instance, the classes numbered $1-4, 7-9$ by Gambier have been united by
Garnier \cite{GarnierThese} into the single class
\begin{eqnarray}
& &
u''=\delta (2 u^3 + x u) + \gamma (6 u^2+x) + \beta u + \alpha,
\end{eqnarray}
a stable equation admitting a second order Lax pair.

The $50$ stable equations (\ref{eqClassODE21}) with $R$ rational in $u$
define $24$ equivalence classes \cite{GambierThese} under the birational group
and less than $24$ 
Cartan equivalence classes \cite{HsuKamran}
under the group of point transformations (\ref{eqGroupFiberPreserving}).

This extremely important result (the discovery of six new functions,
nowadays frequently encountered in physics,
and the exhaustive list of $50+3$ equations)
deserves several comments depending on the field of interest.

\begin{itemize}
\item{} (Practical usage).
Given an algebraic second order, first degree ODE,
either it is transformable by a $T$ transformation 
(\ref{eqGroupHomographicContinuous})
into one of the $50+3$ equations or not.
If it is not, it has movable critical points.
If it is, it is {\it explicitly integrated} and, by looking in the table
of Gambier,
its general solution is a known
finite single valued expression made of the following functions~:
solutions of linear equations of order at most four, Weierstrass, (P1) to (P6).
\item{} (Movable singularities).
The only movable (and of course noncritical) singularities of these ODEs are~:
poles for $50+2$ of them,
in addition a nonisolated essential singularity for $1$ of them.
This result of \Plv\ and Gambier 
(poles are the only movable singularities of stable second order, first degree
equations rational in $(u',u)$ and analytic in $x$)
is often believed to be more general,
leading to wrong definitions for the PP;
it is no more true for third order, degree one,
or even second order, degree higher than one.

\item{} (Fixed critical singularities).
(P1), (P2), (P4) have none, 
(P3) and (P5) have two transcendental critical points ($\infty,0$), 
both removable by the uniformizing transformation $x \to e^x.$
(P6) has three transcendental critical points ($\infty,0,1$).
\index{uniformization}

\item{} (Dependence on the arbitrary constants).
What characterizes the $6$ \Plv\ equations among the $24$ is the 
transcendental (i.e.~not algebraic)
dependence of their general solution on both constants of integration.
The $24-6$ equations whose general solution does not involve 
(P1)--(P6) have either an algebraic dependence on both constants
or a semi-transcendental dependence (algebraic for one, transcendental for the
other one).

\item{} (Confluence).
By a confluence process (Painlev\'e \cite{PaiCRAS1906}, Gambier Th\`ese,
see Mahoux, this volume), 
(P6) generates the five others and (P1) generates the Weierstrass equation,
so up to now algebraic equations have only defined one master function.

\item{} (Monodromy).
(P6) was found independently by R.~Fuchs \cite{FuchsP6}
and Schle\-sin\-ger \cite{SchlesingerP6} in the twenty-first problem of Riemann.
Given the second order linear ODE 
$y''(t) + a_1(t,x,u) y'(t) + a_2(t,x,u) y=0$
with four Fuchsian singular points 
$t=(\infty,0,1,x)$
(see definitions section \ref{sectionFuchs})
and an apparently singular point $u$,
the necessary and sufficient condition for the group of monodromy
(see Mahoux, this volume, for definitions)
to be independent of $x$ is that $u(x)$ satisfies equation (P6).
\end{itemize}

\section{ODEs of higher order or degree}
\indent

Painlev\'e's opinion was that no new function should be expected at third
order and that one should go to fourth order.
In fact, despite huge efforts, no new function has yet been found.

\begin{itemize}
\item{} ODEs of order two, degree higher then one.

Only some subclasses have been studied,
and their classification is nearly finished.
See Chazy (Th\`ese), Bureau (M.~III), 
Cosgrove (1993 \cite{CosScou,CosODE2,Cos1997}).


Those of degree two have the necessary form 
\begin{eqnarray}
& &
\left\lbrack u'' + E_0 u'^2 + E_1 u' + E_2 \right\rbrack^2
\nonumber
\\
& &
=F_0 u'^4 + F_1 u'^3 + F_2 u'^2 + F_3 u' + F_4,
\end{eqnarray}
with $(E_k, F_k)$ rational in $u$ and analytic in $x$.
Its binomial subset ($E_0=E_1=E_2=0$) is classified in Ref.~\cite{CosScou}.

The binomial subset $(u'')^n = F(u',u,x)$ of equations of degree $n\ge 3$
is classified in Ref.~\cite{CosODE2}.

\item{} ODEs of order three, degree one.

The classification is nearly finished.
See Garnier (Th\`ese), Chazy (Th\`ese), Bureau (M.~II, \cite{BureauWhen}),
Cosgrove \cite{Cos1997}.

\item{} ODEs of order four, degree one.

The classification is just started.
See Chazy (Th\`ese), Bureau (M.~II).
\end{itemize}

\medskip

For an account of similar work on PDEs, see Ref.~\cite{Musette}.

\chapter{Construction of necessary conditions. The theory}
\label{sectionTheory}
\indent

The reader only interested in {\it using} the Painlev\'e test
\index{Painlev\'e test}
{\it may} skip this chapter,
whose relevant parts will anyway be referred to in next chapters.
By so doing, however,
his/her confidence in the Painlev\'e test will falter
at the first encounter of one of the innumerable so-called exceptions,
counterexamples, and so on, which are published every year.

This chapter describes all the methods to build necessary conditions (NC) 
for the absence of movable critical points in the general solution.
Most methods are analytic,
and we unify their presentation by describing each of them as a 
perturbation in a small complex parameter $\varepsilon$,
to which can then be applied the theorem of perturbations of Poincar\'e,
itself a generalization of the existence theorem of Cauchy.
One of them is arithmetic and leads to diophantine conditions
on the Fuchs indices of a linear differential equation.

What we try to emphasize is the quite small amount of nonlinear features
in these methods.
Indeed, most of the information is obtained by well known theories concerning
linear equations, whether differential or algebraic.

The detection of singular solutions is first explained in section
\ref{sectionRemovalSS}.
The linear ODEs are then reviewed from the point of view of interest to us.
Then we state the fundamental theorems at the origin of all methods.
For comparison purposes,
two equations are defined which will be later processed by all methods.
Finally, we describe each method and apply it to the two examples.

Unless otherwise stated, the class of DEs considered is made of DEs 
(\ref{eqDEgeneral})
polynomial in $\bfu$ and its derivatives, analytic in $x$,
with $(\bfE,\bfu)$ multidimensional.

\section{Removal of singular solutions}
\label{sectionRemovalSS}
\indent

\index{singular solution}

Since the PP excludes the consideration of singular solutions,
one must discard them as early as possible.

Let us give a practical criterium to detect singular solutions.

{\it Theorem}.
A necessary condition for a solution of an ODE to be singular is the 
existence of a common {\it finite} root $u^{(N)}$
to $E=0$ and its partial derivative with respect to $u^{(N)}$.
If $E(u,x)=0$ depends polynomially on the two highest derivatives 
$u^{(N)},u^{(N-1)}$,
after factorization of this polynomial existence condition in $u^{(N-1)}$
(called discriminant),
it is necessary that the vanishing factor has an odd multiplicity.

{\it Proof}. See e.g.~Chazy (Th\`ese).
The condition is not sufficient,
and details and examples can be found in \cite{SchaumEquaDiff} chap.~10.

Hence the method~:
compute the discriminant,
factorize it,
discard the even factors,
test each odd factor to check if it defines a solution to the equation.

Chazy (Th\`ese p.~358) was the first to notice the absence of correlation
between the structure of singularities of the GS and of the SS.
Here are such examples.

Single valued GS, SS with a movable critical point (Chazy, Th\`ese p.~360)
\begin{eqnarray}
& &
(u''' -2 u' u'')^2 + 4 u''^2 (u'' - u'^2 - 1)=0,\
\nonumber
\\
& &
\hbox{discriminant}=- 16 u''^2 (u'' - u'^2 - 1),\
\nonumber
\\
& &
\hbox{GS : } u=e^{c_1 x + c_2}/c_1 + {c_1^2-4 \over 4 c_1} x + c_3,\
\hbox{SS : } u=C_2 - \Log \cos(x-C_1).
\end{eqnarray}


GS with a movable critical point, single valued SS (Valiron tome II \S 148) 
\begin{eqnarray}
& &
27 u u'^3 - 12 x u' + 8 u=0,\
\nonumber
\\
& &
\hbox{discriminant}=- 12^3 \times 27^2 u^2 (27 u^3 - 4 x^3),
\nonumber
\\
& &
\hbox{GS : } u^3=c (x-c)^2,\
\hbox{SS : } u^3=(4/27) x^3.
\end{eqnarray}

Single valued GS and single valued SS (Painlev\'e BSMF p.~239)
\begin{eqnarray}
& &
u''^2 + 4 u'^3 + 2 (xu'-u)=0,\
\nonumber
\\
& &
\hbox{discriminant}=-8 (2 u'^3 + x u'-u),
\nonumber
\\
& &
\hbox{GS : } u=(1/2)v'^2-2 v^3 - x v,\ v''=6 v^2+x,\
\hbox{SS : } u=C x + 2 C^3.
\end{eqnarray}


\section{Linear equations near a singularity}
\label{sectionLinear}
\indent


Our only interest here is to decide about the local single valuedness
near a singularity $x=x_0$,
put for convenience at the origin by a homographic transformation
($x \to x-x_0$ or $x \to 1/x$ according as $x_0$ is at a finite distance or
not). 

These results are detailed in the course of Reignier, this volume.

Consider the most general linear system,
put in a form solved for all first order derivatives 
(the canonical form of Cauchy)
\begin{equation}
\label{eqFuchs1}
x {\D \bfU \over \D x} = \bfA \bfU,
\end{equation}
with 
$\bfU$ a column vector of $N$ components and $\bfA$ a square matrix
rational in $x$.
This can be the representation of the general scalar ODE (\ref{eqODELinear}),
with $U_k=x^{k} u^{(k)}, k=0,\dots,N-1$ and $b_k=x^{N-k} a_k$, e.~g.~for $N=3$
\begin{equation}
x {\D \over \D x}
\pmatrix{U_0 \cr U_1 \cr U_2 \cr}
=
\pmatrix
{
0 & 1 & 0 \cr 
0 & 1 & 1 \cr 
-b_0 & - b_1 & 2 - b_2 \cr
} 
\pmatrix{U_0 \cr U_1 \cr U_2 \cr}.
\end{equation}

\index{Fuchsian singularity}
\index{nonFuchsian singularity}
{\it Definition}.
The point $x=0$ is called {\it Fuchsian}
iff all solutions of (\ref{eqFuchs1}) have a polynomial growth near it. 
It is called {\it nonFuchsian}
if at least one solution has a nonpolynomial growth.

{\it Example}. 
For the ODE 
$u'+a x^n u=0,\ n$ integer, $a$ nonzero,
the Fuchsian case is $n \ge -1$
and the nonFuchsian one is $n \le -2$.
The solution $u(n)$ for $n=-2,-1,0$ is
$
  u(-2)=e^{a / x},
  u(-1)= x^{-a},
  u( 0)=e^{- a x}
$
and its singularity at $x=0$ is respectively~:
an isolated noncritical essential point,
a critical point or pole or zero (depending on $a$),
a regular point.




{\it Remarks}.
\begin{enumerate}
\item
This definition is the one of modern authors \cite{Beauville}.
It involves a property of the solutions, not of the coefficients of the 
equation.
{\it Fuchsian} denotes at the same time 
a case with the solutions $u=(x, x^2)$ 
(classically called {\it regular point})
and a case with $u=(x^{-1}, x^2)$ 
(classically called {\it singular regular point}).
The motivation for such a definition is the difficulty to recognize it 
on the matricial notation.
While in the scalar case (\ref{eqODELinear}) the canonical form defined
by setting $a_{N-1}=0$ provides an easy criterium to decide about the nature
of the singularity,
in the matricial case the example
\begin{equation}
\bfA / x =
\pmatrix{n/x & 0 \cr x^{-n} & 0 \cr} ,\
\bfU_1=\pmatrix{x^n \cr x \cr} ,\
\bfU_2=\pmatrix{0 \cr 1 \cr}
\end{equation}
shows the difficulty to do so.

\item
We avoid the usual synonyms 
{\it regular singularity} and {\it irregular singularity}
for their built-in conflict.

\end{enumerate}

{\it Definition}.
Given a point $x_0$, a {\it fundamental set of solutions} of a linear ODE
of order $N$ is any set of $N$ linearly independent solutions 
defined in a neighborhood of $x_0$.

\subsection{Linear equations near a Fuchsian singularity}
\label{sectionFuchs}
\indent

\index{Fuchsian singularity}
{\it Definition}.
Given a Fuchsian point $x=0$,
the eigenvalues $i$ of $\bfA(0)$ are called {\it Fuchs indices}.
\index{Fuchs indices}
\index{indicial equation}
The {\it indicial equation} is the characteristic equation of the linear
operator $\bfA(0)$
\begin{eqnarray}
& &
\lim_{x \to 0} \hbox{det }(\bfA(x) - i)=0.
\end{eqnarray}

Near a Fuchsian point $x=0$, 
there exist a fundamental set of solutions
\begin{equation}
x^{\lambda_i} \sum_{j=0}^{m_i} \varphi_{ij}(x) (\Log x)^{j},\
i=1,N
\label{eqFundamentalSetFuchs}
\end{equation}
in which the $\lambda_i$'s are complex numbers 
(the Fuchs indices),
$m_i$ positive integers (their multiplicity),
$\varphi_{ij}$ converging Laurent series of $x$ with finite principal parts.

Series (\ref{eqFundamentalSetFuchs}) are the simplest examples of
$\psi-$series.
\index{$\psi-$series}

The necessary and sufficient condition of local single valuedness of the 
general solution of the linear equation is~: $\lambda_i$ all integer,
no $\Log$ terms.

In the scalar case (\ref{eqODELinear}), the indicial equation is
\begin{eqnarray}
& &
0=\lim_{x \to 0} x^{N-i} E(x,x^i)=
\lim_{x \to 0} \hbox{det }(\bfA(x) - i)
\\
& &
=b_0(0) + b_1(0) i + b_2(0) i (i-1) + \dots + b_N(0) i (i-1) \dots (i-N+1).
\nonumber
\end{eqnarray}

{\it Theorem}.
Given a Fuchsian point of the scalar ODE (\ref{eqODELinear}),
necessary and sufficient conditions for the general solution to be
locally single valued near it are
\begin{description}
\item{$\circ$}
the $N$ indices are distinct integers,

\item{$\circ$}
$N(N-1) / 2$ conditions for the absence of logarithms are satisfied.

\end{description}

{\it Proof}.
See Hille book \cite{Hille} or any other textbook. 

In the matricial case (\ref{eqFuchs1}), these conditions are replaced by
\begin{description}
\item{$\circ$}
the $N$ indices are integers,

\item{$\circ$}
the multiplicity of each index $i$ 
is equal to the dimension of the kernel of $\bfA(0)-i$,

\item{$\circ$}
all conditions for the absence of logarithms are satisfied.

\end{description}

The search for the no-log conditions can be achieved in one loop,
by requiring the existence of a Laurent series
extending from the lowest Fuchs index $i_1$
to $+ \infty$
\begin{equation}
x^{i_1} \sum_{j=0}^{+ \infty} u_j x^{j}
\end{equation}
and containing $N$ arbitrary independent coefficients;
this is a finite process,
which terminates when $j+i_1$ reaches the highest Fuchs index.
Consider for instance the third order ODE admitting the three solutions
\begin{eqnarray}
& &
{\hskip -3 truemm}
u_1= x^{-2},\
u_2= x^{-3} + a x^{-2} \Log x,\
u_3= x^{-4} + b x^{-3} \Log x + {a b \over 2} x^{-2} (\Log x)^2,
\nonumber
\end{eqnarray}
namely
\begin{eqnarray}
& &
\left|\matrix{u & u_1 & u_2 & u_3 \cr 
              u' & u_1' & u_2' & u_3' \cr 
              u'' & u_1'' & u_2'' & u_3'' \cr 
              u''' & u_1''' & u_2''' & u_3''' \cr}\right|=0.
\nonumber
\end{eqnarray}
Its three Fuchs indices $-4,-3,-2$ are simple,
it is sufficient that $j$ runs from $0$ to $2$ with $i_1=-4$,
the condition $b=0$ is found at $j=1$ and the condition $a=0$ at $j=2$.

In some of the next sections,
we will encounter inhomogeneous ODEs in which the rhs is itself a Laurent
series with a finite principal part,
so we will have in addition to express the single valuedness of a
particular solution as well.
This can be incorporated in the single loop described above,
provided it starts from the smallest of the two values $i_1$
and the singularity order of the particular solution,
imposed by the rhs.

\subsection{Linear equations near a nonFuchsian singularity}
\label{sectionNonFuchs}
\indent

\index{nonFuchsian singularity}

Near a nonFuchsian singular point $x=0$, 
there exist $N$ linearly independent solutions
\begin{equation}
e^{Q_i(1/z_i)} x^{s_i} \sum_{j=0}^{m_i} \varphi_{ij}(z_i) (\Log x)^{j},\
z_i=x^{1/q_i},\
i=1,N
\label{eqFundamentalSetNonFuchs}
\end{equation}
in which 
the $q_i$'s are positive integers,
$Q_i$ polynomials,
$s_i$ complex numbers called {\it Thom\'e indices},
\index{Thom\'e indices}
$\varphi_{ij}$ {\it formal} Laurent series with a finite principal part.
The question of local single valuedness of the general solution 
cannot be settled so easily, because formal series are
generically divergent.

\section{The two fundamental theorems
\label{sectionTwoTheorems}}
\indent

{\it Theorem I} (Cauchy, Picard).
Consider an ODE of order $N$, of degree one in the highest derivative,
defined in the canonical form
\begin{equation}
\label{eqODECauchyForm}
{\D \bfu \over \D x}=\bfK[x,\bfu],\ x \in {\cal C},\ \bfu \in {\cal C}^N. 
\end{equation}
Let $(x_0,\bfu_0)$ be a point in ${\cal C} \times {\cal C}^N$ 
and $D$ be a domain containing $(x_0,\bfu_0).$
If $\bfK$ is holomorphic in $D,$
\begin{itemize}
\item{} there {\it exists} a solution $\bfu$ satisfying the initial
condition $\bfu(x_0)=\bfu_0,$
\item{} it is {\it unique},
\item{} it is {\it holomorphic} in a domain containing $(x_0,\bfu_0).$
\end{itemize}

{\it Proof}. See any textbook.
For delicate points on this classical theorem, see {\it Le\c cons} p.~394.
The contribution of Picard is to have moved the holomorphy property from the
hypothesis to the conclusion.

There exists an important complement to the theorem of Cauchy,
due to Poincar\'e~:
the Cauchy solution is also holomorphic in the Cauchy data.

{\it Remark}.
More practically, the canonical form can also be defined as
\begin{equation}
\label{eqODECauchyFormScalar}
{\D^N u \over \D x^N}=\bfK[x,u,u',\dots,u^{(N-1)}].
\end{equation}

The theorem says nothing whenever the holomorphy of $\bfK$ is violated,
as in the following two cases.

Case 1. $\D u / \D x=u / (u-1),$ 
at $u_0=1$, a point of meromorphy for $\bfK$.
The only way to possibly remove this singularity without altering the
structure of singularities is to perform a $T$ transformation
(\ref{eqGroupHomographicContinuous}).
The homography $T:\ 1/(u-1)=U$ yields a new $\bfK$,
defined by $\D U / \D x= -U^2 - U^3$,
which is indeed holomorphic in ${\cal C} \times {\cal C}$,
now making the theorem applicable.
In order to shorten the exposition, this step of an homographic
transformation will be omitted in
the whole chapters \ref{sectionTheory} and \ref{sectionPractice},
and only reminded for the synthesis of all methods into the \Plv\ test
section \ref{sectionTest}.

Case 2. $\D u / \D x=\sqrt{4 (u-e_1) (u-e_2) (u-e_3)},$ 
at $u_0=e_j, j=1,2,3,$ critical points for $\bfK$.

{\it Example}.
$\D u / \D x + u^2=0,$
with the datum $u=u_0$ at $x=x_0.$
The Cauchy solution is represented by the (infinite) Taylor series
$u=u_0 \sum_{j=0}^{+ \infty} [- (x - x_0) u_0]^j,$
a geometric series whose sum depends on one, not two, arbitrary constants,
the arbitrary location $x_1=x_0 - u_0^{-1}$ of the movable simple pole;
it only exists locally,
inside the disk of convergence centered at $x_0$ with radius $\mod{u_0}^{-1}$.
This sum is also represented by the Laurent series $(x-x_1)^{-1}.$
One notices the enormous advantage of the Laurent series~:
it reduces to one term,
and it has a much larger domain of definition
(the whole complex plane but one point).

{\it Lemma} (Poincar\'e, {\it M\'ecanique c\'eleste} \cite{Poincare}).
Consider an ODE of order $N,$ of degree one in the highest derivative,
depending on a small complex parameter $\varepsilon,$
defined in the canonical form
\begin{equation}
\label{eqLemma}
{\D \bfu \over \D x}=\bfK[x,\bfu,\varepsilon],\ 
   x \in {\cal C},\ \bfu \in {\cal C}^N,\ \varepsilon \in {\cal C}. 
\end{equation}
Let $(x_0,\bfu_0,0)$ be a point in 
${\cal C} \times {\cal C}^N \times {\cal C}$ 
and $D$ be a domain containing $(x_0,\bfu_0,0).$
If $\bfK$ is holomorphic in $D,$
the Cauchy solution exists, is unique and holomorphic in a domain containing
$(x_0,\bfu_0,0).$

{\it Proof}. See any textbook.
Note that $\bfK$ may be independent of $\varepsilon.$

{\it Definition}. 
Given $\bfx,$ the application $\bfu \to \bfE(\bfx,\bfu)$ 
and some point $\bfu^{(0)},$
one calls {\it differential} of $\bfE$ at point $\bfu^{(0)}$
the linear application, denoted $\bfE'(\bfx,\bfu^{(0)}),$
defined by
\begin{equation}
 \forall \bfv:\ \bfE'(\bfx,\bfu^{(0)})\bfv= \lim_{\lambda \to 0}
 {\bfE(\bfx,\bfu^{(0)} + \lambda \bfv)-\bfE(\bfx,\bfu^{(0)}) \over \lambda}. 
\end{equation}
This notion is known under various names~:
G\^ateaux derivative,
linearized application, tangent map, Jacobian matrix,
and sometimes Fr\'echet derivative.
\index{linearized equation}

{\it Definition}. 
Given a DE $\bfE(\bfx,\bfu)=0$ and a point $\bfu_0,$
the linear DE 
\begin{equation}
\label{eqAuxiliary}
\bfE'(\bfx,\bfu^{(0)})\bfv=0
\end{equation}
in the unknown $\bfv$ is called
the {\it linearized equation} in the neighborhood of $\bfu^{(0)}$
associated to the equation
$\bfE(\bfx,\bfu)=0$.

This is precisely the {\it \'equation auxiliaire} (\ref{eqAuxiliaryEquation})
of Darboux.
The auxiliary equation of a linear equation is the linear equation itself.


Let us define the formal Taylor expansions
\begin{equation}
   \bfu=\sum_{n=0}^{+ \infty} \varepsilon^n \bfu^{(n)},\
   \bfK=\sum_{n=0}^{+ \infty} \varepsilon^n \bfK^{(n)}. 
\end{equation}
The single equation (\ref{eqLemma}) is equivalent to the infinite sequence
\begin{eqnarray}
\label{eqPerturb0}
         n  =  0:\
 {\D \bfu^{(0)} \over \D x}&=&\bfK^{(0)}=\bfK[x,\bfu^{(0)},0]
\\
\label{eqPerturbn}
 n \ge 1:\ 
 {\D \bfu^{(n)} \over \D x}&=&\bfK^{(n)}=\bfK'[x,\bfu^{(0)},0]
 \bfu^{(n)} 
   + \bfR^{(n)}(x,\bfu^{(0)},\dots,\bfu^{(n-1)}).
\end{eqnarray}

At order zero, the equation is nonlinear. 

At order one, the equation,
in the particular important case when $\bfK$ is independent of $\varepsilon$,
is the linearized equation (without rhs, $\bfR^{(1)}=0$)
canonically associated to the nonlinear equation.
\index{linearized equation}

At higher orders,
this is the same linearized equation with different rhs $\bfR^{(n)}$
arising from the previously computed terms,
and only a particular solution is needed to integrate.
\label{pageTheoremII}

{\it Theorem II}
(Poincar\'e 1890,
Painlev\'e BSMF 1900 p.~208,
Bureau 1939, M.~I).
Take the assumptions of previous lemma.
If the general solution of (\ref{eqLemma}) is single valued in $D$ except 
maybe at $\varepsilon=0,$
then
\begin{itemize}
\item{} $\varepsilon=0$ is no exception, i.e.~the general
solution is also single valued there,
\item{} every $\bfu^{(n)}$ is single valued.
\end{itemize}

{\it Proof}. See BSMF p.~208.
The main difficulty is to prove the convergence of the series.
This theorem remains valid if one replaces 
``single valued'' (Painlev\'e version)
by ``periodic'' (Poincar\'e version) 
or ``free from movable critical points'' (Bureau version).
\smallskip

{\it Remarks}.

\begin{itemize}
\item{}
This feature (one nonlinear equation (\ref{eqPerturb0}),
one linear equation (\ref{eqPerturbn}) with different rhs)
is a direct consequence of perturbation theory,
it is common to all methods aimed at building necessary stability conditions.
The equations may be differential like (\ref{eqPerturb0})--(\ref{eqPerturbn}),
or simply algebraic.
Moreover, all the methods which we are about to describe 
(except the one of Painlev\'e) will reduce the differential
problems to algebraic problems keeping the same feature,
and the overall difficulty will be to solve 
{\it one} nonlinear algebraic equation,
then {\it one} linear algebraic equation with a countable number
(practically, a finite number) of rhs.

\item{}
The two theorems and the lemma express a local property, not a global one,
therefore they cannot serve to prove integrability as defined 
page \pageref{pageToIntegrate}.
Conversely, they can be used to disprove the PP.
In the same spirit, it is generally useless to try and sum the Taylor or 
Laurent series which will be defined.
Indeed, these series only serve as generators of necessary stability
conditions.
The proof of sufficiency is achieved by completely different methods.

\item{}
The two theorems only apply to ODEs written in the canonical form of Cauchy.
\end{itemize}

As a summary, the equations successively involved are

\begin{itemize}
\item
the original DE $\bfE(x,\bfu)=0$ also called unperturbed DE because 
$\varepsilon$ will be introduced into the equation from the outside,

\item
the perturbed DE $\bfE(X,\bfU,\varepsilon)$, obtained from the preceding one 
by some transformation 
$(x,\bfu,\bfE) \to (X,\bfU,\bfE,\varepsilon)$ called perturbation,

\item
a canonical form (it is not unique)
$d \bfU / \D X=\bfK(X,\bfU,\varepsilon)$ of the perturbed equation,
also called abusively perturbed equation,

\item
the infinite sequence (\ref{eqPerturb0})--(\ref{eqPerturbn})
of equations independent of $\varepsilon.$
\end{itemize}
\bigskip

The methods described in next sections 
for establishing necessary stability conditions consist of building
one or two perturbed equations from the original unperturbed equation,
then of applying the theorem II at a point $x_0$ which is {\it movable}.
This movable point can be either regular (method of \Plv)
or singular noncritical (all the others),
which will require its previous transformation to a regular point
(by a transformation close to $u \to u^{-1}$) 
for theorem II to apply.
One is thus led to the equations (\ref{eqPerturbn}),
i.e.~to {\it one linear} DE with a sequence of rhs.
In order to avoid movable critical points in the original equation,
one requires single valuedness in a neighborhood of $x_0$ for~:
the general solution of the linear homogeneous equation,
a particular solution of each of 
the successive linear inhomogeneous equations.

One must therefore express that a very special class of linear inhomogeneous 
DEs has a general solution single valued in a neighborhood of $x_0$.
Their lhs (homogeneous part) is the linearized equation
(\'equation auxiliaire) of a nonlinear equation which has already 
passed the requirement $n=0$ of theorem II.
The rhs (inhomogeneous part) of equation $n$ depends rationally on
$\{\bfu^{(k)}, k=0,\dots,n-1 \}$ and their derivatives,
all single valued near $x_0$ 
since the necessary conditions have been fulfilled until $n-1$.

For the coefficients of the homogeneous linear DE,
the point $x_0$ will appear to be
either a point of holomorphy (method of \Plv)
or a singular noncritical point (the other methods).
In the latter case, 
both situations (Fuchsian, nonFuchsian,
see sections \ref{sectionFuchs} and \ref{sectionNonFuchs})
will occur.

\subsection{Two examples~: complete (P1), Chazy's class III
\label{sectionTwoExamples}}
\indent


{\it Example} 1 (``complete (P1)'') (BSMF p.~224, \cite{Ince} \S 14.311 p.~329,
\cite{BureauMI} p.~267,
\cite{KruskalClarkson1992}). 
\begin{equation}
\label{eqP1complete}
 E \equiv - {\D^2 u \over \D x^2} + c {\D u \over \D x} + e u^2 + f u + g=0,
\end{equation}
with ($c,e,f,g$) analytic in $x,$ and $e$ nonzero.
This equation arises in the systematic study of class (\ref{eqClassODE21})
and has led to the discovery of (P1).

Under a transformation $T(\alpha,\beta;\xi)$
(\ref{eqGroupHomographicContinuous}), equation
(\ref{eqP1complete}) is form-invariant
(\cite{BureauMI} p.~267)
\begin{eqnarray}
& &
- {\D^2 U \over \D X^2} 
+ \left[c - {\xi'' \over \xi'} - 2 {\alpha' \over \alpha}\right]
 \xi'^{-1} {\D U \over \D X}
+ e {\alpha \over \xi'^2} U^2 
+ \left[f + c {\alpha' \over \alpha} + 2 e \beta
   - {\alpha'' \over \alpha} \right] {U \over \xi'^{2}}
\nonumber
\\
& &
+ \left[g + f \beta + c \beta' + e \beta^2 - \beta''\right]
 \alpha^{-1} \xi'^{-2} =0,\
\alpha \xi' \not=0.
\end{eqnarray}
This allows to assign simple predefined values to as many coefficients as
gauges in $T$, i.e.~three.
For any value of $(c,e,f,g)$ it is possible to choose 
for the coefficients of $\D U / \D X,U^2,U$ the values $0,6,0$,
and this requires solving two quadratures and one linear algebraic equation
for $(\alpha,\beta;\xi)$
\begin{eqnarray}
& &
(\Log \alpha)' = {2 \over 5} \left[c - {e' \over 2 e}\right]
\\
\xi'^2
& = &
 {e \alpha \over 6},\
 \beta
= {1 \over 2 e}
 \left[f + c (\Log \alpha)' + (\Log \alpha)'' + (\Log \alpha)'^2\right].
\end{eqnarray}
Consequently, in what follows, one always assumes $c=0, e=6, f=0$ 
in (\ref{eqP1complete}).
 
{\it Example} 2 (Chazy complete equation of class III \cite{ChazyThese}).
\begin{equation}
\label{eqChazyClassIII}
- u_{xxx} + {a \over 2} (2 u u_{xx} - 3 u_x^2) 
+ a_1 u_{xx} + c_1 u u_x + c_0 u_x + d_3 u^3 + d_2 u^2 + d_1 u + d_0 =0,
\end{equation}
where $(a,a_i,c_i,d_i)$ are analytic in $x,$ and $a$ nonzero.
This one has led Chazy to the discovery of his equation (\ref{eqChazyIII}).

Under a transformation $T(\alpha,\beta;\xi)$, equation
(\ref{eqChazyClassIII}) is form-invariant.
For any value of $(a,a_i,c_i,d_i)$ it is possible to choose 
the values $2, 0, 0$
for the coefficients of $U U_{XX}, U_{XX}, U^2$,
and this requires solving the coupled ODEs for $(\alpha,\beta;\xi)$
[notation $\Lambda = {\alpha' / \alpha}$]
\begin{eqnarray}
& &
 2 \Lambda' - \Lambda^2 + 2 a^{-3}[(a^2 c_1 + 18 a d_3) \Lambda
+ a^2 d_2  + 9 a' d_3  - 3 a^2 a_1 d_3] = 0, 
\\
& &
 \xi'
= {a \alpha \over 2},\
 \beta=a^{-2} (6 a \Lambda + 3 a' - a a_1),
\end{eqnarray}
i.e.~one Riccati equation followed by two quadratures. 
Consequently, in what follows, one always assumes $a=2, a_1=0, d_2=0$ 
in (\ref{eqChazyClassIII}).
 
None of these two examples has singular solutions.
\index{singular solution}

\begin{Exercice}
Show the impossibility to cancel $d_3$ 
in (\ref{eqChazyClassIII}) by 
choosing $\alpha,\beta,\xi$.
\end{Exercice}


\section{The method of pole-like expansions}
\label{sectionMethodPole}
\indent

\index{method of pole-like expansions}
This is a reliable version of the meromorphy test given in sections
\ref{sectionCasScalaire} and \ref{sectionCasSysteme}.

Consider a movable singular point $x_0$ of either the general solution
or a particular solution.
Since $\bfu(x_0)$ is not finite,
the theorems of section \ref{sectionTwoTheorems} cannot be applied.
It is nevertheless immediate to check that the perturbation
\begin{equation}
x = x_0 + \varepsilon X,\
u = (\varepsilon X)^p \sum_{n=0}^{+ \infty} (\varepsilon X)^n u^{(n)}(x),\
E = (\varepsilon X)^q \sum_{n=0}^{+ \infty} (\varepsilon X)^n E^{(n)}(x),
\label{eqPerturbationGambier}
\end{equation}
in which the key point is the dependence of $u^{(n)}$ on $x$, not $X$,
generates equations $E^{(n)}=0$ which only differ from the
algebraic equations $E_j=0$ defined by (\ref{eqLaurentE})
by the replacement of $x_0$ by $x$.
The identification is even complete if $\chi$ is defined by $\chi_x=1$ 
instead of $\chi=x-x_0$, see Remark page \pageref{pageRemarkGambier}.

Fortunately, the method we are about to describe has been made by Bureau 
(1939) \cite{Bureau1939}
an application of theorems I and II,
as will be seen in section \ref{sectionMethodBureau}.
This {\it method of pole-like expansions}
is the most widely used in \Plv\ analysis.
Initiated by Paul Hoyer \cite{Hoyer}
and Sophie Kowalevski \cite{Kowa1889,Kowa1890},
it has been formalised by Gambier \cite{GambierThese},
revived by Ablowitz {\it et al.} \cite{ARS1980}
who applied it to wide classes of physical equations,
extended to partial differential equations (PDEs) 
by Weiss {\it et al.} \cite{WTC} (WTC),
with technical simplifications by Kruskal \cite{JKM}
and Conte \cite{Conte1988,Conte1989}.
\Plv\ himself never used 
``le proc\'ed\'e connu de Madame Kowalevski \dots 
dont le caract\`ere n\'ecessaire n'\'etait pas \'etabli''
(Acta pp.~10, 83, O$\!$euvres III pp.~196, 269),
see section \ref{sectionInsufficiency}.

We now rephrase the steps and generated conditions
of sections \ref{sectionCasScalaire} and \ref{sectionCasSysteme}
so as to adapt them to the new objective~: the PP.
The expansion is denoted
\begin{eqnarray}
\bfu=\sum_{j=0}^{+ \infty} \bfu_j \chi^{j+\bfp},\ 
\bfu_0 \not = {\bf 0},\
\chi'=1.
\label{eqMethodPole1}
\end{eqnarray}

{\it First step}.
Determine all possible families $(\bfp,\bfu_0)$.
The necessary condition on      $(\bfp,\bfu_0)$ is
\begin{itemize}
\item {\bf C0}. For each family not describing a singular solution,
all components of $\bfp$ are integer.
\end{itemize}

It there exists no truly singular family
(at least one component of $\bfp$ negative), 
the method stops without concluding.

{\it Remarks}.
\begin{itemize}
\item{}
Some components of $\bfu_0$ can be zero, or even some components of $\bfu$.

\item{}
To avoid missing some family,
one should 
firstly put the ODE under a canonical form of Cauchy
(\ref{eqODECauchyForm}) or (\ref{eqODECauchyFormScalar}),
so as to enumerate all the points $\bfu$ which make inapplicable 
the existence theorem of Cauchy,
secondly for each such point build a transformed ODE under an homography 
making the point regular for the Cauchy theorem,
thirdly determine families of the transformed ODE as 
in section \ref{sectionCasScalaire}.

\item{}
The derivative of order $k$ of $\chi^{p}$ does not behave like $\chi^{p-k}$
if $p$ is positive and $p-k$ negative.
\end{itemize}

{\it Second step}.
For each family, compute the indicial polynomial $\det \bfP$.
None of the conditions {\bf C1}, {\bf C2}, {\bf C3} 
of section \ref{sectionCasSysteme}
is required for the existence of the Laurent series
since we also accept particular solutions and only exclude singular solutions.

{\it Third step}.
Unchanged as compared to section \ref{sectionCasSysteme}.
The condition {\bf C4} is unchanged.

The resulting expansion (\ref{eqMethodPole1}) thus contains as many arbitrary 
coefficients $\bfu_i$ as the sum of the multiplicities of the distinct 
positive indices, 
in addition to the arbitrary location of $x_0$, associated to the index $-1$.

For indices which are not positive integers, the method says nothing,
not even that they should be integers, 
and in such a case the expansion (\ref{eqMethodPole1}) only represents part 
of the general solution,
without indication about some possible multivaluedness in the missing part.

{\it Remarks}.
\begin{itemize}

\item{} 
The semi-infinite Laurent expansion (\ref{eqMethodPole1}) for $u$ 
about the singular point $x_0$ is
equivalent to an expansion for $u^{-1}$ about a regular point,
expansion however different from the Taylor one.
This is used in Bureau's method section \ref{sectionMethodBureau}.

\item{} We prefer the terms ``pole-like singularity'' to ``pole singularity'',
for the actual singularity of the {\it general} solution may not be a pole,
as shown by the example of Chazy's equation (\ref{eqChazyIII}),
for which it is a movable noncritical essential singularity.

\item{} Index $-1$ also corresponds to an arbitrary coefficient but, 
since the general solution cannot depend on more than $N$ such arbitrary
coefficients, some renormalisation occurs.
In the example $\D u / \D x + u^2=0$ already considered in section 
\ref{sectionM1},
the Cauchy solution near the regular point $x_0$ can be reexpanded 
\begin{equation}
u=\sum_{j=- \infty}^{0} (-u_0)^j (x-x_0)^{j-1} 
\end{equation}
[if the example were not so simple, this would be a doubly infinite Laurent
series]
so as to exhibit an arbitrary coefficient at index $j=i=-1,$ 
the only index of this too simple ODE.
Note the ``pole-like'' singularity $x_0,$ which is in fact an apparent,
inessential singularity, in this case a regular point!

\end{itemize}

\subsection{The two examples}
\label{PoleLikeExpansionExamples}
\indent

{\it Example} 1. ``Complete (P1)'' eq.~(\ref{eqP1complete}).
Already handled in section \ref{sectionCasScalaire}.

{\it Example} 2. Chazy's equation (\ref{eqChazyIII}).

{\it First step}.
The dominant terms are among $-u''', 2 u u'' - 3 u'^2, d_3 u^3$,
hence two possible families
\begin{displaymath}
\begin{array}{cll}
(p,q)=(-1,-4)
&
u_0=-6
&
\hat E \equiv u''' - 2 u u'' + 3 u'^2
\\
(p,q)=(-2,-6)
&
d_3 u_0^3=0
&
\hat E \equiv - 2 u u'' + 3 u'^2 + d_3 u^3.
\end{array}
\end{displaymath}
The second family only exists if $d_3=0$.
See section \ref{Miscellaneous perturbations} for a direct proof that $d_3=0$
is a necessary stability condition.

{\it Second step}.
The indicial polynomial of the first family is
\begin{eqnarray}
& &
 \chi^{-(i-4)} [       -  \partial_x^3 
  + 2 u_0    \chi^{-1}    \partial_x^2
  - 6 (u_{0} \chi^{-1})_x \partial_x
  + 2 (u_{0} \chi^{-1})_{xx}] \chi^{i-1}
\nonumber
\\
& = &
-(i-1)(i-2)(i-3) - 12 (i-1)(i-2) - 36 (i-1) + 2 (-1)(-2)
\nonumber
\\
& = & -(i+3)(i+2)(i+1),
\end{eqnarray}
and the indices are $-3,-2,-1$;
the algorithm stops here, due to the absence of positive integer indices.

Provided $d_3=0$, the second family has the indices $-1,0$.

{\it Third step} (only for the second family provided $d_3=0$).
At the index $0$, the condition {\bf C4} $Q_0=0$ is satisfied 
and the algorithm stops.

\begin{Exercice}
Find the families and indices of the following equations.
\begin{equation}
u''-2=0,\ u=(x-a) (x-b)
\end{equation}

\begin{equation}
u u'' - 2 u'^2 = 0,\ u=a(x-x_0)^{-1} 
\end{equation}

\begin{equation}
u'' + 3 u u' + u^3=0,\ u={1 \over x-a} + {1 \over x-b}.
\end{equation}
\end{Exercice}

\subsection{Nongeneric essential-like expansions}
\label{sectionDownwardExpansion}
\indent

Just like (\ref{eqMethodPole1}), the expansion
\begin{equation}
\label{eqMethodPole2}
 \bfu=\sum_{-j=0}^{\infty}  \bfu_j \chi^{j+\bfp},\ \bfu_0 \not= {\bf 0},\
\end{equation}
valid outside a disk centered at $x_0$
i.e.~in a neighborhood of the point $\infty$,
is locally single valued.
From this downward Laurent series,
one could conceive a ``method of essential-like expansions'' quite similar to
the method of pole-like expansions,
in order to generate necessary stability conditions,
this time from the negative integer indices only.

However, for most equations, this method is not applicable.
For instance, with the example $-u''' + 2 u u'' - 3 u'^2 + d_3 u^3=0$ 
(a subset of (\ref{eqChazyIII})),
none of the two expansions (\ref{eqMethodPole2}) with $p=-1$ or $p=-2$ exists,
unless $d_3=0$,
which is {\it not} a reason to conclude that $d_3$ must vanish.

It only applies to the very restricted class of
equations with constant coefficients
invariant under a scaling law $(x,\bfu) \to (k x,k^\bfp \bfu)$,
having at least one pole-like family with a negative integer index other than
$-1$.
Even then, its failure to detect the movable logarithm in numerous
equations which have one makes it of very little use.
Such equations are (\ref{eqOrder2WithLog}), (\ref{eqBureauOrder4})
or 
the equation $u''' - 7 u u'' + 11 u'^2=0$ 
whose single family $p=-1,u_0=-2$
has only negative integer indices $(-6,-1,-1)$.


\section{The $\alpha-$method of Painlev\'e}
\label{sectionMethodAlpha}
\indent

Consider an ordinary differential equation (\ref{eqDEgeneral}),
a regular point $x_0$ (i.e.~a point of holomorphy of the function $\bfK$ when 
(\ref{eqDEgeneral}) is written in the canonical form (\ref{eqODECauchyForm})),
define a small nonzero complex parameter 
(which \Plv\ denoted $\alpha$) and the perturbation
\begin{equation}
\label{eqPerturbationAlpha}
   \alpha \not= 0:\ 
   x=x_0 + \alpha X,\ 
   \bfu=\alpha^\bfp \sum_{n=0}^{+ \infty} \alpha^n \bfu^{(n)}:\
   \bfE=\alpha^\bfq \sum_{n=0}^{+ \infty} \alpha^n \bfE^{(n)}=0,
\end{equation}
where $\bfp$ is a sequence of constant integers to be chosen optimally
(see example below), 
$\bfq$ another sequence of constant integers determined by $\bfp$,
then apply theorem II to the equation for $\bfu(X,\alpha)$.

At perturbation order zero~:
\begin{itemize}
\item{} all the explicit dependence of coefficients on $X$ is 
removed, i.e.~all coefficients of the equation are constant,
\item{} for a suitable choice of $\bfp,$ there only survive a few terms.
\item{} the equation is invariant under the scaling transformation 
\hfill \break
$(X,\bfu^{(0)},\bfE^{(0)}) \to (k X,k^\bfp \bfu^{(0)},k^\bfq \bfE^{(0)})$
(physicists call such an equation scaled, or weighted).
\end{itemize}

{\it Definition}. The {\it simplified equation} (\'equation simplifi\'ee)
\index{simplified equation}
associated to a given perturbation (\ref{eqPerturbationAlpha}) is the equation
of order zero $\bfE^{(0)}(x_0,\bfu^{(0)})=0$ in the unknown $\bfu^{(0)}(X)$.

The simplified equation admits the one-parameter solutions 
$\bfu^{(0)}_0 (X-X_0)^\bfp$ where $\bfu^{(0)}_0$ are constants.
Its above properties usually make it easy to study.

{\it Definition}. The {\it complete equation} (\'equation compl\`ete),
\index{complete equation}
as opposed to the simplified equation, is the equation itself 
(\ref{eqDEgeneral}).

The value $\alpha=0$ is forbidden in (\ref{eqPerturbationAlpha}), 
but theorem II takes care of that.
The constants $\bfp$ and $\bfq$ must be integers, 
chosen so as to satisfy the holomorphy assumption in the small complex 
parameter of the above lemma.
Moreover, since a linear ODE has no movable singularity
and since all successive equations $\bfE^{(n)}=0, n \ge 1,$ are linear,
the only way to have movable singularities,
in order to test their singlevaluedness,
is to select simplified equations which are truly nonlinear.

The successive steps of the $\alpha-$method and the generated necessary
conditions for stability are (BSMF p.~209 \S 7 and footnote 1)

{\it First step}.
Find all sequences $\bfp$ of integers satisfying the holomorphy assumptions of
Theorem II for the perturbation (\ref{eqPerturbationAlpha}).
Retain only those defining a truly nonlinear simplified equation.
For each sequence $\bfp$ perform the next steps.

{\it Remark}.
If the ODE (\ref{eqDEgeneral}) has degree one and order $N$,
the holomorphy assumptions of Theorem II
require that the highest derivative contributes to the simplified equation.

{\it Second step}.
Find the general solution $\bfu^{(0)}$ of the simplified equation.

\begin{itemize}
\item {\bf C0}. Require $\bfu^{(0)}$ to be free from movable critical points.
\end{itemize}

The general solution $\bfv$ of 
the auxiliary equation (\ref{eqAuxiliary}) of the simplified equation
is then (BSMF p.~209 footnote 1)
\begin{equation}
\label{eqMethodAlpha97}
\bfv = \sum_{k=1}^N d_k {\partial \bfu^{(0)} \over \partial_{c_k}},\
d_k \hbox{ arbitrary constants},
\end{equation}
and, since $\bfu^{(0)}$ has no movable critical points,
$\bfv$ has no movable critical points either (theorem {\it Le\c cons} p.~445).

{\it Third step}.
For each $n\ge1,$
define $\bfu^{(n)}$ as a particular solution of equation $\bfE^{(n)}=0$ 
(linear with a rhs),
by the classical method of the variation of constants.

\begin{itemize}
\item {\bf C1}. 
Require each $\bfu^{(n)}$ to be free from movable critical points.
\end{itemize}

These steps amount to require stability for all the sequence of perturbed 
equations,
exactly as formulated in Theorem II.

{\it Remarks}.
\begin{itemize}
\item{} 
In the second step one can take for $\bfu^{(0)}$ either the general solution
or a particular one, but not a singular solution.
The drawback of a particular solution
will be a lesser number of generated necessary stability conditions.
This may be useful when the quadratures of third step 
are difficult with the general solution and easy with a particular solution.

\item{} 
At order $n=1$ equation $\bfE^{(1)}=0$ may contain a rhs,
making it different from the auxiliary equation.

\item{} 
If in the second step only a particular solution has been chosen,
it is better that $\bfu^{(1)}$ be taken as
the general solution of equation $\bfE^{(1)}=0$.

\end{itemize}

Many people have intuitively used the $\alpha-$method,
let us give two recent examples.

{\it Example 1}.
\index{Lorenz model}
In the Lorenz model (\ref{eqLorenz}),
the simultaneous change of variables $(x,y,z) \to (\xi,\eta,\zeta)$ 
and parameters $\bsr \to (b,\sigma,\varepsilon)$ 
defined by 
\begin{equation}
\xi  =\varepsilon          x,\
\eta =\varepsilon^2 \sigma y,\
\zeta=\varepsilon^2 \sigma z,\
\varepsilon^2 \sigma r = 1
\end{equation}
led Robbins \cite{Robbins} to believe to have found a new integrable case,
defined by
\begin{equation}
\sigma\not=0, \varepsilon=0, r=\infty:\
\hbox{first integrals } \xi^2 - 2 \zeta,\ - \xi^2 + \eta^2 + \zeta^2,
\end{equation}
while in fact the new dynamical system is just the simplified of the original
one,
integrable by elliptic functions.

{\it Example 2}.
\index{Lorenz model}
The transformation $t \to t^2 \Log t$ with ``$t \to 0$'' 
from the Lorenz model to the system (24abc) of Ref.~\cite{LevineTabor}
is in fact the $\alpha-$transformation
\hfill \break
$(x,y,z,t)=
 (\varepsilon^{-1} X,\varepsilon^{-2} Y,\varepsilon^{-2} Z,t_0+\varepsilon T)$
resulting in the system
\begin{eqnarray}
{\D X \over \D T} = \sigma (Y-X),\
{\D Y \over \D T} = - Y - X Z,\
{\D Z \over \D T} = X Y,
\end{eqnarray}
whose general solution is elliptic.

\subsection{The two examples\label{sectionMethodAlphaExamples}}
\indent

{\it Example} 1. ``Complete (P1)'' eq.~(\ref{eqP1complete}) 
(BSMF p.~224 \S 15).
The Cauchy form of the perturbed equation is
\begin{equation}
     \alpha^{p-2} E \equiv 
 -                {\D^2 (u^{(0)}+ \dots) \over \D X^2}
 + 6 \alpha^{p+2} [u^{(0)} + \dots]^2
 +   \alpha^{2-p} [g(x_0) + \alpha X g'(x_0) + \dots] 
= 0.
\end{equation}

{\it First step}.
The holomorphy requirement $-2 \le p \le 2, p \in {\cal Z},$ 
selects five values for $p$,
and the requirement of a nonlinear simplified equation only retains $p=-2$,
i.e.~$q=p-2=2p=-4$.
The $g(x_0)$ term, which could vanish,
does not contribute to the simplified equation
\begin{equation}
 E^{(0)} \equiv - {\D^2 u^{(0)} \over \D X^2} + 6 u^{(0)^2} = 0. 
\end{equation}

{\it Second step}.
The general solution of this particular Weierstrass equation is
\begin{equation}
 u^{(0)} = \wp(X-c_0,0,g_3),\ (c_0,g_3) \hbox{ arbitrary}.
\end{equation}

The auxiliary equation of the simplified equation is a Lam\'e equation
\begin{equation}
E'(X,u^{(0)}) v \equiv - {\D^2 v \over \D X^2} + 12 \wp(X-c_0,0,g_3) v = 0
\end{equation}
whose general solution is a linear combination of
$\partial \wp / \partial_{g_3}$ and
$\partial \wp / \partial_{c_0}$
\cite{AbramowitzStegun}
\begin{equation}
v = c_1 (X \wp' + 2 \wp) + c_2 \wp',
\end{equation}
without any movable critical singularity.

{\it Third step}.
The successive linear equations with their rhs are
\begin{eqnarray}
E'(X,u^{(0)}) u^{(1)} 
& = &
0
\\
E'(X,u^{(0)}) u^{(2)} 
& = &
- 6 u^{(1)^2} 
\\
E'(X,u^{(0)}) u^{(3)} 
& = &
- 12 u^{(1)} u^{(2)} 
\\
E'(X,u^{(0)}) u^{(4)} 
& = &
- 12 u^{(1)} u^{(3)} - 6 u^{(2)^2} 
- g_0
\\
E'(X,u^{(0)}) u^{(5)} 
& = &
- 12 (u^{(1)} u^{(4)} + u^{(2)} u^{(3)}) - g_0' X
\\
E'(X,u^{(0)}) u^{(6)} 
& = &
- 12 (u^{(1)} u^{(5)} + u^{(2)} u^{(4)}) - 6 u^{(3)^2} - {1 \over 2} g_0'' X^2
\end{eqnarray}
\par \noindent
with the particular solutions 
\begin{eqnarray}
u^{(n)} 
& = &
0,\ n=1,2,3
\\
u^{(4)} 
& = &
{g_0 \over 24}
\left[2 X \wp \wp' + 2 \wp^2 - \zeta \wp' \right]
\\
u^{(5)} 
& = &
{g_0' \over 24}
\left[2 X^2 \wp \wp' + 2 X \wp^2 + (X \wp' + 2 \wp) \zeta \right]
\\
u^{(6)} 
& = &
{g_0'' \over 48}\big[(X \wp' + 2 \wp) (X^2 + 2 X \zeta - 2 \Log \sigma)
\nonumber
\\
& &
\phantom{{g_0'' \over 48}}
 + (X^3 \wp + X^2 \zeta - 2 X \Log \sigma + 2 \int \Log \sigma dX) \wp'\big],
\end{eqnarray}
where the functions $\zeta$ and $\sigma$
obey $\zeta'=- \wp, \sigma' = \zeta \sigma$.
To prevent movable logarithms at $n=6$
it is necessary that $g''(x_0)=0$.
Since $x_0$ is arbitrary, this condition is $\forall x:\ g''(x)=0$,
and \Plv\ proved it to be sufficient, 
thus defining the (new in the sense of Section \ref{sectionM2}) function (P1)
with the choice $g=x$.

{\it Remarks}.
\begin{itemize}
\item{} 
The reason why $u^{(1)}, u^{(2)}, u^{(3)}$ can be chosen zero is given on
page \pageref{pageTheoremII}.
The reason given in ref.~\cite{KruskalClarkson1992} p.~120 is not correct~:
even if the general solutions $u^{(1)}, u^{(2)}, u^{(3)}$ are meromorphic,
they can in principle (this does not occur for the ODE under study)
generate some multivaluedness further up in the computation.
The theorem proven in {\it Le\c cons} p.~445 is quite profound~:
if a [second order in {\it Le\c cons}] ODE is stable,
its general solution has a single valued dependence on the integration
constants.

\item{} 
Taking the particular solution $u^{(0)}=1/(X-c_0)^2$ 
instead of the general one $\wp$ 
(see first remark section \ref{sectionMethodAlpha})
makes all computations immediate ($\zeta=1/(X-c_0),\sigma=X-c_0$).
For this particular equation,
one would not miss the generation of the only necessary stability condition.
\end{itemize}
{\it Example} 2 (Chazy complete equation of class III (\ref{eqChazyClassIII})).

{\it First step}.
For the canonical form of Cauchy of the perturbed equation
\begin{eqnarray}
& &
-{\D^2 (u^{(0)}+ \dots) \over \D X^2}
+   2 \alpha^{p+1} u^{(0)} {\D^2 u^{(0)} \over \D X^2}
  - 3 \alpha^{p+1}        \left[{\D^2 u^{(0)} \over \D X^2}\right]^2
+ \dots
\nonumber
\\
& &
+ \alpha^{3-p} d_0(x_0) + \dots
=0,
\nonumber
\end{eqnarray}
the holomorphy condition is $-1 \le p \le 3, p \in {\cal Z},$
which the condition for a truly nonlinear simplified equation restricts to
$p=-1,q=-4$.
The value $p=-2$ \cite{FP1991,CFP1993} of the method of pole-like expansions
is therefore forbidden.

{\it Second step}.
The simplified equation is that of Chazy (\ref{eqChazyIII}),
whose general solution $u^{(0)}$ is \cite{Bureau1972,Bureau1987}
an algebraic transform (finite single valued expression)
of the Hermite modular function $y(X)$
\begin{equation}
u^{(0)}=\left[\Log(y_X^3 y^{-2} (y-1)^{-2}) \right]_X, 
\end{equation}
evaluated at the point $(c_1 X + c_2) / (c_3 X + c_4), c_1 c_4 - c_2 c_3=1$ 
and thus obviously depending on three arbitrary constants.

The auxiliary equation of Chazy's simplified equation
\begin{equation}
E^{(0)'} v \equiv 
 [- \partial_X^3 
  + 2 u^{(0)} \partial_X^2
  - 6 u^{(0)}_X \partial_X
  + 2 u^{(0)}_{XX}] v = 0
\end{equation}
has the three independent solutions 
$\partial u^{(0)} / \partial_{c_i},i=1,2,3,$ all single valued.

Instead of the general solution $u^{(0)}$,
which would make the computations rather involved,
let us restrict to the two-parameter particular solution
\begin{equation}
u^{(0)} = - 6 \chi^{-1} + c \chi^{-2},\
\chi = X-x_0,\
(x_0,c) \hbox{ arbitrary constants},
\label{eqChazyIIIPartSol}
\end{equation}
for which the auxiliary equation admits the general solution
\begin{equation}
v = k_2 \chi^{-2} + k_3 \chi^{-3} + k_4 v_4,\
v_4=c^{-2} \chi^{-2} (e^{-2 c / \chi} -1 - 2 c \chi^{-1}),\
\end{equation}
with $(k_2,k_3,k_4)$ arbitrary constants.

{\it Third step}.
The successive linear equations with their rhs are
\begin{eqnarray}
- E^{(0)'} u^{(1)} 
& = &
c_{1,0} u^{(0)} u^{(0)'} + d_{3,0} u^{(0)^3}
\\
- E^{(0)'} u^{(2)} 
& = &
2 u^{(1)} u^{(1)''} - 3 u^{(1)'^2}
+ c_{1,1} X u^{(0)} u^{(0)'} 
+ c_{1,0} u^{(0)'} (u^{(0)} + u^{(1)})
\nonumber
\\
& &
+ d_{3,1} X u^{(0)^3}
+ 3 d_{3,0} u^{(0)^2} u^{(1)}
+ c_{0,0} u^{(0)'}.
\end{eqnarray}

A particular solution 
of the first one is provided by the method of variation of the constants
$u^{(1)}=K_2(X) \chi^{-2} + K_3(X) \chi^{-3} + K_4(X) \chi^{-4}$
\begin{eqnarray}
    K_2' \chi^{-2} +    K_3' \chi^{-3} + K_4' v_4
& = &
0
\\
- 2 K_2' \chi^{-3} -  3 K_3' \chi^{-4} + K_4' v_4'
& = &
0
\\
  6 K_2' \chi^{-4} + 12 K_3' \chi^{-5} + K_4' v_4''
& = &
  c_{1,0} (-6 \chi^{-1} + c \chi^{-2}) (6 \chi^{-2} - 2 c \chi^{-3})
\nonumber
\\
& &
+ d_{3,0} (-6 \chi^{-1} + c \chi^{-2})^3.
\end{eqnarray}
To prevent a movable logarithm in $K_2$ (resp.~$K_3$), 
it is necessary that, in the rhs of last equation,
the coefficients of $\chi^{-5}$ and $\chi^{-6}$ vanish~:
\begin{equation}
\forall (x_0, c):\
- 2 c^2 c_1(x_0) - 18 c^3 d_3(x_0)=0,\
c^3 d_3(x_0)=0,\
\end{equation}
hence the two necessary stability conditions
$\forall x \ d_3(x)=c_1(x)=0,$
obtained at the perturbation order $n=1.$
We leave it as an exercise to check that, after completion of $n=4$,
one has obtained all the conditions ($c_1=c_0=d_3=d_1=d_0=0$)
which Chazy proved to be necessary and sufficient.

\bigskip


\index{index $-1$}
{\it Theorem}.
For any family of the method of pole-like expansions,
the value $i=-1$ is a Fuchs index.

{\it Proof}.
Let $\bfu \sim \bfu_0 \chi^{\bfp}$ be such a family
and $\hat\bfE(x,\bfu)$ be the dominant terms.
The equation
$\hat\bfE(x_0,\bfu)=0$
admits as a particular solution
the monomial $X \to \bfu=\bfu_0(x_0) (X-X_0)^{\bfp}$,
therefore 
the linearized equation at the leading term (\ref{eqAuxiliaryEquation})
admits as a particular solution
its derivative with respect to $X_0$ :
$X \to \hbox{ const }\times \partial_{X_0} (X-X_0)^{\bfp}$.
Since at least one component of $\bfp$ is negative,
the associated component of $\partial_{X_0} (X-X_0)^{\bfp}$ is proportional to
$(X-X_0)^{p-1}$,
therefore $i=-1$ is a root of the indicial equation
(\ref{eqMatriceSysteme}).
$\Box$

\subsection{General stability conditions (ODE of order $m$ and degree 1)
\label{sectionGeneralConditions}}
\indent

Using his method, \Plv\ could obtain quite general necessary stability
conditions for algebraic ODEs of arbitrary order and degree,
cf.~BSMF p.~258, Acta p.~74, Chazy (Th\`ese).
Consider the class, defined in the canonical form of Cauchy
\begin{equation}
  u^{(m)}=R(u^{(m-1)},u^{(m-2)},\dots, u',u,x),
\end{equation}
with $R$ rational in $u$ and its derivatives, analytic in $x$
[for $R$ algebraic, and for arbitrary order and degree, cf.~Acta pp.~73, 77].
{\it Necessary stability conditions} are~:

\begin{description}
\item[C1.]
As a rational fraction of $u^{(m-1)}$,
$R$ is a polynomial of degree at most two
\begin{equation}
 u^{(m)}=A u^{(m-1)^2} + B u^{(m-1)} + C.
\end{equation}

\item[C2.]
As a rational fraction of $u^{(m-2)}$,
$A$ has only simple poles $a_i$
with residues $r_i$ equal to $1 - 1 / n_i$, 
$n_i$ nonzero integers possibly infinite
\begin{equation}
A=\sum\limits_i {1 - 1 /n_i \over u^{(m-2)} - a_i}.
\end{equation}
The above sum is finite. 

For second order $m=2$ the fraction $A$ has at most four simple poles,
and the set of their residues can only take the five values of Table 
\ref{TableI}
\begin{equation}
A=\sum\limits_{i=1}^4 {r_i \over u - a_i},\
  \sum\limits_{i=1}^4 r_i =2,\
r_i=1-{1 \over n_i},\
n_i \in {\cal Z} \hbox{ or } n_i=\infty.
\end{equation}

\begin{table}[h]
\caption[garbage]{
Order two, degree one.
Number of poles (nonzero $r_i$), list of their residues.
The poles may be located at $\infty$ and may not be distinct.
The type numbering convention is that of 
(\protect{\cite{Murphy}} Table I p.~169).
The least common multiplier (lcm) is shown for convenience.
}
\begin{center}
\begin{tabular}{|| c | c | c | c | c | c ||}
\hline
Type
& lcm($r_i$)
& $r_1$
& $r_2$
& $r_3$
& $r_4$
\\
\hline
I
& $n\ge 1$
& $1 + 1/n$
& $1 - 1/n$
& $0$
& $0$
\\
\hline
III
& $2$
& $1/2$
& $1/2$
& $1/2$
& $1/2$
\\
\hline
IV
& $3$
& $2/3$
& $2/3$
& $2/3$
& $0$
\\
\hline
V
& $4$
& $3/4$
& $3/4$
& $1/2$
& $0$
\\
\hline
VI
& $6$
& $5/6$
& $2/3$
& $1/2$
& $0$
\\
\hline
\end{tabular}
\end{center}
\label{TableI}
\end{table}

Note the one-to-one correspondence between Table \ref{TableI} and
the list of powers of the five Briot-Bouquet equations
page \pageref{pageBriotBouquet}.

\begin{Exercice}
For the six equations (Pn), determine the set $(a_i,r_i)$ of simple poles with
their residue.
\end{Exercice}

\Solution.
\begin{eqnarray}
(P6) &  &
(\infty,1/2),\
(0,1/2),\
(1,1/2),\
(x,1/2),\
\\
(P5) &  &
(\infty,1/2),\
(0,1/2),\
(1,1),\
\\
(P4) &  &
(\infty,3/2),\
(0,1/2),\
\\
(P3) &  &
(\infty,1),\
(0,1),\
\\
(P2) &  &
(\infty,2),\
\\
(P1) &  &
(\infty,2).
\end{eqnarray}
For instance, (P4) belongs to type I of Table \ref{TableI},
and it is also a confluent case of types III, V, VI.
$\Box$

The similar finite lists of admissible values of $A$ for any order $m$ can be
found in \Plv\ 
($m=3$   Acta p.~68, \Oeuvres\ vol.~III p.~254;
 $m\ge4$ Acta p.~75, \Oeuvres\ vol.~III p.~261).

\item[C3.]
As rational fractions of $u^{(m-2)},$
$B$ and $C$ have no other poles than those of $A$,
and these poles are all simple.
Writing $B,C$ as rational fractions of $u^{(m-2)}$ whose denominators are
that of $A$,
this implies the degrees limitations
\begin{equation}
(\hbox{order 2, degree 1}):\
\hbox{deg num } B \le 1,\
\hbox{deg num } C \le 3.
\end{equation}
 
\item[C4.]
(Chazy, Th\`ese).
Every ODE $u^{(m-2)} - a_i=0$ (a denominator of $A$) is stable.

\item[C5.]
(\cite{Chazy1918}).
All polynomial degrees 
in $u^{(k)},k=0, \dots, m-2,$
(of the numerator and denominator of $A,B,C$ written 
as irreducible fractions of $u$ and its derivatives)
are limited, except in the ``Fuchsian'' case $n_i=-2,r_i=3/2$
(see Ref.~\cite{Chazy1918} for details).

\end{description}

For additional conditions, see Ref.~\cite{PaiCRAS1900}.


\section{The method of Bureau}
\label{sectionMethodBureau}
\indent

Firstly, this method exhibits a linear differential equation with a Fuchsian
singularity which allows to interpret the indices $i$ in the recursion
relation of Kowalevski as Fuchs indices.
Secondly, it brings rigor to the heuristic method of Kowalevski and Gambier.
However, the generated no-log conditions are identical to those of the method
of pole-like expansions.
\medskip


Consider an $N^{\rm th}$ order ODE $E(x,u)=0$ 
(for simplicity, one assumes $u$ and $E$ unidimensional;
the multidimensional case is handled in \cite{Bureau1992})
and a movable noncritical singular point $x_0$ where the general solution 
behaves like $u \sim u_0 (x-x_0)^p,$ with $p$ a negative integer to be
determined.

The integer $p$ is computed by the method of pole-like expansions
and the highest derivative is required to contribute (M.~II p.~9)
in order to be sure that one deals with the general, not a singular, solution.

One wants to apply the two fundamental theorems.
Since the singularity $x_0$ violates the holomorphy assumption of theorem I,
one defines an equivalent differential system (in fact two systems) 
for which $x_0$ is a point of holomorphy.
These systems will depend on a perturbation parameter $\varepsilon.$

One first defines two new dependent variables $(z,U)$
by the relations
(Gambier Th\`ese p.~50, Bureau 1939)
\begin{equation}
\label{eqMethodBureau0}
 u =s z^p,\ {\D z \over \D x} =1 + U z,\ s \not=0. 
\end{equation}
Elimination of $u$ and the derivatives of $z$ 
(M.~II pp.~13, 77)
\begin{eqnarray}
z^{-p} u
& = &
s 
\label{eqMethodBureau01}
\\
z^{-p+1} {\D u \over \D x}
& = &
p s
+ \left({\D s \over \D x} + p s U \right) z
\\
z^{-p+2} {\D^2 u \over \D x^2}
& = &
p (p-1) s
+ \left(2 p {\D s \over \D x} + p (2 p -1) s U \right) z
\nonumber
\\
& &
+ \left({\D^2 s \over \D x^2} + 2 p {\D s \over \D x} U + p^2 s U^2 
+ p s {\D U \over \D x}
\right) z^2
\label{eqMethodBureau02}
\end{eqnarray}
etc,
transforms $E$ into 
\begin{equation}
\label{eqMethodBureau1}
E \equiv E\big(x,U,{\D U \over \D x},\dots,{\D^{(N-1)}U \over \D x^{N-1}},
           s,{\D s \over \D x},\dots,{\D^{(N)}s \over \D x^{N}},z\big)=0, 
\end{equation}
an equation for $U$ of order $N-1$ polynomial in $z.$ 
For the equivalent system (\ref{eqMethodBureau0}), (\ref{eqMethodBureau1})
made of two ODEs of orders one and $N-1$
in the unknowns $(z,U),$ the point $z=0$ is still a point of meromorphy,
see examples below. 

To remove it, one introduces a dependence in a small nonzero parameter 
$\varepsilon$ 
to obtain a perturbed system to which Theorem II can be applied.
Two such perturbations have been defined (Bureau 1939).

{\it First perturbation of Bureau}
\begin{equation}
   x = x_0 + \varepsilon X,\
   z =       \varepsilon Z,\
   U \hbox{ unchanged}:\
   E \equiv
    (\varepsilon Z)^q \sum_{n=0}^{+ \infty} (\varepsilon Z)^{n} E^{(n)} = 0,
\end{equation}
where the positive integer $-q$ is the singularity order of $E.$
The coefficients must be expanded as Taylor series like in the $\alpha-$method
\begin{equation}
   s(x)=s_0 + (\varepsilon X) s_0' + \dots,\ 
   s_0^{(k)}={d^{(k)}s \over \D x^k}(x_0),\
   a(x)=a_0 + (\varepsilon X) a_0' + \dots .
\end{equation}

Expansions up to order one in $\varepsilon$ for the above derivatives
(\ref{eqMethodBureau01})--(\ref{eqMethodBureau02}) are
\begin{eqnarray}
z^{-p} u
& = &
s_0 + (s_0' X) \varepsilon
+ O(\varepsilon^2) 
\\
z^{-p+1} {\D u \over \D x}
& = &
p s_0
+ \left((p s_0') X + (s_0' + p s_0 U) Z \right) \varepsilon
+ O(\varepsilon^2) 
\\
z^{-p+2} {\D^2 u \over \D x^2}
& = &
p (p-1) s_0
\\
& &
+ p \left(
 (p-1) s_0' X + (2 s_0' + (2 p -1) s_0 U + s_0 Z {\D U \over \D X}) Z
\right) \varepsilon
+ O(\varepsilon^2) 
\nonumber
\end{eqnarray}
etc,
together with 
$\D Z / \D X = 1 + \varepsilon Z U= 1 + O(\varepsilon)$.

Order zero is an algebraic equation $E^{(0)}(x_0,s_0) = 0$
for the nonzero coefficient $s_0$.

Order one is subtle~:
it filters out all terms nonlinear in $U$ and its derivatives
$\D^{(k)} U / \D X^k$,
and it extracts the contribution of 
$\D^{(k)} U / \D X^k$
from the term $z^{k+1}$ in the expansions
(\ref{eqMethodBureau01})--(\ref{eqMethodBureau02}).
This results in 
\begin{equation}
\label{eqMethodBureau3}
E^{(1)} \equiv 
A {X \over Z} + B + \sum_{k=0}^{N-1} c_k Z^k {\D^{(k)} U \over \D X^k} 
= 0,\
(A,B,c_k) \hbox{ constant}.
\end{equation}
Since $\D Z / \D X$ is unity at this order,
this is a linear inhomogeneous ODE of order at most $N-1$ for $U$ 
with constant coefficients,
whose homogeneous part is by construction of Fuchsian type
(exactly one singular point $Z=0$, of the singular regular type)
and even Eulerian type.

In order to be sure of dealing with the general solution of the
original nonlinear ODE,
the linear ODE (\ref{eqMethodBureau3}) must have exactly the order $N-1;$
a necessary stability condition for the nonlinear ODE is the
single valuedness of the general solution of
the linear ODE (\ref{eqMethodBureau3}).
Hence the necessary conditions,
for each value of $(p,s_0)$

\begin{description}
\item[-]
the order of the linear ODE at perturbation order one is exactly $N-1$;
\item[-]
its $N-1$ Fuchs indices are distinct integers;
\item[-]
if $0$ is an index, the rhs vanishes 
($A=B=0$ condition for the particular solution to contain no logarithm).
\end{description}
Since (\ref{eqMethodBureau3}) is Eulerian,
these conditions are sufficient for the general solution of the linear ODE
(\ref{eqMethodBureau3}) to be single valued,
but only necessary for the stability of the nonlinear ODE.

Higher perturbation orders yield no information. 
The reasoning is then that any condition thus found at $x=x_0,$
such as $s_0=1,$ is valid at any $x$ since $x_0$ is arbitrary.

\medskip
{\it Second perturbation of Bureau}
\begin{equation}
   x \hbox{ unchanged},\
   z =       \varepsilon Z,\
   U = \sum_{n=1}^{+ \infty} (\varepsilon Z)^{n-1} U^{(n)}:\ 
   E \equiv 
 (\varepsilon Z)^q \sum_{n=0}^{+ \infty} (\varepsilon Z)^{n} E^{(n)}.
\end{equation}

Expansions for the above derivatives
(\ref{eqMethodBureau01})--(\ref{eqMethodBureau02}) are
\begin{eqnarray}
z^{-p} u
& = &
s
\nonumber
\\
z^{1-p} {\D u \over \D x}
& = &
p s
+ \left(p s U^{(1)} + s' \right) \varepsilon Z
+ p s U^{(2)} (\varepsilon Z)^2
+ p s U^{(3)} (\varepsilon Z)^3
+ O(\varepsilon^4) 
\nonumber
\\
z^{2-p} {\D^2 u \over \D x^2}
& = &
  p (p-1) s
+ p \left((2p-1) s U^{(1)} + 2 s'\right) \varepsilon Z
\nonumber
\\
& &
+ \left(
   p^2 s (2 U^{(2)} + U^{(1)^2}) + 2 p s' U^{(1)} + s''
 + p s {\D U^{(1)} \over \D x}
\right) (\varepsilon Z)^2
\nonumber
\\
& &
+ O(\varepsilon^3) 
\nonumber
\end{eqnarray}
etc,
together with 
$\varepsilon \D Z / \D x = 1 
+  \varepsilon Z    U^{(1)}
+ (\varepsilon Z)^2 U^{(2)}
+ O(\varepsilon^3).$

Equation $E^{(0)}(x,s)=0, s \not= 0,$ is the same algebraic equation as above 
for the unknown $s(x),$ not $s(x_0).$
Each perturbation order $n \ge 1$ defines a linear algebraic equation
\begin{equation}
\label{eqMethodBureau4}
 \forall n\ge 1:\
P(n) U^{(n)} + Q_n(x,U^{(1)}, \dots, U^{(n-1)})=0,
\end{equation}
where $P(n)$ is the indicial polynomial of Fuchsian equation 
(\ref{eqMethodBureau3}),
and $Q_n$ depends on the previously computed coefficients.
Necessary stability conditions $Q_{i}=0$ arise at every value of $i$
which is also one of the $N-1$ Fuchs indices.
These conditions are identical to those of the method of pole-like 
expansions, as proven in section \ref{sectionBureauExpansion}.

The successive steps and generated necessary conditions of the method of 
Bureau are

\begin{description}

\item[Step a.]
Determine all possible $p$ like in the method of pole-like expansions
(details M.~I p.~256, M.~II p.~9).
For all $p$ satisfying ({\bf C0}, {\bf C1}), perform step b.

\item[C0.]
All $p$ are integers.

\item[C1.]
The linear ODE (order one of first perturbation) has exactly order $N-1$
[this holomorphy condition excludes for instance $p=-2$ in Chazy].
This implies the necessity for the highest derivation order to contribute
to the dominant part during the computation of $p.$

\item[Step b.]
Solve the algebraic equation for $s_0$ at order zero of first perturbation.
For all nonzero $s_0$ perform steps c and d.

\item[Step c.]
Solve the linear inhomogeneous Euler equation for $U(Z)$ at order one of
first perturbation.

\item[C2.]
Its $N-1$ Fuchs indices are distinct integers.

\item[C3.]
If $0$ is an index, the inhomogeneous part vanishes.

\item[Step d.]
Solve the linear algebraic equation (\ref{eqMethodBureau4}) (order $n$ of
second perturbation) from $n=1$ to the highest Fuchs index.

\item[C4.]
Whenever the order $n$ in step d is a Fuchs index $i,$
$Q_i$ is zero.

\end{description}

As compared with the $\alpha-$method,
these stability conditions are directly taken at $x,$ not at $x_0.$
However, the method provides no conditions from the negative integer indices.

\subsection{Bureau expansion {\it vs.}~pole-like expansion}
\label{sectionBureauExpansion}
\indent

Let us first prove the existence of a one-to-one correspondence between 
the coefficients $U^{(n)}$ of Bureau (second perturbation)
and those $u_j$ of the method of pole-like expansions.
The relations defining Bureau coefficients are
\begin{eqnarray}
u
& = &
s z^p
\label{eqMethodBureau51}
\\
{\D z \over \D x} 
& = &
1 + U^{(1)} z + U^{(2)} z^2 + O(z^3)
\label{eqMethodBureau52}
\end{eqnarray}
and those defining the pole-like expansion are
\begin{eqnarray}
u
& = &
\chi^p (u_0 + u_1 \chi + u_2 \chi^2 + O(\chi^3))
\label{eqMethodBureau53}
\\
{\D \chi \over \D x}
& = & 
1.
\label{eqMethodBureau54}
\end{eqnarray}

The property $\chi_x = 1$ of $\chi$ first ensures $s=u_0$
[taking $\chi=x-x_0$ would just create useless complications].
The elimination of $u$ between 
(\ref{eqMethodBureau51}) and (\ref{eqMethodBureau53}) yields
\begin{eqnarray}
z
& = &
\chi
\left(1 + {u_1 \over u_0} \chi + {u_2 \over u_0} \chi^2 + O(\chi^3)\right)
^{1/p}
\nonumber
\\
& = & 
\chi
\left(1 + {u_1 \over p u_0} \chi
 + {2 p u_2 + (1-p) u_1^2 \over 2 p^2 u_0^2} \chi^2 + O(\chi^3)\right).
\label{eqMethodBureau55}
\end{eqnarray}
Let us invert this Taylor series $z$ of $\chi$ 
into a Taylor series $\chi$ of $z$
\begin{eqnarray}
\chi
& = &
z
\left(1 - {u_1 \over p u_0} z
 + {-2 p u_2 + (3 + p) u_1^2 \over 2 p^2 u_0^2} z^2 + O(z^3)\right).
\label{eqMethodBureau56}
\end{eqnarray}
One finally substitutes this $\chi$ and $\D z / \D x,$ 
both Taylor series in $z,$ into eq.~(\ref{eqMethodBureau54}) to obtain
\begin{eqnarray}
{\D \chi \over \D x} 
& = &
1
\nonumber
\\
& = &
\left(1 - {2 u_1 \over p u_0} z
 + 3 {-2 p u_2 + (3 + p) u_1^2 \over 2 p^2 u_0^2} z^2 + O(z^3)\right)
\nonumber
\\
& &
\times (1 + U^{(1)} z + U^{(2)} z^2 + O(z^3))
- {1 \over p} {\D \over \D x} \left({u_1 \over u_0}\right) z^2 + O(z^3).
\label{eqMethodBureau57}
\end{eqnarray}
The identification of the lhs and rhs as series in $z$ provides the
correspondence between the two sets of coefficients
\begin{eqnarray}
s
& = &
u_0
\\
U^{(1)}
& = &
{2 u_1 \over p u_0}
\\
U^{(2)}
& = &
{3 u_2 \over p u_0^2}
+\left({2 u_1 \over p u_0}\right)^2
- (3p+1) {u_1^2 \over 2 p^2 u_0^2}
+ {1 \over p} {\D \over \D x} \left({u_1 \over u_0}\right),
\end{eqnarray}
or 

\begin{eqnarray}
u_0
& = &
s
\\
u_1
& = &
{p \over 2} s U^{(1)}
\\
u_2
& = &
{p \over 3} s^2 U^{(2)} 
+ p {3 p + 1 \over 24} s^2 U^{(1)^2}
- {p \over 6} s^2 {\D U^{(1)} \over \D x}.
\end{eqnarray}

This bijection between the coefficients
induces a bijection between 
the equations $E^{(n)}=0$ of the expansion of Bureau (second perturbation)
and the equations $E_j=0$ of the method of pole-like expansions,
hence a bijection between the no-log conditions.

This proves the equivalence between the method of pole-like expansions
and the second perturbation of Bureau.
As to the first perturbation of Bureau, 
it brings quite important information 
{\it not} obtaibable by the method of pole-like expansions.

\subsection{The two examples}
\indent

{\it Example} 1.
``Complete (P1)'' eq.~(\ref{eqP1complete}). See Bureau M.~I eq.~(17.3), (21.2).

The unperturbed equivalent meromorphic system 
(\ref{eqMethodBureau0})--(\ref{eqMethodBureau1}) in $(z,U)$ is, in Cauchy form
\begin{eqnarray}
{\D z \over \D x} 
& = &
1 + U z
\label{eqMethodBureau31}
\\
{\D U \over \D x}
& = & 
6 s^2 z^p - (p-1) z^{-2}
- \left[2 {s' \over s} + (2p-1) U \right] z^{-1}
-{s' p U + s''\over p s}
+ {g \over p s} z^{-p}.
\nonumber
\end{eqnarray}

\begin{description}

\item[Step a.]
The only value is $p=-2$, integer.
The original ODE then reads, by increasing powers of $z$
\begin{eqnarray}
 E 
& \equiv & 6 s (1 - s) z^{-4} 
+ 2 s \left[5 U - {2 \over s} {\D s \over \D x}\right] z^{-3}
\nonumber
\\
& &
- 2 s \left[{\D U \over \D x} + {2 \over s} {\D s \over \D x} U 
       - {1 \over 2 s} {\D^2 s \over \D x^2} - 2 U^2\right] z^{-2} 
- g(x) =0.
\end{eqnarray}

\item[Step b.]
Equation $ E^{(0)} \equiv 6 s_0 (1 - s_0) =0$ has for only nonzero solution
$s_0=1.$

\item[Step c.]
At order one
\begin{equation}
\label{eqBureau2}
   {E^{(1)} \over 2 s_0} \equiv 
 - Z {\D U \over \D X} +5 U -2{s_0' \over s_0} + 3{s_0' \over s_0} {X \over Z}
 =0,\
   {\D Z \over \D X}=1 + O(\varepsilon Z).
\end{equation}

The only Fuchs index is $i=5.$ 
There is no condition on the rhs.

\item[Step d.]
The computation presents no difficulty.
\begin{equation}
{\hskip -3truemm}
   s=1,\ U^{(1)}=U^{(2)}=U^{(3)}=0,\ 
   U^{(4)}={g \over 4},\ 
   U^{(5)}={g'(x) \over 4},\ 
   Q_{ 6 } \equiv - {g''(x) \over 2}=0. 
\end{equation}

\end{description}

{\it Remark}.
On the Cauchy form (\ref{eqMethodBureau31})
with $p=-2,s=1,$
one sees easily how perturbations I and II remove the meromorphy.

{\it Example} 2 (Chazy complete equation of class III (\ref{eqChazyClassIII})).

\begin{description}

\item[Step a.]
The two solutions are $p=-1, p=-2$.
For $p=-2$ the computation of the linear equation (\ref{eqMethodBureau3})
yields a zero coefficient for $\D^2 U / \D X^2$,
thus violating condition {\bf C1}.

For $p=-1$ the original ODE then reads, by increasing powers of $z$
\begin{eqnarray}
 E 
& \equiv &
s (s + 6) z^{-4} 
+ s \left[
6 U - {6 \over s} {\D s \over \D x} + 2 {\D s \over \D x} - c_1 s + d_3 s^2
\right] z^{-3}
\nonumber
\\
& &
+ s \Big[
-2 (s+2) {\D U \over \D x} 
+ ((2 - {9 \over s}) {\D s \over \D x}  - c_1 s) U 
- c_0
\nonumber
\\
& &
+ c_1 {\D s \over \D x}
- {3 \over s} ({\D s \over \D x})^2
+ (2 + {3 \over s}) {\D^2 s \over \D x^2}
+ (7 -s) U^2
\Big] z^{-2} 
\nonumber
\\
& &
+ s \Big[
{\D^2 U \over \D x^2} 
+ 3 {\D s \over \D x} {\D U \over \D x} 
+ ({3 \over s} {\D^2 s \over \D x^2} - c_0) U
+ d_1
+ {c_0 \over s} {\D s \over \D x}
\nonumber
\\
& &
- {1 \over s} {\D^3 s \over \D x^3}
- {3 \over s} {\D s \over \D x} U^2
- 3 U {\D U \over \D x} 
+ U^3
\Big] z^{-1} 
+ d_0.
\end{eqnarray}

\item[Step b.]
Equation $ E^{(0)} \equiv s_0 (s_0 + 6) =0$ has for only nonzero solution
$s_0=-6.$

\item[Step c.]
At order one
\begin{eqnarray}
   {E^{(1)} \over s_0} 
& \equiv &
  Z^2 {\D^2 U \over \D X^2} 
 - 2 (s_0 + 2) Z {\D U \over \D X} 
+ 12 U 
- s_0 c_{1,0} - s_0^2 d_{3,0}
\nonumber
\\
& &
+ (2 - {6 \over s_0}) s_0'
+ (2 + {6 \over s_0}) s_0' {X \over Z}
=0,\
   {\D Z \over \D X}=1 + O(\varepsilon Z).
\end{eqnarray}
The Fuchs indices are $i=-4,-3,$ 
there is no condition on the rhs,
and the algorithm stops here,
due to the absence of positive integer indices.

\end{description}

\section{The Fuchsian perturbative method}
\label{sectionMethodPerturbativeFuchsian}
\indent

It allows to extract the information contained in the negative 
indices \cite{FP1991},
thus building infinitely many necessary conditions
for the absence of movable critical singularities of the logarithmic 
type \cite{CFP1993}.

The perturbation which describes it is close to the identity
\begin{equation}
\label{eqPerturbu}
   x \hbox{ unchanged},\
   \bfu= \sum_{n=0}^{+ \infty} \varepsilon^n \bfu^{(n)}:\
   \bfE= \sum_{n=0}^{+ \infty} \varepsilon^n \bfE^{(n)}=0,
\end{equation}
where, like for the $\alpha-$method, 
the small parameter $\varepsilon$ is not in the original equation.

Then, the single equation (\ref{eqDEgeneral}) is equivalent to the infinite 
sequence
\begin{eqnarray}
\label{eqNL0} 
         n  =  0:\ \bfE^{(0)}& \equiv & \bfE (x,\bfu^{(0)}) = 0
\\
 \forall n \ge 1:\ \bfE^{(n)}& \equiv & \bfE'(x,\bfu^{(0)}) \bfu^{(n)} 
             + \bfR^{(n)}(x,\bfu^{(0)},\dots,\bfu^{(n-1)}) = 0,
\label{eqLinn}
\end{eqnarray}
with $\bfR^{(1)}$ identically zero.
From Theorem II, necessary stability conditions are
\begin{description} 
\item[-]
the general solution $\bfu^{(0)}$ of (\ref{eqNL0}) has no
movable critical points,
\item[-]
the general solution $\bfu^{(1)}$ of (\ref{eqLinn}) has no
movable critical points,
\item[-]
for every $n\ge 2$ there exists a particular solution of (\ref{eqLinn}) 
without movable critical points.
\end{description} 

Order zero is just the complete equation for the unknown $\bfu^{(0)},$
so, to get some information,
one must apply Theorem II for a perturbation different from
(\ref{eqPerturbu}).
Since Bureau has proven that the method of pole-like expansions,
with more rigorous assumptions,
can be casted into an application of the two basic theorems, 
one uses it at order zero,
{\it only} to obtain the leading term
$\bfu^{(0)} \sim \bfu^{(0)}_0 \chi^{\bfp}$ of all the families 
of movable singularities.

{\it First step}.
Determine all possible families $(\bfp,\bfu^{(0)}_0)$
\begin{equation}
\bfu^{(0)} \sim \bfu^{(0)}_0 \chi^{\bfp},\
\bfE^{(0)} \sim \bfE^{(0)}_0 \chi^{\bfq},\
\bfu^{(0)}_0 \not= {\bf 0}
\end{equation}
which do not describe a singular solution,
by solving the algebraic equation
\begin{equation}
\bfE^{(0)}_0 \equiv 
\lim_{\chi \to 0} \chi^{-\bfq} \hat \bfE(x, \bfu^{(0)}_0 \chi^{\bfp})=0.
\end{equation}

\begin{itemize}
\item {\bf C0}. All components of $\bfp$ are integer.
\end{itemize}

If there exists no family which is truly singular (at least one component of
$\bfp$ negative),
the method stops without concluding.

{\it Second step}.
For each family, compute the indicial equation (\ref{eqMatriceSysteme})
and require the necessary conditions~:

\begin{itemize}
\item {\bf C2}. Every zero of $\det \bfP$ (a Fuchs index) is integer.
\item {\bf C3}. Every zero $i$ of $\det \bfP$ has a multiplicity equal to the
 dimension of the kernel of  $\det \bfP(i)$
\begin{equation}
\label{eqRank}
\forall \hbox{ index } i:\
(\hbox{multiplicity of } i)= \hbox{ dim Ker } \bfP(i).
\end{equation}
\end{itemize}

{\it Remark}.
There is no such condition as {\bf C1} on page \pageref{pageC1},
\ie the indicial polynomial may have a degree smaller than $N$.
If the indicial equation has degree $N$,
the conditions {\bf C2} and {\bf C3}
($N$ distinct integers in the one-dimensional case)
are slightly stronger than the conditions in Bureau ($N-1$ distinct integers).

The next step is easily computerizable \cite{ConteAMP,Drouffe}
if one represents 
$\bfu^{(0)},\bfu^{(1)},\dots$,
as Laurent series bounded from below~:
$\bfu^{(0)}$ with powers in the range $(\bfp:+ \infty)$,
$\bfu^{(1)}$ with powers in the range $(\rho+\bfp:+ \infty)$,
where $\rho$ denotes the smallest Fuchs index,
an integer lower than or equal to $-1$,
\dots

Order $n=0$ 
is identical to the method of pole-like expansions
and the Laurent series for $\bfu^{(0)}$
\begin{equation}
 \bfu^{(0)}=\sum_{j=0}^{+ \infty} \bfu^{(0)}_j \chi^{j+\bfp},
\end{equation}
represents a particular solution containing a number of arbitrary coefficients
equal to one (index $-1$) plus the number of positive Fuchs indices, 
counting their multiplicity.

Order $n=1$ is identical to the ``\'equation auxiliaire'' of Darboux
\begin{equation}
 \bfE^{(1)} \equiv \bfE'(x,\bfu^{(0)}) \bfu^{(1)} = 0,
\label{eqLin0}
\end{equation}
and the Laurent series for $\bfu^{(1)}$
\begin{equation}
 \bfu^{(1)}=\sum_{j=\rho}^{+ \infty} \bfu^{(1)}_j \chi^{j+\bfp},
\label{eqMethodPerturbative7}
\end{equation}
represents a particular solution containing a number of arbitrary coefficients
equal to the number of Fuchs indices, counting their multiplicity.
If $\det \bfP(i)$ has degree $N$, it represents the general solution of
(\ref{eqLin0}).

Consequently the sum $\bfu^{(0)} + \varepsilon \bfu^{(1)}$ is already,
in the neighborhood of $(\chi, \varepsilon)=(0,0)$,
a {\it local representation of the greatest particular solution}
of (\ref{eqDEgeneral}) available in this method
(the general solution if $\det \bfP(i)$ has degree $N$),
and this is a Laurent series with a strictly larger extension
$(\rho+\bfp:+ \infty)$ than that for the unperturbed expansion
$(\bfp:+ \infty)$.

At each order $n\ge 2$,
the singularity order of the particular solution of
the linear inhomogeneous equation (\ref{eqLinn}) 
$\bfE^{(n)}=0$ 
is dictated by the
contribution $\bfR^{(n)}$ of the previously computed coefficients~:
it is increased by $\rho$ at each order $n$ and is equal to $ n \rho + \bfp$
\begin{equation}
\forall n \ge 0:\
 \bfu^{(n)}=\sum_{j=n \rho}^{+ \infty} \bfu^{(n)}_j \chi^{j+\bfp}.
\end{equation}

{\it Third step}.
Solve the recurrence relation for $\bfu_j^{(n)}$ for all values of
$(n,j)\not=(0,0)$
\begin{equation}
\label{eqMethodPerturbative8}
 \forall n \ge 0 \ \forall j \ge n \rho,\ (n,j)\not=(0,0):\ 
\bfE_j^{(n)} \equiv \bfP(j) \bfu_j^{(n)}
 + \bfQ_j^{(n)}(x,\{\bfu^{(n')}_{j'} \}) = 0.
\end{equation}

The generated necessary stability conditions are
\begin{itemize}
\item {\bf C4}.
\begin{equation}
\label{eqOrtho}
\forall n \ge 0 \ \forall \hbox{ index } i,\
(n,i) \not= (0,0):\
\bfQ_i^{(n)} \hbox{ orthogonal to Ker adj } \bfP(i).
\end{equation}
These orthogonality conditions must be satisfied whatever be the
previously introduced arbitrary coefficients.
For a single equation, the condition {\bf C4} is simply $Q_i^{(n)}=0$.
\end{itemize}

$\bfQ_j^{(n)}$ depends on all $\bfu^{(n')}_{j'}$
with $n' \le n, j'-n'\rho \le j-n\rho, (n',j')\not= (n,j),$
and this is the only ordering to be respected during the resolution.
The costless ordering on $(n,j)$ is the one which generates stability 
conditions the sooner, and it depends on the structure of indices of the DE 
under study.

In order to avoid introducing more arbitrary coefficients than $N$,
the precise rule is~:
\begin{itemize}
\item
if $n=0$ or ($n=1$ and $i<0$), assign arbitrary values to mult($i$) components
of $\bfu^{(n)}_i$ defining a basis of Ker $\bfP(i)$,
\item
if ($n=1$ and $i\ge 0$) or $n\ge 2$, assign the value $0$ to mult($i$)
components of $\bfu^{(n)}_i$ defining a basis of Ker $\bfP(i)$.
\end{itemize}

The resulting double expansion 
(Taylor in $\varepsilon$, Laurent in $\chi$ at each order in $\varepsilon$)
can be rewritten as a
{\it Laurent series in $\chi$ extending to both infinities}
\begin{eqnarray}
 \forall n \ge 0:\
 \bfu^{(n)}
&=& \sum_{j=n \rho}^{+ \infty} \bfu^{(n)}_j \chi^{j+\bfp},\
 \bfE^{(n)}
 =  \sum_{j=n \rho}^{+ \infty} \bfE_j^{(n)} \chi^{j+\bfq},
\\
 \bfu
&=& \sum_{n=0}^{+ \infty} \varepsilon^n
  \left[\sum_{j=n \rho}^{+ \infty} \bfu_j^{(n)} \chi^{j+\bfp}\right]
 = \sum_{j=- \infty}^{+ \infty} \bfu_j \chi^{j+\bfp}.
\label{eqMethodPerturbative29}
\end{eqnarray}

{\it Remarks}.
\begin{enumerate}
\item
The Fuchsian perturbative method (as well as the nonFuchsian one which will
be seen section \ref{sectionMethodPerturbativeNonFuchsian})
is useful if and only if
the zeroth order $n=0$ fails to describe the general solution.
This may happen for two reasons.
The most common one is a negative Fuchs index in addition to $-1$ counted 
once,
the second, less common one is a multiplicity higher than one for some
family, as in the example of section \ref{sectionIndex0}.

\item
We do not know of an upper bound for $n$, but there exists a lower bound.
Indeed, in the linear inhomogeneous ODE (\ref{eqLinn}),
logarithms can arise only when some precise powers of $\chi$,
only depending on the homogeneous part,
are present in the rhs Laurent series $R^{(n)}$.
The lower bound $n$ results from the condition that the lowest Fuchs index
and the highest one, once forced to interfere by the nonlinear terms,
start to contribute to such dangerous powers.
An example of such a condition is given in section \ref{sectionOrderSeven}.

Even if all Fuchs indices are positive (except $-1$ counted once),
the lower bound on $n$ may be greater than $0$, as in the example of
section \ref{sectionIndex0}.

\item
\index{index $-1$}
{\it Remark on index $-1$.}
In the case of a single equation, since indices must be distinct integers,
the condition $Q_{\rho}^{(1)}= 0$ at the smallest Fuchs index $i=\rho$ is
identically satisfied.
Nevertheless, the frequently encountered statement 
``resonance $-1$ is always compatible'' is erroneous,
and numerous nonzero stability conditions $Q_{-1}^{(n)}= 0$ can be found
in the examples of \cite{CFP1993}.
Indeed, even at first perturbation order, the stability condition at index
$-1$ may not be satisfied~:
just like Fuchs index $i=\rho$ provides an identically satisfied stability
condition, \Plv\ ``resonance'' $-1$ has the same property if and only if
$\rho=-1,$ i.e.~if $-1$ is the smallest integer index.
 
If $\rho$ is different from $-1$,
\Plv\ resonance $-1$ {\it seems to} be satisfied,
but it is only because
a Laurent series ranging from power $p$ to $+ \infty$ cannot represent the
{\it general} solution,
thus preventing the building of \Plv\ stability condition $Q_{-1}^{(0)}=0.$
 
\end{enumerate}
\subsection
{Fuchs indices, Painlev\'e ``resonances'' or Kowalevski exponents?}
\indent

Given a nonlinear algebraic DE of order $N$,
one can define three sets of at most $N$ numbers associated to it~:
\begin{enumerate}
\item 
the Fuchs indices of the auxiliary equation of Darboux 
(section \ref{sectionMethodPerturbativeFuchsian}),
\item
the ``resonances'' of the nonlinear equation
(section \ref{sectionMethodPole}),
\item
the Kowalevski exponents,
only defined if the nonlinear equation in invariant under a 
scaling transformation
$(x,\bfu,\bfE) \to (k X,k^\bfp \bfu,k^\bfq \bfE)$.
\end{enumerate}

We have already seen the identity of the first two sets,
defined for each family of movable singularities.
Let us prove that the third notion is not distinct.
The third set is defined as follows 
(for an introduction, see \cite{Bessis1990}).
The invariance implies the particular solution (``scaling solution'')
$\bfu^{(0)}=\hbox{const }(x-x_0)^{\bfp}$,
which is identical to a family of movable singularities.
The Kowalevski exponents $\rho$ are defined as the characteristic exponents
of the linearized system near this solution,
which proves the identity of the three notions.

Said differently, all these numbers are Fuchs indices,
and this link to the theory of {\it linear} DEs proves the uselessness
of the notions of Kowalevski exponents and Painlev\'e resonances.

\subsection{Understanding negative Fuchs indices}
\indent

The ODE with a meromorphic general solution \cite{CFP1993}
\begin{equation}
E \equiv u_{xx} + 3 u u_x + u^3 = 0,\
 u={1 \over x-a} + {1 \over x-b},\ a \hbox{ and } b \hbox{ arbitrary},
\label{eqTwopoles}
\end{equation}
has two families,
\begin{description}
\item (F1) $p=-1,u_0^{(0)}=1$, indices $(-1,1)$,
\item (F2) $p=-1,u_0^{(0)}=2$, indices $(-2,-1)$,
\end{description}
and this provides a clear comprehension of negative Fuchs indices,
since the index $-2$ must coexist with the meromorphy.
Indeed, the representation of the general solution (\ref{eqTwopoles})
as a Laurent series of $x-x_0$
is the sum of two copies of an expansion of $1/(x-c)$,
and there exist two expansions of $1/(x-c)$
\begin{eqnarray}
(x-c)^{-1}
& = &
\sum\limits_{j=- \infty}^{-1} (c-x_0)^{-1-j} (x-x_0)^{j},\
\mod{c-x_0} < \mod{x-x_0}
\\
& = &
\sum\limits_{j=0}^{+ \infty} - (c-x_0)^{1-j} (x-x_0)^{j},\
\mod{x-x_0} < \mod{c-x_0}.
\end{eqnarray}
The family (F1) corresponds to the sum
(first expansion with $c=a=x_0$) plus (second expansion $c=b$),
while 
the family (F2) corresponds to the sum
(first expansion with $c=a$) plus (second expansion with $c=b$).
This can be checked by a direct application of the algorithm,
which for family (F2) gives \cite{CFP1993}
\begin{eqnarray}
u
& = & 2 \chi^{-1} 
    + \varepsilon (A_1 \chi^{-3} + B_1 \chi^{-2}) 
    + \varepsilon^2
      ({A_1^2 \over 2} \chi^{-5} + {3 A_1 B_1 \over 2} \chi^{-4}) 
\nonumber
\\
& &
    + \varepsilon^3
      ({A_1^3 \over 4} \chi^{-7} + {5 A_1^2 B_1 \over 4} \chi^{-6}
       +A_1 B_1^2 \chi^{-5}-{1 \over 2} B_1^3 \chi^{-4})
    + O(\varepsilon^4)
\\
& = & 2 \chi^{-1}
    + \varepsilon B_1 \chi^{-2}
    + \varepsilon A_1 \chi^{-3}
   +({3 \over 2}\varepsilon^2A_1 B_1 -{1 \over 2}\varepsilon^3 B_1^3)\chi^{-4}
    + O(\chi^{-5})
\\
& = & {2 \chi - \varepsilon B_1 \over
 \chi^2 - \varepsilon B_1 \chi + {1 \over 2} (- \varepsilon A_1 +
 \varepsilon^2 B_1^2)},
\end{eqnarray}
where $A_1$ and $B_1$ are the arbitrary coefficients at order one.
The simple pole $\chi=0$ with residue $2$ has been ``unfolded''
by the perturbation
into two simple poles with residue $1$, at the two arbitrary locations
${1 \over 2}
\left[\varepsilon B_1\pm\sqrt{2 \varepsilon A_1- \varepsilon^2 B_1^2}\right]$,
both close to $0$.

For other examples, see \cite{PickeringThesis} and conference proceedings
referenced in \cite{CFP1993}.

\subsection{The simplest constructive example}
\label{sectionIndex0}
\indent

The equation
\begin{equation}
u''+4 u u' + 2 u^3=0
\label{eqOrder2WithLog}
\label{eqdoublefamily}
\end{equation}
is the simplest constructive example, because

\begin{enumerate}
\item
there exists a movable logarithm,
as shown by the $\alpha-$method (BSMF \S 13, p 221),

\item
the method of pole-like expansions fails to find it,

\item
the assumption of a ``descending'' Laurent series (\ref{eqMethodPole2})
fails to find it,

\item
the Fuchsian perturbative method finds it after a very short computation,
as we now show.

\end{enumerate}

There exists a single family
\begin{eqnarray}
& &
p=-1,\
E_0^{(0)}= u_0^{(0)} (u_0^{(0)}-1)^2=0,\
\hbox{indices } (-1,0),
\end{eqnarray}
with the puzzling fact that $u_0^{(0)}$ should be at the same time 
equal to $1$ according to the equation $E_0^{(0)}=0$,
and arbitrary according to the index $0$.
The application of the method provides
\begin{eqnarray}
u^{(0)}
& = & 
\chi^{-1} \hbox{ (the series terminates)}
\\
E'(x,u^{(0)}) 
& = &
\partial_x^2 + 4 \chi^{-1} \partial_x + 2 \chi^{-2}
\\
u^{(1)}
& = &
u_{0}^{(1)} \chi^{-1},\
u_{0}^{(1)}) \hbox{ arbitrary},\
\\
E^{(2)}
& = &
E'(x,u^{(0)}) u^{(2)} + 6 u^{(0)} u^{(1)^2} + 4 u^{(1)} u^{(1)'}
\nonumber
\\
& = &
\chi^{-2} (\chi^2 u^{(2)})'' + 2 u^{(0)^2} \chi^{-3}
=0
\\
u^{(2)}
& = &
- 2 u_{0}^{(1)^2} (\chi^{-1} \Log \chi - \chi^{-1}).
\end{eqnarray}
The movable logarithmic branch point is therefore detected in a systematic way
at order $n=2$ and index $i=0$.

The necessity to perform a perturbation arises from the multiple root
of the equation for $u_0^{(0)}$,
responsible for the insufficient number of arbitrary parameters in
the zeroth order series $u^{(0)}$.

\subsection{The two examples}
\label{sectionMethodPerturbativeExamples}
\indent

{\it Example} 1. ``Complete (P1)'' eq.~(\ref{eqP1complete}).
The method is useless.

{\it Example} 2 (Chazy complete equation of class III (\ref{eqChazyClassIII})).

For the {\it second family} (in case $d_3=0$), the method is useless.

For the {\it first family}, since all indices are negative, one must start the
perturbation process $n\ge 1$.

To obtain the stability conditions up to a given order $n\ge 1$, 
we only need to compute the first $3n$ coefficients of each element~:
\begin{equation}
u^{(r)}=\sum_{j=-3r}^{-3r+3n-1}u_j^{(r)}\chi^{j-1}, r=0,\dots ,n,
\end{equation}
i.e.~ 
\begin{equation}
   u_{0    :3n-1}^{(0  )},\   u_{-3   :3n-4}^{(1  )},\ \dots,
   u_{-3n+3:   2}^{(n-1)},\   u_{-3n  :  -2}^{(n  )}, 
\end{equation}
where $j_1:j_2$ denotes a range of $j$ values.
The most efficient way to perform the double loop on $(n,j)$ is
to perform the outside loop in the variable $k=j - n \rho,$ with $\rho=-3$,
and the precise double loop is~:
for $k=0$ to $k_{\hbox{max}}$ do 
for $n=(\hbox{if } k=0 \hbox{ then } 1 \hbox{ else } 0)$ to $n_{\hbox{max}}$ 
do solve the linear algebraic equation (\ref{eqMethodPerturbative8})
for $u_{j}^{(n)}, j=k + n \rho$.
 
Let us compute all the stability conditions at first and second order.
The computer printout reads (full details are given in Ref.~\cite{CFP1993})~:
\begin{eqnarray}
  k=0:
& &
  Q_{-3}^{(1)} \equiv 0,\
  u_{-3}^{(1)} \hbox{ arbitrary}
\\
  k=1:
& &
  Q_{-2}^{(1)} \equiv -6 (5 c_1 + 42 d_3) u_{-3}^{(1)} = 0,\
  u_{-2}^{(1)} \hbox{ arbitrary}
\\
  k=2:
& &
 Q_{-1}^{(1)} \equiv 
  -12 (c_1 + 9 d_3) u_{-2}^{(1)} + 18 (c_1 - 8 d_3) u_{-3}^{(1)'} 
\nonumber
\\
& &
 +{6 \over 5} (2 c_0 - 3 c_1^2- 117 c_1 d_3 - 594 d_3^2
 + 18 c_1' + 108 d_3') u_{-3}^{(1)} =0,
\\
& &
  u_{-1}^{(1)} \hbox{ arbitrary},\ 
\\
  k=3:
& &
  Q_{-3}^{(2)} \equiv - {66 \over 5} d_1 u_{-3}^{(1)^2} = 0,
\\
k=4:
& &
  Q_{-2}^{(2)} \equiv {1 \over 7} (8 d_0 + 57 d_1') u_{-3}^{(1)^2}
               - 12 d_1 u_{-3}^{(1)} u_{-2}^{(1)} = 0
\\
k=5:
& &
  Q_{-1}^{(2)} \equiv - {1 \over 35} (18 d_0' + 99 d_1'')
         u_{-3}^{(1)^2} - {24 \over 5} d_1 u_{-2}^{(1)^2}
\nonumber
\\
& &
  + {3 \over 35} (16 d_0 + 72 d_1') u_{-3}^{(1)} u_{-2}^{(1)} = 0.
\end{eqnarray}

Five conditions are obtained, three at order one, equivalent to 
$ d_3=c_1=c_0=0$, and two at order two, equivalent to $ d_1=d_0=0$
[in order to simplify expressions, we have put the first order conditions in
the above expressions for $ k = j + 3 n \ge 3 $],
after seventeen values of $(n,j)$.
These conditions were given without any detail by Chazy\cite{ChazyThese}.
They restrict the complete ODE (\ref{eqChazyClassIII}) 
to the simplified ODE (\ref{eqChazyIII}),
{\it modulo} (\ref{eqGroupHomographicContinuous}).
 
Chazy proved the general solution of (\ref{eqChazyIII}) to be single valued 
inside or outside a circle whose centre and radius depend on the choice of the
three arbitrary constants; it is holomorphic in this domain, and the only
singularity is a movable natural boundary (``coupure essentielle'')
defined by this circle.
He also gave a parametric representation of the general solution $u(x)$
in terms of two solutions of the (linear) hypergeometric equation,
but single valuedness is not at all apparent on this representation.
 
The direct explicit solution of Bureau \cite{Bureau1972,Bureau1987}
is given section \ref{sectionMethodAlphaExamples}.

\subsection{An example needing order seven to conclude}
\label{sectionOrderSeven}
\indent

The following equation, isolated by Bureau (\cite{BureauMII} p.~79),
\begin{equation}
 u'''' + 3 u u'' - 4 u'^2 = 0
\label{eqBureauOrder4}
\end{equation}
possesses the two families
\begin{eqnarray}
p
& = & -2, u_0^{(0)}=-60, \hbox{ ind. } (-3,-2,-1,20),\
  {\hat K}=u'''' + 3 u u'' - 4 u'^2, 
\\
p
& = & -3, u_0^{(0)} \hbox{ arbitrary}, \hbox{ indices } (-1,0),\
  {\hat K}= 3 u u'' - 4 u'^2. 
\label{eqBureau4p3}
\end{eqnarray}
 
The {\it second family} has a Laurent series $(p:+ \infty)$ which happens to
terminate \cite{CFP1993}
\begin{equation}
 u^{(0)}=c (x-x_0)^{-3}-60(x-x_0)^{-2},\ (c,x_0) \hbox{ arbitrary}.
\label{eqBureau4PartSol}
\end{equation}
The Fuchsian perturbative method is useless,
for the two arbitrary coefficients corresponding to the two Fuchs indices
are already present at zeroth order.
 
The {\it first family} provides, at zeroth order, only a two-parameter
expansion and,
when one checks the existence of the perturbed solution
\begin{equation}
 u=\sum_{n=0}^{+ \infty} \varepsilon^n
 \left[\sum_{j=0}^{+ \infty} u_j^{(n)} \chi^{j-2-3n}\right],
\end{equation}
one finds that coefficients
$u_{20}^{(0)}, u_{-3}^{(1)}, u_{-2}^{(1)}, u_{-1}^{(1)}$
can be chosen arbitrarily,
and, at order $n=7$, one finds two violations \cite{CFP1993}
\begin{equation}
   Q_{-1}^{(7)} \equiv u_{20}^{(0)}   u_{-3}^{(1)^7} = 0,
   Q_{20}^{(7)} \equiv u_{20}^{(0)^2} u_{-3}^{(1)^6} u_{-2}^{(1)} = 0,
\end{equation}
implying the existence of a movable logarithmic branch point.
 
{\it Remark}.
The value $n=7$ is the root of the linear equation 
$n (i_{\rm min}-p) + (i_{\rm max}-p)=-1$,
with $p=-2,i_{\rm min}=-5,i_{\rm max}=18$,
linking the pole order $p$ in the Fuchsian case $c=0$,
the smallest and the greatest Fuchs indices.
It expresses the condition for the first occurence of a power $\chi^{-1}$,
leading by integration to a logarithm,
in the r.h.s.~$R^{(n)}$ of the linear inhomogeneous equation
(\ref{eqLinn}),
r.h.s.~created by the nonlinear terms $3 u u'' - 4 u'^2 $.

\subsection{Closed-form solutions of the Bianchi IX model}
\label{sectionBianchiClosedForms}
\indent

In this example, the no-log conditions are used in a constructive way,
in order to isolate all possible single valued solutions.

The Bianchi IX cosmological model 
\cite{LandauLifshitzTheorieChamps}
is a system of three second order ODEs
\begin{equation}
(\Log A)'' = A^2 - (B-C)^2
\hbox{ and cyclically},\
'=\D / \D \tau,
\label{eqBianchi1}
\end{equation}
or equivalently
\begin{equation}
(\Log \omega_1)''
= \omega_2^2 + \omega_3^2 - \omega_2^2 \omega_3^2 / \omega_1^2,\
A= \omega_2 \omega_3 / \omega_1,\
\omega_1^2=B C
\hbox{ and cyclically}.
\end{equation}

One of the families \cite{CGR1993,LMC1994}
\begin{eqnarray}
& &
A= \chi^{-1} + a_2 \chi + O(\chi^3),\ \chi=\tau-\tau_2,
\nonumber
\\
& &
B= \chi^{-1} + b_2 \chi + O(\chi^3),
\label{eqLaurentFamilyTwoOrder0}
\\
& &
C= \chi^{-1} + c_2 \chi + O(\chi^3),
\nonumber
\end{eqnarray}
has the Fuchs indices $(-1,-1,-1,2,2,2)$.
The Fuchsian perturbative method 
\begin{eqnarray}
& &
A=\chi^{-1} \sum_{n=0}^N \varepsilon^n
\sum_{j=-n}^{2+N-n} a_{j}^{(n)} \chi^{j},\
\chi=\tau-\tau_2,\
\hbox{ and cyclically},
\label{eqBianchi4}
\end{eqnarray}
then gives a failure of condition {\bf C4}
at $(n,i)=(3,-1)$ and $(5,-1)$ \cite{LMC1994},
and the satisfaction of these no-log conditions generates the three solutions~:
\begin{eqnarray}
& &
(b_{ 2}^{(0)}=c_{ 2}^{(0)} \hbox{ and } b_{-1}^{(1)}=c_{-1}^{(1)})
\hbox{ or cyclically}
\label{eqSelectTaub}
\\
& &
a_{2}^{(0)}=b_{2}^{(0)}=c_{2}^{(0)}=0,\
\label{eqSelectHalphen}
\\
& &
a_{-1}^{(1)}=b_{-1}^{(1)}=c_{-1}^{(1)}.
\label{eqSelectEuler}
\end{eqnarray}
These are constraints which reduce the number of arbitrary coefficients
to, respectively, four, three and four,
thus defining particular solutions which may have no movable critical points.
The question is~: do they define additional solutions to what is known?

The only three closed-form solutions which are known are single valued,
they are defined as the general solution of the following three subsystems.
\begin{enumerate}

\item
The 4-dim axisymmetric case $B=C$ \cite{Taub},
whose general solution (\ref{eqTaubNoMetric}) is trigonometric.

\item
The 3-dim {\it Darboux-Halphen system} \cite{Darboux1878,Halphen1881}
\begin{equation}
\omega_1' 
= \omega_2 \omega_3 - \omega_1 \omega_2 - \omega_1 \omega_3,\ 
\hbox{ and cyclically}.
\label{eqDarboux}
\end{equation}

\item
The 3-dim {\it Euler system} (1750) \cite{BGPP},
describing the motion of a rigid body around its center of mass
\begin{equation}
\omega_1' = \omega_2 \omega_3,
\hbox{ and cyclically},
\label{eqEuler}
\end{equation}
whose general solution is elliptic \cite{BGPP},
see (\ref{eqDiagonalHoyer}) and (\ref{eqEulerPoinsot}).

\end{enumerate}

The first constraint (\ref{eqSelectTaub})
implies the equality of two of the components $(A,B,C)$ at every order
and thus represents the four-parameter solution of Taub (\ref{eqTaubNoMetric}).

The second constraint (\ref{eqSelectHalphen}) 
represents the three-parameter solution of the Darboux-Halphen system 
(\ref{eqDarboux}).

The third and last constraint (\ref{eqSelectEuler})
represents an extrapolation to four parameters of
the three-parameter solution ot the Euler system described by
$a_{2}^{(0)}+b_{2}^{(0)}+c_{2}^{(0)}= 0$.
This would-be four-parameter, global, closed form, single valued
exact solution has not yet been found.

\section{The nonFuchsian perturbative method}
\label{sectionMethodPerturbativeNonFuchsian}
\indent

\index{nonFuchsian singularity}

Whenever the family under study has a number of Fuchs indices 
smaller than the differential order $N$,
the Fuchsian perturbation method fails to build a
representation of the general solution,
thus possibly missing some stability conditions.
Examples are (\ref{eqBureau4p3})
and the second family of (\ref{eqChazyClassIII}) in the case $d_3=0$.
The missing solutions of the auxiliary equation (\ref{eqLin0})
are then nonFuchsian solutions, see section \ref{sectionNonFuchs}.

There is no difficulty to algorithmically compute the nonFuchsian expansions
(\ref{eqFundamentalSetNonFuchs}),
but these are of no immediate help, due to their generic divergence.

There is one situation where some stability conditions can be generated
{\it algorithmically}
(indeed, we are not interested in computations adapted to a given equation, 
only in computerizable methods).
It occurs when the two following conditions are met \cite{MC1995}.

\begin{enumerate}
\item
There exists a particular solution $\bfu=\bfu^{(0)}$
which is known globally, meromorphic and has at least one 
movable pole at a finite distance denoted $x_0$.

\item 
The only singular points of the linearized equation $\bfE^{(1)}=0$
are $x=x_0$, nonFuchsian,
and $x=\infty$, Fuchsian.

\end{enumerate}

Then, the property that a fundamental set of solutions $\bfu^{(1)}$
be locally single valued near $\chi=x-x_0=0$ 
is equivalent to the same property near $\chi=\infty$.
This is the global nature of $\bfu^{(0)}$ which allows the study of the point 
$\chi=\infty$, easy to perform with the Fuchsian perturbation method.

An important technical bonus is the lowering of the differential order $N$
of equation $\bfE^{(1)}=0$ by the number $M$ of arbitrary parameters $c$ which
appear in $\bfu^{(0)}$.
Indeed, again since $\bfu^{(0)}$ is closed form,
its partial derivatives $\partial_c \bfu^{(0)}$ are closed form and are
particular solutions of $\bfE^{(1)}=0$,
which allows this lowering of the order.

At each higher perturbation order $n \ge 2$,
one similarly builds particular solutions $\bfu^{(n)}$ 
as expansions near $\chi=\infty$ and one requires the same properties.

\subsection{An explanatory example~: Chazy's class III ($N=3,M=2$)
\label{sectionChazy}}
\indent

The simplified equation (\ref{eqChazyIII}),
which possesses the PP \cite{ChazyThese} 
and therefore for which no $u^{(n)}$ is multivalued,
is quite useful just to understand the method.
This equation admits the global two-parameter solution
(\ref{eqChazyIIIPartSol})
$u^{(0)}=c \chi^{-2} - 6 \chi^{-1}$.
The linearized equation
\begin{equation}
E^{(1)} \equiv E'(x,u^{(0)}) u^{(1)} \equiv 
 [              \partial_x^3 
  - 2 u^{(0)}   \partial_x^2
  + 6 u^{(0)}_x \partial_x
  - 2 u^{(0)}_{xx}] u^{(1)} = 0
\label{eqChazyLin}
\end{equation}
possesses the two single valued global solutions
$\partial_{x_0} u^{(0)}$ and $\partial_c u^{(0)}$,
i.e.~$u^{(1)}=\chi^{-3},\chi^{-2}$,
and it has only two singular points $\chi=0$ (Fuchsian)
and $\chi=\infty$ (nonFuchsian with Thom\'e rank two).
The lowering by $M=2$ units of the order of the linearized equation 
results from the change of function
\begin{equation}
u^{(1)}=\chi^{-3} v:\
E^{(1)} \equiv \chi^3 [\partial_x + 3 \chi^{-1} - 2 c \chi^{-2}] v''=0,
\end{equation}
and the study of the Fuchsian point $\chi=\infty$ yields an integer Fuchs 
index,
which proves the {\it global} single valuedness of the general solution
$u^{(1)}$.

{\it Remarks}.
\begin{itemize}

\item
The local study of $\chi=0$
provides a formal expansion (\ref{eqFundamentalSetNonFuchs}) which happens
to terminate, a nongeneric situation, thus providing the
fundamental set of {\it global} solutions at perturbation order $n=1$
\begin{eqnarray}
\forall \chi\
\forall c:\
u^{(1)} & = & \chi^{-2},\
              \chi^{-3},\
              (e^{-2 c / \chi} - 1 + 2 c \chi^{-1}) \chi^{-2} / (2 c^2).
\label{eqChazySGOrder1}
\end{eqnarray}
This proves the existence of an essential singularity at $\chi=0$
(ref.~\cite{Ince} chap.~XVII).

\item
Going on with the formalism of Painlev\'e's lemma at higher orders
constitutes the rigorous mathematical framework of the local representation
of the general solution obtained by Joshi and Kruskal \cite{JoshiKruskal1993}
\begin{equation}
u= - 6 \chi^{-1}
+ c \chi^{-2} 
\left(1 + z  - {z^2 \over 8}  + {z^3 \over 144} - 7 {z^4 \over 13824}
+ O(\varepsilon^5)\right),\
z={\varepsilon \over c} e^{- 2 c / \chi}.
\end{equation}
This representation reduces to the one given by Chazy (Taylor series in 
$1/ \chi$)
if one starts from the Fuchsian family $u \sim -6 \chi^{-1}$.


\end{itemize}

\subsection{The fourth order equation of Bureau ($N=4,M=2$)
\label{eqsectionBureau}}
\indent

In section \ref{sectionOrderSeven}, 
the fourth order equation (\ref{eqBureauOrder4}) has been proven to be
unstable after a computation practically untractable without a computer.
Let us now prove this result without computation at all \cite{MC1995}.
For the global two-parameter solution (\ref{eqBureau4PartSol}),
the linearized equation
\begin{equation}
E^{(1)} = E'(x,u^{(0)}) u^{(1)} \equiv 
 [              \partial_x^4 
  + 3 u^{(0)}   \partial_x^2
  - 8 u^{(0)}_x \partial_x
  + 3 u^{(0)}_{xx}] u^{(1)} = 0
\label{eqBureauLin}
\end{equation}
has only two singular points $\chi=0$ (nonFuchsian) and $\chi=\infty$
(Fuchsian),
it admits the two global single valued solutions
$\partial_{x_0} u^{(0)}$ and $\partial_c u^{(0)}$,
i.e.~$u^{(1)}=\chi^{-4},\chi^{-3}$.
The lowering by $M=2$ units of the order of the linearized equation
(\ref{eqBureauLin}) is obtained with
\begin{equation}
u^{(1)}=\chi^{-4} v:\
[\partial_x^2 -16 \chi^{-1} \partial_x +3 c \chi^{-3} - 60 \chi^{-2}] v'' = 0,
\end{equation}
and the local study of $\chi=\infty$ is unnecessary,
since one recognizes the confluent hypergeometric equation.
The two other solutions in global form are
\begin{eqnarray}
c \not=0:\
v_1''
& = &
\chi^{-3} {}_{0} F_{1} (24;-3c/\chi)
=
\chi^{17/2} J_{23}(\sqrt{12 c/\chi}),
\\
v_2''
& = &
\chi^{17/2} N_{23}(\sqrt{12 c/\chi}),
\end{eqnarray}
where the hypergeometric fonction ${}_{0} F_{1} (24;-3c/\chi)$
is single valued and possesses an isolated essential singularity at $\chi=0$,
while the fonction $N_{23}$ of Neumann is multivalued because of a
$\Log \chi$ term.

{\it Remark}.
The local study of (\ref{eqBureauLin}) near $\chi=0$
provides the formal expansions (\ref{eqFundamentalSetNonFuchs}) for the 
two nonFuchsian solutions
\begin{equation}
\chi \to 0,\
c \not=0:\
u^{(1)}=e^{\pm \sqrt{-12c/ \chi}}\chi^{31/4} (1 + O(\sqrt{\chi})),
\label{eqBureau4Formal}
\end{equation}
detecting the presence in (\ref{eqBureauLin}) 
of an essential singularity at $\chi=0$,
but the generically null radius of convergence of the formal series forbids
to conclude to the multivaluedness of $u^{(1)}$.
A nonobvious result is the existence, as seen above, 
of a linear combination of the two formal expansions (\ref{eqBureau4Formal})
which is single valued.

\subsection{An example in cosmology~: Bianchi IX ($N=6,M=4$)
\label{sectionBianchi}}
\indent

The Bianchi IX cosmological model in vacuum (\ref{eqBianchi1})
does not possess the PP \cite{CGR1994,LMC1994}.
Let us prove it rapidly \cite{LMC1994,MC1995}.
Taub \cite{Taub} found the general solution of the axisymmetric case of
two equal components,
a meromorphic expression (\ref{eqTaubNoMetric})
depending on the four arbitrary parameters $(k_1,k_2,\tau_1,\tau_2)$.
The linearized system generated by the perturbation
\begin{eqnarray}
& &
A= A^{(0)} (1 + \varepsilon A^{(1)} + O(\varepsilon^2))
\hbox{ and cyclically}
\end{eqnarray}
has the differential order $N=6$,
which is then lowered by $M=4$ units by the change of function
dictated by the symmetry of the system~:
$P^{(1)}=B^{(1)}+C^{(1)},M^{(1)}=B^{(1)}-C^{(1)}$
\begin{eqnarray}
& &
A^{(1)''} - 2 A^{(0)^2} A^{(1)} = 0,\
\label{eqTaub1A}
\\
& &
P^{(1)''} - 2 A^{(0)} B^{(0)} P^{(1)}
 = 4 (A^{(0)} B^{(0)} - A^{(0)^2}) A^{(1)},\
\label{eqTaub1P}
\\
& &
M^{(1)''} + 2 (A^{(0)} B^{(0)} - 2 B^{(0)^2}) M^{(1)} = 0.
\label{eqTaub1M}
\end{eqnarray}
Indeed, the four single valued global solutions
\begin{equation}
(A^{(1)},P^{(1)})=\partial_c (\Log A^{(0)},\Log (B^{(0)}+C^{(0)})),\
c=k_1,k_2,\tau_1,\tau_2,
\end{equation}
are those of the equations (\ref{eqTaub1A})--(\ref{eqTaub1P}),
\begin{equation}
M^{(1)}=0,
(A^{(1)},P^{(1)} + 2 A^{(1)})=
\left\{
\begin{array}{l}
(
(\tau - \tau_1) \coth k_1 (\tau - \tau_1) - 1/k_1
,
0
),
\\
(
0
,
(\tau - \tau_2) \coth k_2 (\tau - \tau_2) - 1/k_2
),
\\
(
\coth k_1 (\tau - \tau_1)
,
0
),
\\
(
0
,
\coth k_2 (\tau - \tau_2)
),
\end{array}
\right.
\nonumber
\end{equation}
and there only remains to study the equation (\ref{eqTaub1M}).
It has a countable infinity of singular points~:
$\tau-\tau_2=i m \pi /k_2, m \in {\cal Z}$ 
(nonFuchsian, of Thom\'e rank two),
accumulating at $\tau=\infty$.
This uneasy situation can be overcome by taking the limit $k_1=k_2=0$;
it is indeed sufficient to exhibit a movable logarithm in this limit,
for it will persist for $(k_1,k_2)\not=(0,0)$. In this limit
\begin{eqnarray}
k_1=k_2=0:\
& &
{\D^2 M^{(1)} \over \D t^2}
 + \left({2 \over t^2} - {4 (t-1)^2 \over t^4} \right) M^{(1)}=0,\
t={\tau - \tau_2 \over \tau_1 - \tau_2},
\end{eqnarray}
the only singular points are $t=0$ (nonFuchsian) and $t=\infty$ (Fuchsian), 
the optimal situation.
The Fuchs indices being $-2$ and $1$,
the computation of three terms is sufficient to exhibit a logarithm,
and this proves the absence of the Painlev\'e property
for the Bianchi IX model in vacuum.

{\it Remarks}
\begin{enumerate}
\item
The two solutions are globally known \cite{LMC1994}~:
\begin{eqnarray}
k_1=k_2=0:\
& &
M^{(1)}=
e^{-2/t} t^{-1},\
e^{-2/t} t^{-1} \int^{1/t} z^{-4} e^{4 z} \D z,
\end{eqnarray}
which shows the presence of a logarithmic branch point at $t=0$, 
or at $t=\infty$ as well.

\item 
The two formal non-Fuchsian solutions are 
\begin{equation}
\tau - \tau_2 \to 0:\
M^{(1)}=
e^{\alpha / (\tau - \tau_2)}
\sum_{k=0}^{+ \infty} \lambda_k (\tau-\tau_2)^{k+s},\
\lambda_0 \not=0,
\end{equation}
with
\begin{equation}
\alpha=  \pm 2 k_1^{-1} \sinh k_1 (\tau_2 - \tau_1),\
s     =1 \mp 2          \cosh k_1 (\tau_2 - \tau_1).
\end{equation}
The two generically irrational values for the Thom\'e exponents $s$
allow to conclude
only if the divergent series $\lambda_k (\tau-\tau_2)^{k}$ can be summed.

\end{enumerate}

\section{Miscellaneous perturbations}
  \label{Miscellaneous perturbations}
\indent

The differential complexity of the $\alpha-$method explains why
it usually succeeds in case of failure of all the other methods,
which only have an algebraic complexity.
Consider the ODEs, none of which admits a power-law leading behaviour
\begin{eqnarray}
& &
-2 u u'' + 3 u'^2 + d_3 u^3=0,\ d_3 \not=0,
\label{eqNoDomOrder2}
\\
& &
u''' + u u'' - 2 u'^2 = 0,
\label{eqNoDomOrder3}
\\
& &
u'''' + 2 u u'' - 3 u'^2 = 0,
\label{eqNoDomOrder4}
\end{eqnarray}
and let us prove that each of them has movable logarithms.
The first one is extracted from Chazy's class III (\ref{eqChazyClassIII})
by the perturbation $u=\varepsilon^{-1} U, x=x_0 + \varepsilon X$,
and it represents its second family, see section
\ref{PoleLikeExpansionExamples}.
The second and third ones were considered by Chazy
\cite{Chazy1912c,Chazy1918}
who had to establish a special theorem, using divergent series,
to exhibit the movable logarithms.
Having degree one, none of these ODEs admits singular solutions.

The first equation (\ref{eqNoDomOrder2}) is classically processed by the
$\alpha-$method
\begin{eqnarray}
u &= &
    \varepsilon^{-1} \sum_{n=0}^{+ \infty} \varepsilon^n u^{(n)},\
E = \varepsilon^{-4} \sum_{n=0}^{+ \infty} \varepsilon^n E^{(n)},\
x=x_0 + \varepsilon X,
\end{eqnarray}
resulting in
\begin{eqnarray}
E^{(0)} & \equiv & - 2 u^{(0)} u^{(0)''} + 3 u^{(0)'^2} =0 
\\
u^{(0)} & = & c (X-X_0)^{-2},\ (X_0,c) \hbox{ arbitrary},
\\
E^{(1)} & \equiv &
     c (X-X_0)^{-5} [-2 ((X-X_0)^3 u^{(1)})'' + c^2 d_3/(X-X_0)] = 0,
\\
u^{(1)} & = & c^2 d_3 (X-X_0)^3 [(X-X_0) \Log (X-X_0) - (X-X_0)]/2.
\end{eqnarray}
and proving the instability at perturbation order one.

For equations (\ref{eqNoDomOrder3}) and (\ref{eqNoDomOrder4}),
there exists no perturbation satisfying the assumptions of Theorem II
page \pageref{pageTheoremII},
there only exist singular perturbations,
\ie which discard the highest derivative.
Since they however give the correct information, it would be desirable to
extend Theorem II in that direction.

Equation (\ref{eqNoDomOrder3}) is handled by the singular perturbation
\begin{eqnarray}
u &= &
    \varepsilon^{-1} \sum_{n=0}^{+ \infty} \varepsilon^n u^{(n)},\
E = \varepsilon^{-2} \sum_{n=0}^{+ \infty} \varepsilon^n E^{(n)},\
\label{eqPerturmationMisc1}
\end{eqnarray}
which excludes $u'''$ from the simplified equation
\begin{eqnarray}
E^{(0)} & \equiv & u^{(0)} u^{(0)''} - 2 u^{(0)'^2} =0 
\\
u^{(0)} & = & c \chi^{-1},\ \chi=x-x_0,\ (x_0,c) \hbox{ arbitrary},
\\
E^{(1)} & \equiv & c (\chi^{-3} (\chi^2 u^{(1)})'' - 6 \chi^{-4}) = 0,
\\
u^{(1)} & = & 6 \chi^{-1} (\Log \chi-1).
\end{eqnarray}

This same perturbation (\ref{eqPerturmationMisc1}) solves the case of
the equation (\ref{eqNoDomOrder4})
\begin{eqnarray}
E^{(0)} & \equiv & 2 u^{(0)} u^{(0)''} - 3 u^{(0)'^2} =0 
\\
u^{(0)} & = & c \chi^{-2}
\\
E^{(1)} & \equiv & c (2 \chi^{-5} (\chi^3 u^{(1)})'' + 120 \chi^{-6}) = 0
\\
u^{(1)} & = & -60 \chi^{-2} (\Log \chi-1).
\end{eqnarray}

\section{The perturbation of the continuum limit of a discrete equation}
\indent

Discrete equations can be considered as
functional equations linking the values taken by some
field variable $u$ at a finite number $N+1$ of points,
either arithmetically consecutive~: $x + k \pas, k-k_0=0,1, \dots, N$,
or      geometrically consecutive~: $x \pasq^k,  k-k_0=0,1, \dots, N$,
where $\pas$ or $\pasq$ is the lattice stepsize,
assumed to lay in some neighborhood of, respectively, $0$ or $1$,
and $k_0$ is just some convenient origin.

\index{discrete Painlev\'e property}
{\it Definition} \cite{CM1996}.
A discrete equation is said to possess the
{\it discrete Painlev\'e property}
if and only if
there exists a
neighborhood of $\pas=0$
at every point of which
the general solution $x \to u(x,\pas)$ 
has no movable critical singularities.

Consider an arbitrary discrete equation (\ref{eqDiscretexpas}),
\begin{eqnarray}
& &
\forall x\ \forall \pas\ :\
E(x,\pas,\{u(x+k \pas),\ k-k_0=0,\dots,N\})=0
\label{eqDiscretexpas}
\end{eqnarray}
algebraic in the values of the field variable,
with coefficients analytic in $x$, the stepsize and some parameters $a$.
Let $(x, \pas, u, a) \to (X, \Pas, U, A, \varepsilon)$
be an arbitrary perturbation admissible by the suitable extension of 
the theorem of Poincar\'e to discrete systems.
Two such perturbations are well known,
the {\it autonomous limit}
\begin{eqnarray}
& &
x=x_0 +\varepsilon X,\
\pas = \varepsilon H,\
u    = U,\
a    = \hbox{ analytic }(A, \varepsilon),
\end{eqnarray}
and the {\it continuum limit}
\begin{eqnarray}
& &
x      \hbox{ unchanged},\
\pas = \varepsilon,\
u    = U,\
a    = \hbox{ analytic }(A, \varepsilon).
\end{eqnarray}
The latter can be extended into
a {\it perturbation of the continuum limit} \cite{CM1996}
\index{perturbation of the continuum limit}
\begin{eqnarray}
& &
x      \hbox{ unchanged},\
\pas = \varepsilon,\
u    = \sum_{n=0}^{+ \infty} \varepsilon^n u^{(n)},\
a    = \hbox{ analytic }(A, \varepsilon),
\label{eqPerturbationLimiteContinue}
\end{eqnarray}
entirely analogous to
the Fuchsian (section \ref{sectionMethodPerturbativeFuchsian})
or nonFuchsian (section \ref{sectionMethodPerturbativeNonFuchsian})
perturbative method of the continuous case.

It generates an infinite sequence of (continuous) differential equations
$E^{(n)}=0$
whose first one $n=0$ is the continuum limit.
The next ones $n\ge 1$, which are linear inhomogeneous,
have the same homogeneous part $E^{(0)'} u^{(n)}=0$ independent of $n$,
defined by the derivative of the equation of the continuum limit,
while their inhomogeneous part $R^{(n)}$ (``right-hand side'')
comes at the same time from the nonlinearities and the discretization.

Let us just handle the Euler scheme for the Bernoulli equation 
\begin{equation}
E \equiv (\overline{u} - u) / \pas + u^2 = 0
\label{eqdRiccatiEuler}
\end{equation}
(notation is $u=u(x),\overline{u}=u(x+ \pas)$),
i.e.~the logistic map of Verhulst,
a paradigm of chaotic behaviour.
Let us expand the terms of (\ref{eqdRiccatiEuler})
according to the perturbation (\ref{eqPerturbationLimiteContinue})
up to an order in $\varepsilon$ sufficient to build the first 
equation $E^{(1)}=0$ beyond the continuum limit $E^{(0)}=0$
\begin{eqnarray}
u & = &
u^{(0)} + u^{(1)} \varepsilon + O(\varepsilon^2)
\\
u^2 & = &
u^{(0)^2}
 + 2 u^{(0)} u^{(1)} \varepsilon 
 + O(\varepsilon^2)
\\
\overline{u} & = &
u + u' \pas + (1/2) u'' \pas^2 + O(\pas^3)
\\
{\overline{u} - u \over \pas}
& = &
u^{(0)'} + (u^{(1)'} + (1/2) u^{(0)''}) \varepsilon + O(\varepsilon^2).
\end{eqnarray}
The equations of orders $n=0$ et $n=1$ are written as
\begin{eqnarray}
E^{(0)} & = &
u^{(0)'} + u^{(0)^2} =0
\\
E^{(1)} & = &
E^{(0)'} u^{(1)} + (1/2) u^{(0)''} = 0,\
E^{(0)'} =\partial_x + 2 u^{(0)}.
\end{eqnarray}
Their general solution is
\begin{eqnarray}
u^{(0)} & = &
\chi^{-1}, \chi=x-x_0,\ x_0 \hbox{ arbitrary}
\\
u^{(1)} & = &
u_{-1}^{(1)} \chi^{-2} - \chi^{-2} \Log \psi,\ 
\psi=x-x_0,\
u_{-1}^{(1)} \hbox{ arbitrary},
\end{eqnarray}
and the movable logarithm proves the instability as soon as order $n=1$,
at the Fuchs index $i=-1$.

\section{The diophantine conditions}
\label{sectionDiophantine}
\indent

In fulfilling the systematic programme of Painlev\'e,
one encounters the following kind of diophantine equations
\begin{equation}
\sum_{k=1}^p {1 \over n_k} = {1 \over n},
\end{equation}
with $n$ and $p$ given integers,
whose unknowns $(n_k)$ are Fuchs indices,
which must therefore be either integer or infinite.
They admit a finite set of solutions,
which allows all cases to be further examined.
Details can be found in Bureau 1964, M.I, M.II.

Such a diophantine condition always arises when there exists more than
one family,
as the constraint that, {\it simultaneously},
all Fuchs indices of all families be integer.
Let us just give one example \cite{CFP1993}.
The Hamiltonian H\'enon-Heiles system \index{H\'enon-Heiles system}
\cite{HenonHeiles}
in two coupled variables $(q_1,q_2)$ 
\begin{eqnarray}
H
& \equiv & 
 (1/2) (q_{1,x}^2 + q_{2,x}^2 + c_1 q_1^2 + c_2 q_2^2)
    + \alpha q_1 q_2^2 - (1/3) \beta q_1^3=E,
\label{eqHHH}
\\
& & q_{1,xx} + c_1 q_1 - \beta q_1^2 + \alpha q_2^2 = 0 
\label{eqHH1}
\\
& & q_{2,xx} + c_2 q_2 + 2 \alpha q_1 q_2 = 0
\label{eqHH2}
\end{eqnarray}
defines by elimination the fourth order ODE in $v=q_1$ \cite{Fordy1991}
\begin{eqnarray}
& &
v_{xxxx} + (8 \alpha - 2 \beta) v v_{xx} - 2 (\alpha + \beta) v_x^2 
 - (20/3) \alpha \beta v^3
\nonumber
\\
& &
  +(c_1 + 4 c_2) v_{1,xx}
  + (6 \alpha c_1 - 4 \beta c_2) v^2 + 4 c_1 c_2 v + 4 \alpha E = 0,
\label{eqHH4}
\end{eqnarray}
with $(\alpha,\beta,c_1,c_2,E)$ constants.
Let us restrict here to $c_1=c_2=0$.

There exist two families
\begin{eqnarray}
p
& = &
-1, v_0={3 \over \alpha}, \hbox{ indices } (-1,10,r_1,r_2),\ 
\\
p
& = &
-1, v_0=-{6 \over \beta,} \hbox{ indices } (-1,5,s_1,s_2),\ 
\end{eqnarray}
in which $r_i$ and $s_i$ satisfy the equations
\begin{equation}
   r^2 - 5 r + 12 + 6 \gamma=0,\
   s^2 - 10 s + 24 + 48 \gamma^{-1}=0,\
   \gamma=\beta / \alpha.
\label{eqFiftha}
\end{equation}
 
 
The diophantine equations to be solved are
\begin{equation}
(r_1-r_2)^2=(2k-1)^2, (s_1-s_2)^2=(2l)^2,\
(r_i+1)(r_i-10)(s_i+1)(s_i-5)\not=0,
\end{equation}
with $k$ and $l$ two strictly positive integers.
Making use of $(r_1-r_2)^2=(2r-5)^2, (s_1-s_2)^2=4(s-5)^2$,
the elimination of $\gamma$ between (\ref{eqFiftha}) yields 
\begin{equation}
 \gamma={48 \over 1-l^2},\ l^2=1 + {1152 \over 23 + (2k-1)^2},
\end{equation}
and this provides sharp bounds for $l:$
$1 < l^2 \le 49$.
One thus obtains the four solutions for
$\beta / \alpha, (k,l), (r_1,r_2), (s_1,s_2)$
\begin{eqnarray}
- 1:\ & & ( 1,7),\ ( 2, 3),\ (-2,12)\ \hbox{ (SK)},   \\
- 2:\ & & ( 3,5),\ ( 0, 5),\ ( 0,10),\                \\
- 6:\ & & ( 6,3),\ (-3, 8),\ ( 2, 8)\ \hbox{ (KdV5)}, \\
-16:\ & & (10,2),\ (-7,12),\ ( 3, 7)\ \hbox{ (KK)}.
\end{eqnarray}

Three of them restrict the ODE to the stationary reduction of
well-known soliton equations, thus proving the PP~:
Sawada-Kotera (SK \cite{SK1974}),
higher-order \KdV\ (KdV5, \cite{Lax1968})
and Kaup-Kupershmidt (KK, \cite{Kaup1980,FG1980})
equations.

The case $\beta=-2 \alpha$ is similar to that of
the ODE in section \ref{sectionIndex0}~:
$v_0$ is a double root of its
algebraic equation and is not arbitrary although $0$ is an index.
The results of the Fuchsian perturbative method are also similar;
listed by increasing cost (number of needed values of $(n,i)$),
the first stability conditions $Q_i^{(n)}=0$ are
\begin{eqnarray}
Q_{ 0}^{(1)} & \equiv & 0,\                                 \hbox{cost}=2 \\
Q_{ 0}^{(2)} & \equiv & - 40 \alpha u_0^{(1)^2} = 0,\       \hbox{cost}=5 \\
Q_{10}^{(0)} & \equiv & - 30 \alpha^3 u_5^{(0)^2}=0,\       \hbox{cost}=10\\
Q_{ 5}^{(1)} & \equiv & -120 \alpha u_5^{(0)} u_0^{(1)}=0,\ \hbox{cost}=12.
\end{eqnarray}
To detect the instability,
the method of pole-like expansions is here sufficient 
but the Fuchsian perturbative method is much cheaper.
 

\chapter{Construction of necessary conditions. The Painlev\'e test}
\label{sectionPractice}
\indent

\index{Painlev\'e test}
This chapter makes the synthesis of all the methods of chapter
\ref{sectionTheory} in order to define a usable end product 
which makes obsolete the meromorphy test of section \ref{sectionCasScalaire}.
This end product is widely known as the {\it Painlev\'e test}.
Before detailing the steps of this algorithm in section \ref{sectionTest},
for ODEs as well as for PDEs,
some prerequisite technical developments are needed~:
implementation of physicists' desiderata (section \ref{sectionPhysical}),
technicalities to simplify the computations
(section \ref{sectionTechnicalities})
and the quite important feature of the invariant Painlev\'e analysis
(sections \ref{sectionThreeODEs},
\ref{sectionOptimalChoice} and 
\ref{sectionUnifiedInvariant}).

\section{Physical considerations}
\label{sectionPhysical}
\indent

Some DEs encountered in physics are unstable,
although integrable or partially integrable in some obvious physical sense.
It is then extremely important not to discard them;
this is achieved by relaxing some of the mathematical requirements.

Firstly, nonpolynomial DEs can be made polynomial by transformations on $u$
like in section \ref{sectionCartan},
necessarily outside the groups of invariance of the PP defined in sections
\ref{sectionGroupHomographic} and
\ref{sectionGroupBirational}.

{\it Example} 1 (sine-Gordon).
\begin{equation}
 \hbox{(sine-Gordon) } u_{xt}=\sin u,\ e^{i u}= v,\
2 (v v_{xt} - v_x v_t) - v^3 + v=0.
\end{equation}

{\it Example} 2 (Benjamin-Ono). The nonlocal, nonpolynomial PDE
\begin{equation}
 u_t + u u_{x} + {\rm H}(u_{xx})=0,\
{\rm H}(v)= {1 \over \pi}
 {\rm pp} \int_{- \infty}^{+ \infty} {v(x',t) \over x' - x} \D x',
\end{equation}
in which ${\rm H}$ is the Hilbert transform,
and pp the Cauchy principal value distribution,
is equivalent \cite{GDR1984,RGB1989} to the local and polynomial system
\begin{equation}
 u_t + u u_x + u_{xy},\ u_{xx} + u_{yy}=0
\end{equation}
in one additional independent variable $y$.

Secondly, unstable polynomial DEs may be made stable and polynomial
by transformations like (\ref{eqGroupPoint}).

{\it Example} 3 (parity invariance).
The Ermakov-Pinney ODE \cite{Ermakov,Pinney}
\begin{equation}
\label{eqErmakovPinney}
 u_{xx} - \alpha^2 u + \beta^2 u^{-3}=0. 
\end{equation}
is unstable (algebraic branch point $p=1/2$) and invariant by
parity on $u$~:
the transformation $u \to u^2$ or $u^{-2}$ preserves its polynomial form
and makes it stable.

\section{Technicalities}
\label{sectionTechnicalities}
\indent

A careful {\it choice of the dependent variables}
can save many computations.

\index{Lorenz model}
{\it Example} 1 (dynamical systems).
These systems of first order ODEs sometimes possess an {\it equivalent}
scalar ODE.
This is the case of the Lorenz model (\ref{eqLorenz}), equivalent to
\cite{SenTabor}
\begin{equation}
\label{eqLorenzOrder3}
 x x''' - x' x'' + x^3 x'
+(b + \sigma +1) x x'' + (\sigma + 1) (b x x' - x'^2) + \sigma x^4
+ b (1 - r) \sigma x^2=0
\end{equation}
and of the H\'enon-Heiles Hamiltonian system in $(q_1,q_2)$ (\ref{eqHHH})
which implies the fourth order ODE in $q_1$ only (\ref{eqHH4}).
This offers two advantages.
The first one is to reduce the matricial recurrence relation to a scalar one.
The second one is much more interesting~:
the scalar ODE has a number of families lower than or equal to that of the DS,
which saves a lot of useless cases to consider;
thus, in the HH system, the leading powers for $(q_1,q_2)$ 
are $(-2,-2), (-2,-1),(-2,0),$
while the equivalent fourth order ODE for $q_1$ has only one leading power
$-2$.

Choosing an {\it integrated dependent variable} for the computations 
saves a lot. 
The principle is that, if a DE for $u$ is to be stable,
this allows the presence of {\it one} movable logarithm in its primitive
$v=\int u\ \D x.$
If changing $u$ to $v_x$ allows the DE to be integrated once or more,
expressions are shortened.

\section{Equivalence of three fundamental ODEs}
\label{sectionThreeODEs}
\indent

Let $ S $ be a given analytic function of a complex variable $ x$,
and let us consider 
the three differential equations in $\varphi, \chi, \psi$~:
\smallskip
\begin{eqnarray}
{\varphi_{xxx} \over \varphi_x} - {3 \over 2} \left({\varphi_{xx} 
\over \varphi_ x} \right) ^2  
& = & S
\label{eqSchwarz}
\\
\omega=\chi^{-1},\ -2 \omega_x -2 \omega^2 
& = & S
\label{eqRiccatiN} 
\\
-2 {\psi_{xx} \over \psi}     
& = & S
\label{eqSturm}
\end{eqnarray}
\smallskip
The first one is the Schwarz equation; if read backwards, it defines $ S $ as 
the Schwar\-zian $\lbrace \varphi ; x \rbrace $ of $ \varphi$.
The second one is the Riccati equation in its normalized form 
(equations for $\omega$ or $\chi$ are equivalent and both of Riccati type).
The third one is the second order linear Sturm-Liouville ODE in its normalized
form.

\index{Schwarz equation}
\index{Sturm-Liouville equation}
\index{Riccati equation}
Each of these three ODEs possesses a fundamental uniqueness property. 
As shown by S.~Lie,
the Schwar\-zian is the unique elementary homographic differential invariant 
of a function $ \varphi,$ i.e.~the unique elementary function of
the derivatives $ D \varphi $ of $ \varphi,$ excluding $\varphi $ itself,
invariant under the 6-parameter group $ {\cal H} $ (or M\"obius group, or 
PSL(2,{\cal C})) of homographic transformations~:
\begin{equation}
 {\cal H}:
\varphi \rightarrow {a \varphi + b \over c \varphi + d},\ (a,b,c,d)
\hbox{ arbitrary complex constants},\ a d - b c = 1. 
\end{equation}

Among nonlinear first order ODEs in the class~:
\begin{equation}
u'=R(u,x)
\end{equation}
where $ R $ is rational in $ u $ and analytic in $ x,$ 
the Riccati equation is the unique one
whose general integral has no movable critical points.
As to equation (\ref{eqSturm}), its uniqueness lies in its linear form.

It is a classical result due to \Plv\
(1895, Le\c{c}ons p.~230, O$\!$euvres I)
that the three ODEs 
(\ref{eqSchwarz}), (\ref{eqRiccatiN}), (\ref{eqSturm}) are equivalent~: 
it is sufficient to integrate one in order to integrate the two others.
Consequently, any ODE reducible to one of these three ODEs can be considered 
as {\it explicitly linearizable}.
The six ODEs obtained by elimination of $ S $ between any two of the three 
ODEs have the general solution~:
\begin{eqnarray}
\omega (\varphi)
& = & {c_1 \varphi_x \over c_1 \varphi + c_2}
                   - {\varphi_{xx} \over 2 \varphi_x}
\label{eqRiSC}
\\
\psi   (\varphi)
& = & (c_1 \varphi + c_2) \varphi_x^{-{1 \over 2}}
\label{eqStSC}
\\
\varphi(\omega )
& = & {c_1 (\omega_2 - \omega_1) + c_2 (\omega_3 - \omega_1)
               \over  c_3 (\omega_2 - \omega_1) + c_4 (\omega_3 - \omega_1)}
,\ c_1 c_4 - c_2 c_3 = 1
\\
\psi   (\omega )
& = & c_1 \psi_1 + c_2 \psi_2,\ 
\psi^2_1 = {\omega_2 - \omega_3 \over 
(\omega_2 - \omega_1) (\omega_3 - \omega_1)},\
\psi_2 = \psi_1 {\omega_3 - \omega_1 \over \omega_3 - \omega_2}
\\
\varphi(\psi)
& = & {c_1 \psi_1 + c_2 \psi_2 \over c_3 \psi_1 + c_4 \psi_2},\ 
c_1 c_4 - c_2 c_3 = 1
\\
\omega (\psi)
& = & {c_1 \psi_{1,x} + c_2 \psi_{2,x} \over c_1 \psi_1 + c_2 \psi_2}
\end{eqnarray}
where the $ c_i \hbox{ 's}$ are arbitrary constants, 
$ \omega_i $ and $\psi_i $ particular solutions of (\ref{eqRiccatiN}) and 
(\ref{eqSturm}).

Only two of these six solutions, namely 
$\chi(\varphi)$ and $\psi(\varphi),$ eq.~(\ref{eqRiSC})--(\ref{eqStSC}),
are expressed with a single function;
therefore, among the three equivalent functions, $\varphi $ is the most 
elementary one, and we are going to see that the two others,
$ \chi(\varphi) \hbox{ and } \psi(\varphi),$ 
are the basic
building blocks of the invariant Painlev\'e analysis of both PDEs and ODEs.

If the space of independent variables is multidimensional, for each additional
independent variable $t$ let us define a function $C(x,t,\dots)$ by~:
\begin{equation}
 - {\varphi_ t \over \varphi_ x} = C. 
\end{equation}
As seen from eq.~(\ref{eqRiSC})--(\ref{eqStSC}),
the $t$ dependence of the three equivalent functions
is then characterized by the three equivalent {\it linear} PDEs
(two homogeneous, one inhomogeneous)~:
\begin{eqnarray}
\label{eqSchwarzT}
& &\varphi_t + C \varphi_x = 0
\\
& &\omega=\chi^{-1},\ \omega_t + (C \omega - {1 \over 2} C_x)_x = 0
\label{eqomegat}
\\
& &\psi_t + C \psi_x - {1 \over 2} C_x \psi = 0.
\label{eqSturmT}
\end{eqnarray}
The linearity of these PDEs, as well as the invariance of $C$ under the
change of function $ \varphi \rightarrow F(\varphi), \ F 
\hbox{ arbitrary},$ show that all independent variables but one give rise to
{\it linear} equations.

Systems (\ref{eqSchwarz})--(\ref{eqSturm}) 
and (\ref{eqSchwarzT})--(\ref{eqSturmT}) 
require the cross-derivative condition~:
\begin{eqnarray}
& & \varphi^{-1}_x \left((\varphi_{xxx})_t - (\varphi_t)_{xxx} \right)
= 2 ((\chi^{-1})_t)_x - 2 ((\chi^{-1})_x)_t 
\\
& & = 2 \psi^{-1} ((\psi_t)_{xx} - (\psi_{xx})_t)
= S_t + C_{xxx} + 2 C_x S + C S_x = 0.
\end{eqnarray}

\section{Optimal choice of the expansion variable}
\label{sectionOptimalChoice}
\indent

\index{singular manifold}
A PDE has movable singularities which are not isolated,
on the contrary to an ODE,
but which lay on a codimension one manifold
\begin{eqnarray}
& &
\varphi(x,t,\dots) - \varphi_0=0,
\label{eqPDEManifold}
\end{eqnarray}
in which $\varphi$ is an arbitrary function of the independent variables 
and $\varphi_0$ an arbitrary movable constant.
Even in the ODE case,
the movable singularity can be defined as $\varphi(x) - \varphi_0$,
since the implicit functions theorem allows this to be inverted to
$x-x_0=0$;
this provides a gaude freedom to be used later on in chapter 
\ref{chapterSufficiency}.

\index{expansion variable $\chi$}
The singular manifold and the expansion variable play two different roles,
and there is no {\it a priori} reason to confuse them,
so let us denote $\varphi$ the function which defines the movable singular
manifold $\varphi - \varphi_0=0,$
and $\chi$ the expansion variable.
The only requirement on $\chi$ is that it must vanish as $\varphi - \varphi_0$
and be a single valued function of $\varphi - \varphi_0$ and its derivatives.

The Laurent series for $ u $ and $E$ are defined as
\begin{eqnarray}
u
& = & \sum^{ +\infty}_{j=0} 
       u_j \chi^{ j+p}, \
       -p \in {\cal N}
\\
E
& = & \sum^{ +\infty}_{j=0} E_j \chi^{j+q},\
      -q \in {\cal N}^*
\end{eqnarray}

To illustrate our point, let us take as an example the \KdV\ equation
\begin{eqnarray}
E
& \equiv & -u_t + u_{xxx} + 6 u u_x=0
\label{eqKdVcons}
\end{eqnarray}
(this is one of the very rare locations where this equation can be taken as
an example;
indeed, usually, things work so nicely for KdV that it is hazardous to draw 
general conclusions from its single study).

With the choice $ \chi =\varphi-\varphi_0$ \cite{WTC},
the coefficients $(u_j,E_j)$ 
are invariant under the two-parameter group of translations
$\varphi \to \varphi + b,\ b$ arbitrary complex constant,
and therefore they only depend on the differential invariant $\varphi_x$
of this group and its derivatives~:
\begin{equation}
\label{eqKdVLaura}
 u=-2 \varphi_x^2 \chi^{-2}
+2 \varphi_{xx} \chi^{-1}
+             {\varphi_t \over 6 \varphi_x}
- {2 \over 3} {\varphi_{xxx} \over \varphi_x}
+ {1 \over 2} \left[{\varphi_{xx} \over \varphi_x}\right]^2 + O(\chi).
\end{equation}
[The quantity $C=-{\varphi_t \over \varphi_x}$ is invariant under
$\varphi \to F(\varphi),\ F$ arbitrary function, 
and therefore is uninteresting for the moment.]

With the choice $ \chi =(\varphi-\varphi_0)/ \varphi_x$,
always possible since the gradient of $\varphi$ has at least one nonzero
component,
the invariance is extended to the four-parameter group of affine
transformations
$\varphi \to a \varphi + b,\ (a,b)$ arbitrary complex constants,
with accordingly a dependence on the differential invariant
$\varphi_{xx} / \varphi_x$ and its derivatives~:
\begin{equation}
 u=-2 \chi^{-2}
+2 {\varphi_{xx} \over \varphi_x} \chi^{-1}
+                   {\varphi_t \over 6 \varphi_x}
- {2 \over 3} \left[{\varphi_{xx} \over \varphi_x}\right]_x
- {1 \over 6} \left[{\varphi_{xx} \over \varphi_x}\right]^2 + O(\chi).
\end{equation}

Let us extend this invariance to the six-parameter homographic group.

Eliminating $\varphi_0$ between $\chi$ and $\chi_x$ for each of the two
choices of $\chi$,
one obtains the ODEs of order one for $\chi$
\begin{equation}
\chi_x - \varphi_x=0,\ \chi=\varphi- \varphi_0 
\end{equation}
\begin{equation}
 {1 - \chi_x \over \chi} - {\varphi_{xx} \over \varphi_x} = 0,\ 
\chi={\varphi - \varphi_0 \over \varphi_x},
\end{equation}
whose coefficients only depend on the respective differential invariants.
Now, one knows since S.~Lie
the differential invariant of the homographic group
\begin{equation}
 S=\{\varphi;x\}=
 \left[{\varphi_{xx} \over \varphi_x}\right]_x
- {1 \over 2} \left[{\varphi_{xx} \over \varphi_x}\right]^2 
\end{equation}
and {\it ipso facto} the associated ODE of order one.
Its general solution leads,
by taking the homographic transform which vanishes as $\varphi - \varphi_0$,
to {\it the good expansion variable} 
\begin{equation}
 \chi
={\varphi-\varphi_0 \over \varphi_x - {\varphi_{xx} \over 2 \varphi_x}
(\varphi-\varphi_0)}
=\left[{\varphi_x \over \varphi-\varphi_0}
- {\varphi_{xx} \over 2 \varphi_x}\right]^{-1}.
\end{equation}

Check~:
due to the homographic dependence of $ \chi $ on
$ \varphi,\hbox{ grad }\chi $
is a polynomial of degree two in $ \chi $ with coefficients 
homographic invariants.
Denoting $ t $ an arbitrary independent variable,
possibly equal to $ x$, one obtains
\begin{eqnarray}
\label{eqchit}
\chi_t
& = & - C + C_x \chi  - {1 \over 2} (C S + C_{xx}) \chi^2
\\
\label{eqchix}
\chi_x
& = & 1 + {S \over 2} \chi^2
\\
(\hbox{Log } \psi)_t
& = & - C \chi^{-1} + {1 \over 2} C_x 
                      = - C (\hbox{ Log }\psi)_x + {1 \over 2} C_x
\end{eqnarray}

Again, eq.~(\ref{eqchit})--(\ref{eqchix}) are not different from 
eq.~(\ref{eqRiccatiN}), (\ref{eqomegat}).

The only price to pay for invariance is to privilege some coordinate $ x$.

For our KdV example, the final Laurent series,
to be compared with the initial one (\ref{eqKdVLaura}),
is remarkably simple~:
\begin{equation}
 u=-2 \chi^{-2} - {C \over 6} - {2 S \over 3} + O(\chi).
\end{equation}

The successive values of $\chi$ and the corresponding subgroup items are
gathered in the following table.
\begin{table}[h]
\begin{tabular}{||l | c |l |l ||}
\hline
Group
& Invariant $I$
& Riccati($\chi$)
& Solution for $\chi$
\\
\hline
  $\varphi + b$
& $\varphi_x$
& $\chi_x=I$
& $\varphi - \varphi_0$
\\
\hline
  $a \varphi + b$
& $\varphi_{xx} / \varphi_x$
& $\chi_x=1 - I \chi$
& $(\varphi - \varphi_0) / \varphi_x$
\\
\hline
  $(a \varphi + b) / (c \varphi + d)$
& $\{\varphi;x\}$
& $\chi_x=1 + (I/2) \chi^2$
& ${\varphi-\varphi_0 \over \varphi_x - {\varphi_{xx} \over 2 \varphi_x}
   (\varphi-\varphi_0)}$
\\
\hline
\end{tabular}
\end{table}

\noindent {\it Kruskal's choice}
\indent

Kruskal \cite{JKM} indicated the very simple choice
$ \chi =x-f(t,\dots) $ of expansion variable to make the practical
computations as short as possible.
This choice is equivalent in our formalism to a choice of gauge, namely
$ S=0, C_x=0,$
and $ \varphi $ is then an arbitrary homographic function of $ x-f(t,\dots) $
with constant coefficients.
The choice of Kruskal is really a choice of the expansion variable, {\it not}
of the singular manifold, i.e.~$ \chi_x=1, \hbox{ not } \varphi_x=1.$

Caution~: this choice should only be used at the stage of
building necessary conditions for the PP,
and never at the stage of sufficiency because of the constraints put on
$(S,C)$.

\section{Unified invariant \Plv\ analysis (ODEs, PDEs)}
\label{sectionUnifiedInvariant}
\indent

This is a reference section containing all the items of that version of
Painlev\'e analysis which is common to ODEs and PDEs and which generates the
simplest possible expressions, due to its built-in invariance.

Consider a DE 
\begin{equation}
\bfE(\bfu,\bfx)=0
\end{equation}
polynomial in $\bfu$ and its derivatives, analytic in $\bfx$
($\bfE,\bfu,\bfx$ multidimensional),
and the Laurent series for $\bfu$ and $\bfE$ around the movable singular 
manifold $ \varphi-\varphi_0=0:$

\begin{eqnarray}
\bfu 
& = &
 \bfu_{-\bfp,1} \Log \psi + \sum_{j=0}^{+\infty} \bfu_j \chi^{ j+\bfp},\
       -\bfp \in {\cal Z}
\\
\bfE 
& = &
 \sum_{j=0}^{+\infty} \bfE_j \chi^{j+\bfq},\
      -\bfq \in {\cal Z}
\end{eqnarray}
The coefficient $\bfu_{-\bfp,1} $ can be nonzero only if $\bfE$ does not 
explicitly depend on $\bfu$.
Let us denote $x$ any independent variable such that $ \varphi_x \not= 0$.
In order to establish the most general formulae, we need two other independent
variables, $t$ and $y$.
The gradient of expansion variables $\chi$ and $\psi$ is
(auxiliary notation is $\omega=\chi^{-1}$)~:
\begin{eqnarray}
\chi_x 
& = &
 1 + {S \over 2} \chi^2
\label{eqChiX}
\\
\chi_t 
& = &
 - C + C_x \chi  - {1 \over 2} (C S + C_{xx}) \chi^2
\label{eqChiT}
\\
\chi_y 
& = &
 - K + K_x \chi  - {1 \over 2} (K S + K_{xx}) \chi^2
\\
(\Log \psi)_x 
& = &
 \chi^{-1}
\label{eqPsiX}
\\
(\Log \psi)_t 
& = &
 - C \chi^{-1} + {1 \over 2} C_x = - C (\Log \psi)_x + {1 \over 2} C_x
\label{eqPsiT}
\\
(\Log \psi)_y 
& = &
 - K \chi^{-1} + {1 \over 2} K_x = - K (\Log \psi)_x + {1 \over 2} K_x
\\
\omega_x 
& = &
 - \omega^2 - {S \over 2}
\\
\omega_t 
& = &
 C \omega^2- C_x \omega + {1 \over 2} (C S + C_{xx})
 = (-C \omega + {1 \over 2} C_x)_x
\\
\omega_y 
& = &
 K \omega^2- K_x \omega + {1 \over 2} (K S + K_{xx}) 
 = (-K \omega + {1 \over 2} K_x)_x
\label{eqOmegaY}
\end{eqnarray}
(note that eq.~(\ref{eqChiT}) generates the eight others)
where $ S, C, K $ are elementary homographic differential invariants linked by 
the cross-derivative conditions~:

\begin{eqnarray}
& & \varphi^{ -1}_x \left((\varphi_{xxx})_t - (\varphi_t)_{xxx} \right) 
 = 
 S_t + C_{xxx} + 2 C_x S + C S_x = 0
\label{eqCrossXT}
\\
& & \varphi^{ -1}_x \left((\varphi_{xxx})_y - (\varphi_y)_{xxx} \right)
 = 
 S_y + K_{xxx} + 2 K_x S + K S_x = 0
\label{eqCrossXY}
\\
& & \varphi^{ -1}_x \left((\varphi_y)_t - (\varphi_t)_y \right)
 = 
C_y - K_t + C_x K - C K_x = 0.
\label{eqCrossXYT}
\end{eqnarray}

Kruskal's choice is implemented by putting
$ S=0, C=f_t, K=f_y, \dots $ 
in eq.~(\ref{eqChiX})--(\ref{eqOmegaY}),
thus reducing each rhs to one term and making 
eq.~(\ref{eqCrossXT})--(\ref{eqCrossXYT}) useless.

The function $ \varphi-\varphi_0 $ never appears in the above formulae.
Similarly, the explicit expressions of $ \chi, \psi, S, C, K $ as functions of
$ \varphi - \varphi_0 $ are {\it not} needed during the computations.
We recall them here only for reference~:
\begin{eqnarray}
\chi 
& = &
 \left({\varphi_x \over \varphi - \varphi_0}
       - {\varphi_{xx} \over 2 \varphi_x} \right)^{-1}
\label{eqchi}
\\
\psi 
& = &
 (\varphi -\varphi_0) \varphi_x^{-{1 \over 2}}
\\
S 
& = & 
\lbrace \varphi ;x\rbrace
 =  
{\varphi_{xxx} \over \varphi_x} - {3 \over 2} \left({\varphi_{xx} 
\over \varphi_ x} \right) ^2
 =  
\left({\varphi_{xx} \over \varphi_x}\right)_x 
- {1 \over 2} \left({\varphi_{ xx} \over \varphi_ x} \right) ^2
\\
& = & 
- 2 \left({\varphi_ x \over \varphi  - \varphi_ 0}
- {\varphi_{ xx} \over 2 \varphi_ x} \right)_x
  -2 \left({\varphi_ x \over \varphi  - \varphi_ 0}
- {\varphi_{ xx} \over 2 \varphi_ x} \right) ^2  , \
\\
C 
& = & 
- {\varphi_t \over \varphi_x}
\\
K 
& = & 
- {\varphi_y \over \varphi_x}.
\end{eqnarray}

In some applications, it is necessary to choose for $\chi$ the most general
homographic transform of (\ref{eqchi}) which vanishes as 
$ \varphi - \varphi_0$

\begin{eqnarray}
 \grad \chi   
& = & 
  {\bf X}_0 + {\bf X}_1 \chi   + {\bf X}_2 \chi^2 \ ,\
\\
 \grad \omega 
& = &
 - {\bf X}_2 - {\bf X}_1 \omega - {\bf X}_0 \omega^2 
\\
 \grad \Log \psi 
& = &
 {\bf X}_0 \chi^{-1} + {1 \over 2} {\bf X}_1.
\end{eqnarray}

The vectorial coefficients $ {\bf X}_i $ depend on $(S,C,K,\dots)$ and two
additional arbitrary functions.
The auxiliary expansion variable $ \psi $ is defined
by its logarithmic gradient and by the condition that it should vanish as
$ \varphi - \varphi_0$.


\section{The \Plv\ test\label{sectionTest}}
\indent

The synthesis of the different methods to generate necessary conditions 
for the PP produces the following algorithm, called ``\Plv\ test''.

Consider a DE (\ref{eqDEgeneral}) of order $N,$
already transformed if necessary,
see sections \ref{sectionPhysical} and \ref{sectionTechnicalities},
so as to be polynomial in $\bfu$ and its derivatives, analytic in $x$. 
The \Plv\ test is made of the following steps.

\begin{description}
\item[Step 0.]
Perform a transformation (\ref{eqGroupHomographicContinuous}) in order to
reduce the number of terms in the equation
(details in section \ref{sectionTwoExamples} and ref.~M.I and, for PDEs,
\cite{CosPDEpara}).

Ex.~: 
equation $ u_x + u_t +u_{xxt} + u_x u_t = 0$,
under the translation $u=U-x-t$, 
becomes $U_{xxt} + U_x U_t + 1 = 0$.

\item[Step 1.]
Require the satisfaction of 
the very general necessary conditions obtained by \Plv\
(details in section \ref{sectionGeneralConditions}).

Ex.~\cite{KruskalSainteAdele}~: 
$- 3 u^2 u' u''' + 5 u^2 u''^2 - u u'^2 u'' - u'^4 = 0$. 
Unstable for $5/3$ has not the required value $1 - 1/n$.

Ex.~\cite{Clarkson1985a}~: 
$ (1+u^2) u_{xx} - 2 u u_x^2 + u_t^2=0.$
The ODE obtained by the reduction $(x,t) \to x-ct$ has an $A$ with two simple
poles $u=\pm i$ and residues $1 \pm i c^2 / 2.$ 
The ODE is unstable, and so is the PDE.

\item[Step 2.]
If the degree is greater than one,
establish the ODE satisfied by the singular solutions
(details in section \ref{sectionRemovalSS}).

\item[Step 3.]
Put the DE under a canonical form of Cauchy;
find all the exceptional points where the Cauchy theorem fails;
for each such point, define a homographic transformation
(\ref{eqGroupHomographicContinuous}) allowing the Cauchy theorem to apply
(details in sections \ref{sectionTwoTheorems} and \ref{sectionMethodAlpha}).
For each DE (the original one and all these homographic transforms),
perform step 4.

Ex.~: (P5) has the exceptional points $u=1$ and $u=0$ 
(poles of $A$, see section \ref{sectionGeneralConditions}).
These points are regular for the ODE in $(u-1)^{-1}$ and $u^{-1}$.

Ex.~: the reduced three-wave interaction dynamical system
\begin{eqnarray*}
& &
x'=- 2 y^2 + z + \gamma x + \delta y,\
y'= 2 x y      + \gamma y - \delta x,\
z'=- 2 x z - 2 z
\end{eqnarray*}
has the exceptional point $y=\delta/2$ \cite{BRGD},
not so evident on the system itself but easily unveiled by
considering the equivalent third order ODE for $y(t)$.

\item[Step 4.]
Find all the families $\bfu \sim \bfu^{(0)}_0 \chi^{\bfp}$
($\bfu^{(0)}_0\not={\bf 0}$)
(details in section \ref{sectionMethodPole}).
Discard those families which are also families of the ODE for singular 
solutions established at step 2.
Require all components of remaining $\bfp$'s to be integer.
Discard all families having all components of $\bfp$ positive.
For each remaining family,
perform step 5 and at least one of the steps 6 and 7.

Ex.~: (P5) has six families of movable simple pole-like singularities
\begin{equation}
 u         \sim \pm (  2 \alpha)^{-{1 \over 2}} x \chi^{-1},\
 u^{-1}    \sim \pm (- 2 \beta )^{-{1 \over 2}} x \chi^{-1},\
(u-1)^{-1} \sim \pm (- 2 \delta)^{-{1 \over 2}}   \chi^{-1}.
\end{equation}

Ex.~: the reduced three-wave interaction has the families \cite{BRGD}
\begin{eqnarray}
& &
(x,y,z) \sim (-(1/2) \chi^{-1},(i/2) \chi^{-1}, z_0),\
z_0 \hbox{ arbitrary},\
\hbox{indices } (-1,0,2)
\nonumber
\\
& &
(x,y,z) \sim (\chi^{-1},\delta/2, - \chi^{-2}),\
\hbox{indices } (-1,2,2).
\nonumber
\end{eqnarray}
The first one will pass the test while the second will generate at index $2$
the conditions $\gamma \delta=0, \gamma (\gamma+1)=0$.

In case the DE has too many terms,
this step is worth being programmed on a computer,
by fear of missing some families. 

{\it Warning}. If one is unsure about some component $u$ of $\bfu$ behaving
like a positive integer power $p$ of $\chi$,
it may be safer to switch to the DE for $u^{-1}$.

\item[Step 5.]
From the auxiliary equation of the simplified equation, 
compute the linear operator $\bfP(i)$
eq.~(\ref{eqMatriceSysteme}) and the indicial equation 
(\ref{eqMatriceSysteme}) $\det \bfP(i)=0$
(details in section \ref{sectionMethodPole}).
Compute its zeroes (the Fuchs indices). 
Require each index to be integer (details in section \ref{sectionDiophantine})
and to satisfy the rank condition (\ref{eqRank}).

Ex.~(\cite{CFP1993} example 5.B). 
These are two coupled PDEs with a single family
whose linear operator $\bfP(i)$ is
\begin{equation}
 \bfP(i)=
\pmatrix{- {1 \over 3} (i+2)^2 & {1 \over 3} (i+2) \cr - (i+2) & i^2 \cr}.
\end{equation}
The indices are the zeroes of its determinant $(-2,-2,-1,1)$.
For the double index $i=-2$, the rank of $\bfP(i)$ is one,
so the system of PDEs is unstable.

Ex.~\cite{ChazyThese,BureauMII,FP1991}~:
the equation $ u_{xxx} - 7 u u_{xx} + 11 u_x^2 = 0 $
has only one family with three indices~:
$ p=-1, u_0^{(0)}=-2,$ indices $(-6,-1,-1).$ 
The double index $-1$ immediately proves the instability.

\item[Step 6.]
(NonFuchsian case).
If the degree of the indicial polynomial is strictly lower than $N$,
and if a particular solution is known in closed form,
apply the NonFuchsian perturbative method
(details in section \ref{sectionMethodPerturbativeNonFuchsian}).

\item[Step 7.]
(Fuchsian case).
Denote $\rho$ the smallest integer Fuchs index, lower than or equal to $-1$.
Define two positive integer upper bounds $k_{\hbox{max}}$ and $n_{\hbox{max}}$
representing the cost of the computation to come, see advice below.
Solve the linear algebraic system (\ref{eqMethodPerturbative8}) in the unknown
$\bfu^{(n)}_j, (j,n) \not=(0,0),$
for the successive values 
$k=0$ to $k_{\hbox{max}},n=0$ to $n_{\hbox{max}}$ with $j=k + n \rho$;
whenever $j$ is an index $i$ of multiplicity mult($i$),
\begin{description}
\item[-]
require the orthogonality condition (\ref{eqOrtho}) to be satisfied for any
value of the previously introduced arbitrary coefficients,
\item[-]
if $n=0$ or ($n=1$ and $i<0$), assign arbitrary values to mult($i$) components
of $\bfu^{(n)}_i$ defining a basis of Ker $\bfP(i)$,
\item[-]
if ($n=1$ and $i\ge 0$) or $n\ge 2$, assign the value 0 to mult($i$)
components of $\bfu^{(n)}_i$ defining a basis of Ker $\bfP(i)$.
\end{description}
Details in section \ref{sectionMethodPerturbativeFuchsian}.

Advice for choosing $k_{\hbox{max}}$ and $n_{\hbox{max}}$~:
if the order $n=0$ fails to describe the general solution,
take at least $n_{\hbox{max}}=2$;
take $k_{\hbox{max}}$ so as to test the greatest Fuchs index for 
$n=n_{\hbox{max}}$
(all details in the remarks at the end of section 
\ref{sectionMethodPerturbativeFuchsian}).

\end{description}

\medskip
This ends the test.
Step 6 has been put before step 7 because in all our examples it allows to
conclude sooner.

Let us again stress that these sets of conditions may not be sufficient~:
\Plv\ gave the counterexample of the second order ODE whose general
solution is $\pm \sn[\lambda \Log(c_1 x + c_2);k)]$, 
with $(c_1,c_2)$ arbitrary,
for which no local test can generate the necessary and sufficient stability
condition that $2 \pi i \lambda$ be a period of the elliptic function $\sn$.
For advanced features,
see section \ref{Miscellaneous perturbations} and \cite{Cos1997}.

\section{The partial Painlev\'e test}
\label{sectionTestPartial}.
\indent

In the search for the tiniest piece of integrability,
the physicist, see section \ref{sectionPhysicist},
will perform the above Painlev\'e test to its end, \ie without stopping
even in case of failure of some condition,
so as to collect a bunch of necessary conditions.

Turning to sufficiency, 
these conditions will then be examined separately in the hope of finding
some global element of integrability, most often a Darboux eigenvector.

\index{Lorenz model}
For instance,
the Lorenz model (\ref{eqLorenz}) has two families
\begin{equation}
x \sim 2 i \chi^{-1},\
y \sim - (2 i / \sigma) \chi^{-2},\
z \sim - (2   / \sigma) \chi^{-2},\
i^2=-1,
\end{equation}
with the same indices $(-1,2,4)$,
which generate the no-log conditions \cite{Segur,ConteMusette1991}
\begin{eqnarray*}
& &
Q_2 \equiv (8/3) (b - 2 \sigma) (b + 3 \sigma -1) = 0
\\
& &
Q_4 \equiv 
- 4 i (b - \sigma -1) (b - 6 \sigma + 2) x_2
+ (8/3) (b-1) (b - 3 \sigma + 1) S
\nonumber
\\
& &
-4 b \sigma (b - 3 \sigma +5) r + f(b,\sigma) = 0,
\end{eqnarray*}
in which $x_2$ is arbitrary, $S$ is the Schwarzian of the invariant analysis,
and $f$ a polynomial irrelevant for what we want to emphasize.
Performing a logical {\it or} operation on these conditions instead of the 
logical {\it and} of the mathematician,
one obtains the condition on $(b,\sigma)$
\begin{eqnarray}
& &
(b - 2 \sigma) (b + 3 \sigma -1) 
(b - \sigma -1) (b - 6 \sigma + 2)
(b-1) (b - 3 \sigma + 1)
= 0.
\end{eqnarray}
What is remarkable is that {\it all} known analytic results on this model
(first integrals \cite{Kus},
particular solutions \cite{ConteMusette1991},
Darboux eigenvectors \cite{Dryuma,LC1996b})
belong to one of these six cases.
Conversely, to each of the six factors there corresponds such a result,
although sometimes only for a finite set of values of $(b,\sigma)$.

{\it Remarks}.
\begin{enumerate}
\item
With the restriction $S=0$ one would miss two of the six factors.

\item
First integrals $P(x,y,x) e^{\lambda t}$,
with $P$ polynomial and $\lambda$ constant,
should not be searched for with the assumption $P$ the most general polynomial
in three variables.
Indeed, $P$ must be an entire function of $t$ \ie have no singularities
at a finite distance.
The generating function of such polynomials 
is built from the singularity degrees of $(x,y,z)$ \cite{LevineTabor}
\begin{equation}
{1 \over (1 - \alpha x) (1 - \alpha^2 y) (1 - \alpha^2 z)} 
\end{equation}
and it provides the basis, ordered by singularity degrees
\begin{eqnarray}
& &
(1),\
(x),\
(x^2,y,z),\
(x^3,x y, x z),\
(x^4,x^2 y, x^2 z, y z, z^2, y^2),\
\dots
\end{eqnarray}
Thus, $P_2$ should be searched for as a linear combination of
$(1,x,x^2,y,z)$.
All known first integrals are found at the $P_4$ level \cite{Kus}.

\item
The case $b=1 - 3 \sigma$ is on an equal footing with the case $b=2 \sigma$
which admits the first integral $(x^2 - 2 \sigma z) e^{2 \sigma t}$,
but finding its first integral is still an open problem.

\end{enumerate}

\chapter{Sufficiency~: explicit integration methods}
\label{chapterSufficiency}
\indent

We review the algorithmic methods which {\it may} perform the explicit 
integration,
with emphasis on ODEs. The PDE case is handled in another part of this volume
\cite{Musette}.

We assume that the application of the \Plv\ test
(necessary conditions for the PP)
has led either to no failure or to a minor failure,
corresponding respectively to a presumption of integrability in the \Plv\
sense or of partial integrability.
If perturbative methods have been used,
one has to decide to give up at some perturbation order $n$
(remember the counterexample of Painlev\'e).
The goal is then
either to prove the sufficiency (integrability)
or to build particular solutions (partial integrability).

If the DE belongs to one of the fully studied classes enumerated in chapter
\ref{sectionClassical},
the question is solved.
Indeed, either it is possible,
by some homographic transformation (\ref{eqGroupHomographicContinuous}),
to bring the DE back to a normalized (``classified'') DE,
in which case the integration is finished,
or this is impossible,
in which case the DE has not the PP.

For a DE which has not been classified,
if one excludes the case where the DE is an ODE and defines a new function 
(a quite improbable event which has not occured since 1906),
the explicit proof of sufficiency amounts to
(the cases below are not mutually exclusive)
\begin{itemize}
\item{}
either (ODE case) express the general solution as a finite expression of a 
finite number of
elementary functions (solutions of linear equations, the Weierstrass $\wp$
function, the six \Plv\ functions),
\item{}
or (PDE case) find a Lax pair.
\index{Lax pair}
\end{itemize}

In the partial integrability situation, one tries to obtain degeneracies of
these results~: a particular solution or a pair of linear operators able to
generate a subclass of solutions.

The methods to handle both cases are the same,
and they again only rely on the singularity structure.
Their basic common idea is 
that the singular part of the Laurent expansions (of a {\it local} nature)
contains all the information for a {\it global} knowledge of the solution.

The two existing methods are known as 
the {\it singular part transformation}
and the {\it truncation method}.
Before describing them, let us give a few definitions and explain how
\Plv\ proved the sufficiency for the six equations (P1)--(P6).


\section{Sufficiency for the six \Plv\ equations
\label{sectionSufficiencySix}}
\indent

\Plv\ introduced the concept of ``int\'egration parfaite'' and used it to 
solve the question of sufficiency for the six equations discovered by himself
and Gambier.
The idea is to perform a finite (in the sense of Poincar\'e~: 
finite expression) single valued transformation from (Pn) to another ODE
which has no more movable singularities although it may still have fixed
critical singularities.
Such an ODE has qualitatively the same singularities than a linear ODE,
and \Plv\ says that its integration is then ``parfaite'' (achieved)
(BSMF p.~205)~:
given any initial conditions, its solution can be computed with an
arbitrary accuracy
(by e.g.~the sequence of coefficients of convergent Taylor series)
since one knows in advance where the remaining (fixed) singularities are
located.
The movable singularities of the original ODE are then totally under control.
The equations with fixed critical points therefore constitute
a natural extension to the linear equations.

\Plv\ defined such transformations (nowadays called ``singular part
transformations'') for each of the six equations (P1)--(P6).
These transformations, {\it via} logarithmic derivatives,
transform (P1)--(P6) into equations for $\psi$ without movable singularities
((P1) Acta p.~14,
 (P2) Acta p.~15,
 (P3) Acta p.~16,
 (P4,P5,P6) CRAS 1906, \Oeuvres\ III p.~120)
\begin{eqnarray}
{\rm (P1)}
& &
u=-\partial_x^2 \Log \psi
\label{eqP1SP}
\\
{\rm (P2)}
& &
u=\partial_x \Log \psi_1 - \partial_x \Log \psi_2
\label{eqP2SP}
\\
{\rm (P3)}
& &
u=e^{-x} (\partial_x \Log \psi_1 - \partial_x \Log \psi_2)
\\
{\rm (P4)}
& &
u=\partial_x \Log \psi_1 - \partial_x \Log \psi_2
\\
{\rm (P5)}
& &
u=x e^{-x} (2 \alpha)^{-1/2}
(\partial_x \Log \psi_1 - \partial_x \Log \psi_2).
\\
{\rm (P6)}
& &
u=x (x-1) e^{-x} (2 \alpha)^{-1/2}
(\partial_x \Log \psi_1 - \partial_x \Log \psi_2).
\end{eqnarray}

\index{Lax pair}
The Lax pairs of (P1)--(P6) can be found in 
Ref.~\cite{GarnierThese} and \cite{JimboMiwaII}.

The two methods developed in next sections 
(\ref{sectionSingularPartTransfo}) and (\ref{sectionTruncation})
rely on this result.

The {\it logarithmic derivative} plays a privileged role,
as generator of a movable simple pole with a residue generically unity.
A prerequisite to the algorithmic derivation of a transformation
from $u$ to $\psi$ such as (\ref{eqP1SP}) is the introduction
of a free gauge function which we denote $\varphi$.

Such a gauge naturally arises if one thinks of an ODE as the canonical
reduction of a PDE defined by suppressing the dependence upon all
independent variables but $x$.
This is the function $\varphi$ used in the description of 
the movable singularities 
by (\ref{eqPDEManifold}) rather than $x-x_0=0$.
Useless at the stage of building necessary conditions (the \Plv\ test),
this feature is the key to the algorithmic explicit integration methods.

\section{The singular part(s)}
\indent

{\it Definition}.
The {\it singular part} of one of the families of movable singularities 
of a given DE is the finite sum of the Laurent series 
restricted to the nonpositive powers in the method of pole-like expansions
\begin{equation}
\label{eqSingularPart}
u_T=\sum_{j=0}^{-p} u_j \chi^{j+p}.
\end{equation}
Synonyms are~: truncation, truncated expansion.

Given $\varphi,$ the singular part $u_T$ is a one-parameter ($\varphi_0$)
family of expressions $u_T(\varphi_0)$,
and the two particular values $\varphi_0=0$ and $\varphi_0=\infty$
are of special interest.
For the example of KdV (\ref{eqKdVcons})
\begin{eqnarray}
u_T(0)
& = &
-2 \left[{\varphi_x \over \varphi}- {\varphi_{xx} \over 2 \varphi_x}\right]^2
 + {\varphi_t \over 6 \varphi_x}
 - {2 \over 3} \left({\varphi_{xxx} \over \varphi_x}
           - {3 \over 2} \left[{\varphi_{xx} \over \varphi_x}\right]^2\right)
\\
u_T(\infty)
& = &
-2 \left[\phantom{{\varphi_x \over \varphi}}
- {\varphi_{xx} \over 2 \varphi_x}\right]^2
 + {\varphi_t \over 6 \varphi_x}
 - {2 \over 3} \left({\varphi_{xxx} \over \varphi_x}
           - {3 \over 2} \left[{\varphi_{xx} \over \varphi_x}\right]^2\right).
\end{eqnarray}

{\it Definition}.
The {\it singular part operator} ${\cal D}$ of a family is defined by
\begin{equation}
\Log \varphi \to {\cal D} \Log \varphi=u_T(0)-u_T(\infty).
\label{eqdefD}
\end{equation}

{\it Example} 1 (KdV). The operator ${\cal D}$ is linear and equal to 
$2 \partial_x^2$.
This linearity is strongly linked with the Darboux transformation
\cite{Musette}.

{\it Example} 2.
The single family of (P1) and the two families of (P2) have the singular
parts
\begin{eqnarray}
{\rm (P1)}:\ u_T
& = & \chi^{-2} + {S \over 3},\
{\cal D}=- \partial_x^2
\\
{\rm (P2)}:\ u_T
& = & \pm \chi^{-1},\
{\cal D}=\pm \partial_x.
\end{eqnarray}

\section{Method of the singular part transformation}
\label{sectionSingularPartTransfo}
\indent

This is the method used by \Plv\ and outlined in previous section
\ref{sectionSufficiencySix}.
It consists of transforming the DE for $u$ into a DE for $\varphi$ by the
nonlinear transformation
\begin{equation}
u={\cal D} \Log \varphi,
\end{equation}
where ${\cal D}$ is the singular part operator associated to one of the
families of the equation for $u$.

If the transformed equation for $\varphi$ can be integrated,
so is the original equation.

{\it Example} 1 (linearization).
The unique first order first degree ODE with the PP,
namely the Riccati equation (\ref{eqRiccatiG}),
has a ${\cal D}$ operator equal to $ -a_2^{-1} \partial_x$,
computable from the basic formulae
(\ref{eqChiX}),
(\ref{eqchi}) and
(\ref{eqdefD}).
The transformation $u=-a_2^{-1} \partial_x \Log \varphi$ from $u$ to $\varphi$
leads to the second order linear equation (\ref{eqRiccatiL}) for $\varphi$.
It is then sufficient to know two particular solutions
$\varphi_1$ and $\varphi_2$ (which {\it are} functions)
of this linear equation to have a global knowledge of the general solution of
the Riccati equation by the formula
\begin{equation}
u=-a_2^{-1} \partial_x \Log (c_1 \varphi_1 + c_2 \varphi_2).
\end{equation}

Similarly, the transformation $\wp=- \partial_x^2 \Log \sigma$ associates to 
the Weierstrass elliptic function $\wp$ a function $\sigma$ which is
an entire function, solution of a nonlinear ODE.

{\it Example} 2 
(simplified equation of one of the 50 stable ODEs (\ref{eqClassODE21})).
The ODE
\begin{equation}
E \equiv u'' + u u' - u^3=0
\end{equation}
possesses two families of movable simple poles $u_0=1$ and $u_0=-2$,
with the one-parameter particular solutions $u_0 / (x-x_0)$.
The first family operator is ${\cal D}=\partial_x$ and it transforms it into
\begin{equation}
u=\partial_x \Log \varphi,\
E \equiv \varphi \left({\varphi'' \over \varphi^2}\right)'=0,
\end{equation}
which integrates as 
$\varphi=a \wp(x-x_0,0,g_3)$ with $(a,x_0,g_3)$ arbitrary and provides the
general solution.
The two families of movable simple poles for $u$ correspond to
the movable simple zeroes of $\wp$ (residue $u_0=1$) and to
the movable double poles of $\wp$ (residue $u_0=-2$).

{\it Example} 3 (indirect linearization).
The Ermakov-Pinney equation (\ref{eqErmakovPinney}),
after the transformation $u^{-2}=v$ removing its algebraic singularity
\begin{equation}
\label{eqErmakovPinney2}
E \equiv 
- {1 \over 2} v v_{xx} + {3 \over 4} v_x^2 - \alpha^2 v^2 + \beta^2 v^4=0,
\end{equation}
has two families $v \sim \pm (2 \beta)^{-1} \chi^{-1}$,
and the transformed ODE under 
\hfill \break
$v= (2 \beta)^{-1} \partial_x \Log \varphi$
\cite{Conte1992a}
\begin{equation}
{\varphi_{xxx} \over \varphi_x} - {3 \over 2} \left({\varphi_{xx} 
\over \varphi_ x} \right) ^2  
 = - 2 \alpha^2
\end{equation}
is a Schwarz ODE (\ref{eqSchwarz}).
This integrates the Ermakov-Pinney equation {\it via} a finite two-valued
expression.

Well suited to DEs possessing only one family
(Riccati, Weierstrass, (P1), KdV),
this transformation must be adapted,
following the \Plv\ formulae for (P2)--(P6) in 
section \ref{sectionSufficiencySix},
to suit DEs with more than one family (Jacobi elliptic equation, (P2) to (P6)).
This is done in the course on PDEs \cite{Musette}.

\section{Method of truncation (Darboux transformation)}
\label{sectionTruncation}
\indent

Perfectly adapted to PDEs \cite{Musette},
this method is rather poor for ODEs,
for an intrinsic reason which is the absence in this case
of a B\"acklund transformation
\index{B\"acklund transformation}
(link between two different solutions of the same DE introducing at each
iteration at least one more arbitrary parameter in the solution).
It nevertheless succeeds, at least partially, in many situations.

The idea \cite{WTC} is to consider the singular part (\ref{eqSingularPart}) 
of one family (or the sum $u=\sum_f {\cal D} \Log \psi_f$
of the singular parts of several families)
as a {\it parametric representation} of a solution in terms of 
one function $\psi$ linked to $\chi$ by $\chi^{-1}=\partial_x \Log \psi$
(or several functions $\psi_f$, one per family $f$).
Every function $\psi_f$, which defines a singular manifold $\psi_f=0$,
is required to be an entire function,
and for instance to satisfy the {\it same} linear system of two PDEs 
$L_1=0, L_2=0$ with some adjustable coefficients.

The method consists of identifying to zero the lhs $E(u_T)$ considered as a
polynomial of $\psi_f$ and its independent derivatives modulo the constraint
that each $\psi_f$ satisfies the linear system.
This generates an overdetermined set of {\it determining equations} whose
unknowns are the coefficients $u_j$ of (\ref{eqSingularPart}) and the
coefficients of the linear system.
The remarkable fact is that the determining equations are easy to solve.

The result is some class of exact solutions,
and this class is easily interpreted.
If the commutator $[L_1,L_2]$ is identically zero
(which is always the case if the linear system has constant coefficients),
the solutions are particular ones (PDE case) or any kind
(particular or general) (ODE case).
If this commutator is zero only when some coefficient of $(L_1,L_2)$ satisfies
some PDE,
quite probably $(L_1,L_2)$ define a Lax pair.

Again, the ODE case to which we restrict is much less rich than the PDE case
\cite{Musette} to which we refer the reader.


\subsection{One-family truncation}
\indent

This is the celebrated WTC truncation procedure \cite{WTC}.
Applicable to any DE with any number of families,
it consists of selecting one of the families $\psi=0$,
in which $\psi$ obeys the linear system of the invariant analysis
\begin{eqnarray}
& &
\psi_{xx} + {S \over 2} \psi=0,
\\
& &
\psi_t + C \psi_x - {C_x \over 2} \psi=0.
\end{eqnarray}
The functions $S$ et $C$ are adjustable fonctions,
only constrained by the cross-derivative condition
(\ref{eqCrossXT}).
Consider for instance the Ermakov-Pinney ODE \cite{Ermakov,Pinney}
\begin{eqnarray}
E & \equiv &
- {1 \over 2} v v_{xx} + {3 \over 4} v_x^2 - \alpha^2 v^2 + \beta^2 v^4=0.
\end{eqnarray}
The infinite Laurent series is
$ v = (2 \beta)^{-1} \chi^{-1} + v_1 + O(\chi)$
with $v_1$ arbitrary and $\beta$ one of the two square roots of $\beta^2$.
Thanks to the gauge $\varphi$,
the coefficient $v_1$ is not a constant but a function.

The method consists of assuming that a solution $v$ can be represented
by the truncation
\begin{eqnarray}
& &
v=v_T={1 \over 2 \beta} \chi^{-1} + v_1 = {\cal D} \Log \psi + v_1
\end{eqnarray}
implying for the lhs $E$ of the DE the similar truncated expansion
\begin{eqnarray}
& &
E \equiv \sum_{j=0}^4 E_j \chi^{j-4}.
\end{eqnarray}
This generates, in this example, five equations $E_j=0$ in the unknowns
$(v_1,S)$.
Among them, $E_0$ is zero since the coefficient $v_0$ of the series for $v$
is already the good one.
$E_1$ is zero since $1$ is a Fuchs index whose orthogonality condition is
satisfied.
Denoting $v_1=- V_1 / (2 \beta)$, there remain the three equations
\begin{eqnarray}
& &
16 \beta^2 E_2 \equiv
- 4 \alpha^2 + S + 6 V_1^2 + 6 V_{1,x}=0,
\\
& &
16 \beta^2 E_3 \equiv
8 \alpha^2 V_1 + 2 S V_1 - 4 V_1^3 + S_x + 2 V_{1,xx}=0,
\\
& &
16 \beta^2 E_4 \equiv
{3 \over 4} S^2 - 4 \alpha^2 V_1^2 + V_1^4 - V_1 S_x
 + 3 S V_{1,x} + 3 V_{1,x}^2 - 2 V_1 V_{1,xx}=0.
\end{eqnarray}

The algebraic elimination (\ie\ without differentiation) of
$V_{1,x}$ and $V_{1,xx}$ among these three equations yields
$(S - s)^2=0$, with $s=-2 \alpha^2$,
then $V_1$ is found to satisfy the Riccati equation
\begin{eqnarray}
& &
- 2 V_{1,x} - 2 V_1^2=s.
\end{eqnarray}
Hence the particular solution 
\begin{eqnarray}
& &
v={1 \over 2 \beta} (\chi^{-1} - V_1),
\end{eqnarray}
in which each variable $\chi^{-1}$ and $V_1$ satisfies the same Riccati
equation and depend on one arbitrary parameter.
This is the general solution,
which can be written as
$v=(2 \beta)^{-1} (\partial_x \Log \psi_1 - \partial_x \Log \psi_2)$
in agreement with the structure of singularities, cf.~(\ref{eqP2SP}).

{\it Remark}.
The class of particular solutions generically found by this method
is the class of polynomials in $\tanh$,
which correspond to a constant value for $S$.

Another example is the (P2) equation,
for which the one-family truncation $u=\chi^{-1} + u_1$
provides the one-parameter particular solution
$u_1=0,S=x$ on the condition $\alpha=1/2$,
\ie\ an algebraic transform of the Airy equation.

\subsection{Two-family truncation}
\indent

When a DE admits two families with opposite principal parts,
such as (\ref{eqErmakovPinney2}),
it is natural to seek particular solutions described by two singular manifolds
\cite{ConteMusette1993}
\begin{equation}
\label{eqErmakovPinney3}
v= {1 \over 2 \beta} [\partial_x \Log \psi_1 - \partial_x \Log \psi_2 + v_0],\
\end{equation}
in which $(\psi_1,\psi_2)$ is a basis of the two-dimensional space of
solutions of some ODE whose general solution is entire,
e.g.~the second order linear equation with constant coefficients
\begin{eqnarray} 
\label{eq7a}
& &
 \psi_{xx} - {k^2 \over 4} \psi = 0
\\
& &
\Psi_2
 =  C_1 e^{{k \over 2} x} + C_2 e^{-{k \over 2} x}
 = C_0 \cosh {k \over 2} (x-x_0),
\label{eq7c}
\\
& &
 \psi_1(x)=\Psi_2(x+a),\ \psi_2(x)=\Psi_2(x-a),\
a \hbox{ arbitrary},
\end{eqnarray}

Substituting (\ref{eqErmakovPinney3}) into (\ref{eqErmakovPinney2}) and
eliminating any derivative of $(\psi_1,\psi_2)$ of order higher than or equal
to two in $x$
results into a polynomial in the two variables
$\psi_{1,x}/ \psi_1, \psi_{2,x} / \psi_2$.
Before identifying it to the null polynomial, one must take account of the
first integral $\mu_0$, the ratio of two constant Wronskians
\begin{equation}
\label{eqErmakovPinney5}
{\psi_{1,x} \over \psi_1} {\psi_{2,x} \over \psi_2} 
= {k^2 \over 4} - \mu_0 {k \over 2} 
\left({\psi_{1,x} \over \psi_1} - {\psi_{2,x} \over \psi_2} \right),\
\mu_0=\cotanh k a,
\end{equation}
which splits the polynomial of two variables into the sum of two
polynomials in one variable~:
\begin{eqnarray} 
\label{eqErmakovPinney6}
16 \beta^2 E
& \equiv &
(k^2 - 4 \alpha^2 + 6 v_0^2 + 6 k \mu_0 v_0)
( \left({\psi_{1,x} \over \psi_1} \right)^2
+ \left({\psi_{2,x} \over \psi_2} \right)^2)
\nonumber
\\
& &
+ (k \mu_0 (k^2 - 4 \alpha^2 + 6 v_0^2) 
+ 2 (3 k^2 \mu_0^2 - k^2 - 4 \alpha^2 + 4 v_0^2) v_0)
\nonumber
\\
& &
\times
( {\psi_{1,x} \over \psi_1}
- {\psi_{2,x} \over \psi_2})
\nonumber
\\
& &
+ 2 \alpha^2 k^2 - {k^4 \over 2} - 3 k^3 \mu_0 v_0 - 4 \alpha^2 v_0^2 
- 3 k^2 v_0^2 + v_0^4.
\end{eqnarray}

This defines three different algebraic equations in the unknowns 
$(k,v_0,\mu_0)$; their two solutions
\begin{eqnarray} 
& &
k^2 = 4 \alpha^2,\
v_0 = 0,\
\mu_0 \hbox{ arbitrary},
\\
& &
k^2 = 4 \alpha^2,\
v_0 = 2 \alpha,\
k \mu_0 = - 2 \alpha
\end{eqnarray}
are just two different representations \cite{ConteMusette1993}
of a solution of (\ref{eqErmakovPinney2}) depending on two
arbitrary constants $(\mu_0,x_0)$~:
with this simple assumption, we have obtained the general solution
\begin{equation}
\label{eqErmakovPinney7}
u^{-2} = v =
{1 \over 2 \beta} 
\left[{\psi_{1,x} \over \psi_1} - {\psi_{2,x} \over \psi_2} \right]
= 
{\alpha \over \beta}  {\sinh k a        \over \cosh k(x-x_0) + \cosh k a}.
\end{equation}

In particular, with $\mu_0=0$ one thus obtains immediately the class of 
solutions polynomial in the two variables $\tanh$ and $\sech$
\cite{ConteMusette1992},
thus augmenting the class indicated at the end of previous section.
Evidently, if the DE has only one family, no dependence on $\sech$ can be
found.

\chapter{Conclusion}
\indent

The solution of an ODE cannot escape the structure of singularities of
the ODE.
Such a structure can be studied on the equation itself,
without any {\it a priori} knowledge of the solution,
providing a deep insight on the possibility or not to perform the
explicit integration.

Two levels of integrability have been defined~:
the \Plv\ property (the most elementary level)
and the integrability in the sense of Poincar\'e
(the practical level).

A first series of methods (globally called ``the \Plv\ test'')
provide {\it necessary} conditions for a differential equation to have the 
\Plv\ property,
without any guarantee on the sufficiency.
In case of a negative answer from these first methods,
there exist other methods (\Lecons\ 8, 9, 10, 13, 19),
not developed here,
to provide necessary conditions for the general solution to have only a
finite amount of movable branching,
which implies the integrability in the sense of Poincar\'e,
a weaker property than the PP.

In case of a positive answer,
the DE {\it may} have the PP,
i.e.~a general solution free from movable critical singularities.
Then, a second series of methods are available to perhaps constructively 
prove the PP by explicitly building the general solution or some
equivalent information (Lax pair).
In case of failure of these second methods,
the only remaining tool is human ability.

There exists another approach to DEs which is not based on
the study of singularities,
this is the method of infinitesimal symmetries \cite{Olver,Ovsiannikov}.
It provides reductions of PDEs to ``smaller'' PDEs or to ODEs,
and it may provide first integrals of ODEs.
However, the PDEs or ODEs left over after its completion still require
to be integrated,
and the only methods to do so are those based on singularities.
For instance, with the ODE (P1),
the method of symmetries cannot provide any information
(existence or not of a first integral, single valuedness or multivaluedness).



\printindex 

\vfill \eject
\end{document}